\documentclass[oldversion]{aa}  
\usepackage{color}
\usepackage{graphicx}
\usepackage{amsmath}
\usepackage{txfonts}
\usepackage{natbib}
\usepackage{url}
\usepackage{xspace}
\bibpunct{(}{)}{;}{a}{}{,}
\usepackage{lscape}

\newcommand{\teff}{$T_{\rm eff}$}
\newcommand{\gta}{\lower 0.5ex\hbox{$ \buildrel>\over\sim\ $}}
\newcommand{\lta}{\lower 0.5ex\hbox{$ \buildrel<\over\sim\ $}}
\newcommand{\nhe} {$N$({\rm He})/$N$({\rm H})}
\newcommand{\msun}{$M_{\rm \odot}$}

\newcommand{\omcen}{\object{$\omega$\,Cen}}
\newcommand{\forsmb}{FORS\textunderscore MB}
\newcommand{\kms}{km\,s$^{-1}$}
\newcommand{\omc}{\mbox{$\omega$ Cen}} 

\usepackage{lineno}
\usepackage{ulem} 

\begin{document} 

\title{SHOTGLAS I: The ultimate spectroscopic census of extreme horizontal branch stars in $\omega$\,Centauri 
\thanks{Based on observations collected at the European Organisation for Astronomical Research in the Southern Hemisphere, Chile (proposal IDs 076.D-0810 (\forsmb), 075.D-0280(A), 077.D-0021(A) (FLAMES), 386.D-0669, 091.D-0791 (FORS2.6), 093.D-0873(A), 095.D-0238(A) (VIMOS), 081.D-0139(A) (FORS1.6)} }

\author{Marilyn Latour\inst{1,2}, Suzanna K. Randall\inst{3},  Annalisa Calamida\inst{4}, Stephan Geier\inst{5}, and
            Sabine Moehler\inst{3}}
     
\institute{ Institute for Astrophysics, Georg-August-University, Friedrich-Hund-Platz 1, 37077 G\"{o}ttingen, Germany, \email{marilyn.latour@uni-goettingen.de}
\and Dr. Karl Remeis-Observatory \& ECAP, Astronomical Institute, Friedrich-Alexander University Erlangen-N\"{u}rnberg, 
Sternwartstr. 7, 96049 Bamberg, Germany, 
\and ESO, Karl-Schwarzschild-Str. 2, 85748 Garching bei M\"{u}nchen, 
Germany 
\and Space Telescope Science Institute, 3700 San Martin Drive, 
Baltimore, MD 21218, USA
\and Institut f\"{u}r Physik und Astronomie, Universit\"{a}t Potsdam, Karl-Liebknecht-Str. 24/25, 14476, Potsdam, Germany
}

\date{Received 30 March 2018 ; accepted 19 June 2018}

\abstract{
The presence of extreme horizontal branch (EHB) and blue hook stars in some Galactic globular clusters (GGCs) constitutes one of the remaining mysteries of stellar evolution. While several evolutionary scenarios have been proposed to explain the characteristics of this peculiar population of evolved stars, their observational verification has been limited by the availability of spectroscopic data for a statistically significant sample of such objects in any single GGC. We recently launched the SHOTGLAS project with the aim of providing a comprehensive picture of this intriguing stellar population in terms of spectroscopic properties for all readily accessible GGCs hosting an EHB.

In this first paper, we focus on \omcen, a peculiar, massive GGC that hosts multiple stellar populations. We use non-LTE model atmospheres to derive atmospheric parameters (\teff, log $g$ and $N$(He)/$N$(H)) and spectroscopic masses for 152 EHB stars in the cluster. This constitutes the largest spectroscopic sample of EHB stars ever analyzed in a GGC and represents $\approx$~20\% of the EHB population of \omcen.  
We also search for close binaries among these stars based on radial velocity variations. 

Our results show that the EHB population of \omcen\ is divided into three spectroscopic groups that are very distinct in the \teff\ $-$ helium abundance plane. 
The majority of our sample 
consists of sdOB stars that have roughly solar or super-solar atmospheric helium abundances. It is these objects that constitute the blue hook at $V >$ 18.5 mag in the \omcen\ color-magnitude diagram. Interestingly, the helium-enriched sdOBs do not have a significant counterpart population in the Galactic field, indicating that their formation is dependent on the particular environment found in \omcen\ and other select GGCs. 

Another major difference between the EHB stars in \omcen\ and the field is the fraction of close binaries. From our radial velocity survey we identify two binary candidates, however no orbital solutions could be determined. We estimate an EHB close binary fraction of $\approx$~5\% in \omcen. This low fraction is in line with findings for other GGCs, but in sharp contrast to the situation in the field, where around 50\% of the sdB stars reside in close binaries.

Finally, the mass distribution derived is very similar for all three spectroscopic groups, however the average mass (0.38~\msun) is lower than that expected from stellar evolution theory. While this mass conundrum has previously been noted for EHB stars in \omcen, it so far appears to be unique to that cluster.
}

\authorrunning{M. Latour et al.}  
         
\titlerunning{The ultimate spectroscopic census of Extreme Horizontal Branch stars in \omcen}
 \maketitle
%

\section{Introduction}
Galactic globular clusters (GGCs) were long considered simple stellar populations. However, in the last decade, spectroscopic and photometric studies demonstrated that these stellar systems are much more complex. 
Almost all the GGCs studied so far host multiple generations of stars, showing more or less pronounced light-element abundance enhancements and anti-correlations (O-Na, Mg-Al) \citep{carretta2010,gratton2012}. 

The peculiar GGC \omc~ (NGC\,5139) is the most massive known in our Galaxy, with $M = 2.5 \times 10^6$ \msun\ \citep{vandeven2006}. It not only shows the light-element abundance enhancements and anti-correlations typical of a cluster, but also hosts (at least) three separate stellar populations with a large undisputed spread in metallicity \citep{norris_dacosta1995, norris1996, suntzeff1996, kayser2006, villanova2007, calamida2009, johnson2010}. Another peculiar property of \omc~ is the splitting of the main-sequence (MS). Hubble Space Telescope (HST) photometry revealed that the \omc~ MS bifurcates into two main components, the so called blue-MS (bMS) and the red-MS (rMS) \citep{anderson2002,bedin2004}. A spectroscopic follow-up study by \citet{piotto2005} showed that bMS stars are more metal-rich than rMS stars. These authors then suggested that bMS stars constitute a helium-enhanced sub-population in the cluster due to their bluer colors compared to the more metal-poor rMS stars. The presence of stellar sub-populations with different metallicities and helium abundances in a cluster also affects more advanced stellar evolutionary phases, and influences the morphology of the core-helium-burning horizontal branch (HB) and its blue extension, the extreme horizontal branch (EHB). The occurrence of EHB stars in \omc~ (as well as in other massive GGCs with complex multiple populations, such as NGC~2808) cannot be explained by canonical evolution \citep{dcruz96}. 

EHB stars are low-mass objects ($M$ $\sim$0.5 \msun) with effective temperatures \teff\ $\ga$ 20\,000~K.
Among the Galactic field population, they are usually referred to as hot subdwarf (or hot subluminous) stars. They are classified according to spectral type into two main categories, sdB and sdO. The transition between the types takes place at effective temperatures around 38\,000~K, above which the \ion{He}{ii} lines prominently appear in the optical spectra. The stars in this transition region are sometimes referred to as sdOB (see also Sect. 2.2 of \citealt{heb09} for more details on the spectral classification of hot subdwarfs). The high temperature of these stars is associated with the fact that we are essentially observing a naked He-burning core, since the surrounding hydrogen-rich envelope is extremely thin ($M$ $<$ 0.02 \msun, not massive enough to sustain significant hydrogen shell burning). 

A key aspect in the evolutionary history of any hot subdwarf star is the loss of almost all of its hydrogen envelope. The generally accepted scenario is that this important mass loss happens during the red giant branch (RGB) phase prior to the helium flash \citep{fau72}. However, the physical reasons behind the unusually high mass loss are not fully understood. One possible way to strip the outer hydrogen layers of an EHB progenitor is via binary interactions (including a common-envelope and/or Roche-lobe overflow phase), during which the companion accretes a significant amount of the hot subdwarf progenitor's mass \citep{han02,han03}. Such scenarios are successful at explaining the origin of a large fraction of field sdBs; indeed, about half of them are found in close binary systems \citep{max01,nap04,cop10}, and others ($\sim$20$-$30\%) reside in wider systems \citep{ferg84,stark03} with periods longer than 500 days \citep{vos17}.

However, the situation is more complicated for single sdB stars. Evolutionary models can produce such objects when the mass loss on the RGB is artificially increased,\footnote{This is done by increasing the mass-loss efficiency parameter $\eta$ in the Reimers formula $\dot{M}=-4 \times 10 ^{-13} \eta \frac{RL}{M}$.} but the physical mechanisms able to produce the necessary strong mass loss are not well understood \citep{dcruz96,bro01}. Identifying a suitable mechanism for enhanced mass loss in single stars is especially important when considering the formation of EHB stars in globular clusters, where binaries are notoriously rare \citep{moni10}. So far, only one binary candidate has been spectroscopically confirmed in NGC~6752 \citep{moni15}. Single sdB stars can also be formed via the merger of two low mass He-WDs \citep{han02,han03}. This formation channel is particularly relevant for hot subdwarfs in globular clusters, as it is expected to be dominant in old ($>$ 10 Gyr) stellar populations \citep{han08}.

\omc~ hosts a large and complex EHB population. The HB not only extends to very blue colors, but also to magnitudes fainter than that of the canonical EHB in the color-magnitude diagram (CMD) \citep{whit94,dcruz00,bro01}. The special population found below the EHB is termed the ``blue hook" due to its characteristic shape in the ultraviolet CMD. Spectroscopic observations of blue hook stars in \omc~ revealed that they have an atmosphere enriched in helium, as well as carbon \citep{moe07, moe11,lat14}. The origin of this peculiar population is the subject of many debates and various scenarios have been proposed to explain its existence.

One scenario suggests that the blue hook stars in \omc~ are the progeny of a second generation of stars enriched in helium (Y $\approx$ 0.4) \citep{lee05,dan05,dan10}. The progenitors of these stars should populate the cluster bMS, a possibly helium enhanced stellar sub-population. The higher helium content of this generation of stars could explain the atmospheric helium enhancement of the blue hook objects as well as their lower luminosity. However, \citet{yar17} showed that even when taking into account the proposed helium enhancement, an increased mass loss on the RGB is additionally needed in order to populate the very hot end of the EHB. 

Another proposed scenario predicts that some stars experience the helium flash only after having evolved away from the RGB \citep{cas93,dcruz96,lanz04}. A consequence of this delayed flash is extra mixing between the helium-rich material in the core and the hydrogen-rich superficial layers, producing an atmosphere not only enriched in helium, but also in carbon. The carbon enrichment observed in the spectra of blue hook stars in \omcen\ supports this scenario. However, the late-flash models predict surface abundances of helium and carbon higher than those measured \citep{bro01,cass03,mil08}, and diffusion effects such as gravitational settling must be taken into account in order to reconcile the observed and predicted abundance \citep{ung05}. 

The main deficiency of both scenarios is that they do not explain the physical mechanism behind the enhanced mass loss required on the RGB. Lei et al. (\citeyear{lei13,lei15,lei16}) investigated possible mechanisms and showed that if the star is initially part of a wide-binary system (P $\sim$900-4600 d), the mass loss on the RGB can be tidally enhanced sufficiently to delay the helium-flash. \citet{lei16} estimated that binaries with periods below 1400 d could have survived dynamical encounters during the evolution of \omcen. Dynamical encounters could themselves produce tidal stripping through Roche-Lobe overflow \citep{pas14}.
\citet{sok98} suggested that the presence of planets (within $\la$ 5 AU of the star) may also lead to enhanced mass-loss due to interactions with the envelope during the RGB phase.
An additional factor was considered by \citet{swei97l} and more recently investigated in detail by \citet{tailo15}, whereby rotation increases the core mass of EHBs and the mass-loss along the RGB, shifting the EHB stars toward brighter magnitudes and bluer colors. This happens because rotation tends to cool down the interior of the star and subsequently slow down the evolution along the RGB and delay the Helium flash. Invoking a He-enhanced (0.35 $\leq Y \leq$ 0.38), metal poor (0.0006 $\leq Z \leq$ 0.001) progenitor population making up 24\% of the cluster population, \citet{tailo16} were then able to use population synthesis models to reproduce the EHB and blue hook of \omcen\ quite successfully.

Understanding the formation of EHB stars in GGCs remains a challenge. It is now important to improve the observational constraints on EHB stars in GGCs, and to obtain statistically significant samples of EHBs with both photometric and spectroscopic data available. We recently started a long-term project, SHOTGLAS\footnote{Spectroscopy of HOT GLobular cluster Aging Stars}, aimed at characterizing the origin of EHB stars in GGCs. In this first SHOTGLAS paper we paint the most comprehensive picture to date of the properties of EHB (and blue hook) stars in the peculiar GGC \omc. While atmospheric parameters for EHB stars in \omc~ have already been derived in previous studies, the number of objects included in those samples was rather limited (between 35 and 45 individual stars in \citealt{moe11}, \citealt{moni12}, \citealt{lat14}, and \citealt{ran16}). Of course, the results of these individual studies can be pooled to yield a larger total sample, however the latter will be somewhat inhomogeneous since the parameters derived suffer from systematics caused by the use of different spectrographs and model atmospheres \citep{moni12,lat14}.

In this work, we present our analysis of previously unpublished spectra obtained with the FORS1 and VIMOS spectrographs at the Very Large Telescope (VLT, ESO). We supplement these new samples with a re-analysis of all relevant previously published spectroscopic samples of EHB stars in \omcen\ (see above), which comprise spectra obtained with the FORS2 and FLAMES spectrographs at the VLT. While we cannot avoid the systematics caused by the use of different instruments, by doing the analysis with the same model atmospheres and fitting technique we ensure that the results are as consistent as possible. This way we derive atmospheric parameters for the largest sample of EHB stars ever analyzed in a single GGC. We note that the term EHB is used quite loosely in this paper to refer to the bluest and faintest morphological part of the CMD, including the blue hook region. Our EHB sample includes canonical EHB stars in their helium-core burning evolutionary phase, as well as late-flasher and post-EHB objects.

In addition to the spectroscopic study, we search for close binaries among the EHB stars in \omcen\ observed with the VIMOS spectrograph. Our VIMOS observations were explicitly designed for a radial velocity (RV) analysis, and therefore the spectra were obtained over multiple epochs. While similar RV surveys have been conducted before among HB stars in a few globular clusters (e.g., NGC~6752, NGC~5986, NGC~2808, and M~80; Moni Bidin et al. \citeyear{moni06,moni09,moni11binary}), searches for EHB binaries in \omcen\ \citep[e.g.,][]{moe11} have so far been inconclusive. However, \citet{kal07} found a peculiar post-common-envelope eclipsing binary in \omcen, slightly redder than the blue HB in the cluster CMD, where the secondary less luminous component is a very low mass (0.14 \msun) pre-helium-core white dwarf.

The paper is structured as follows: the spectroscopic observations are described in Sect. 2, while the methods used to derive radial velocities, atmospheric parameters, and stellar masses are explained in Sect. 3. This is followed by our results on the atmospheric parameters, mass and radial velocity distributions, and the binary fraction in Sect. 4. We then discuss the results in Sect. 5 before summarizing and concluding.

\section{Observational material}

\begin{figure*}
\begin{center}
\includegraphics{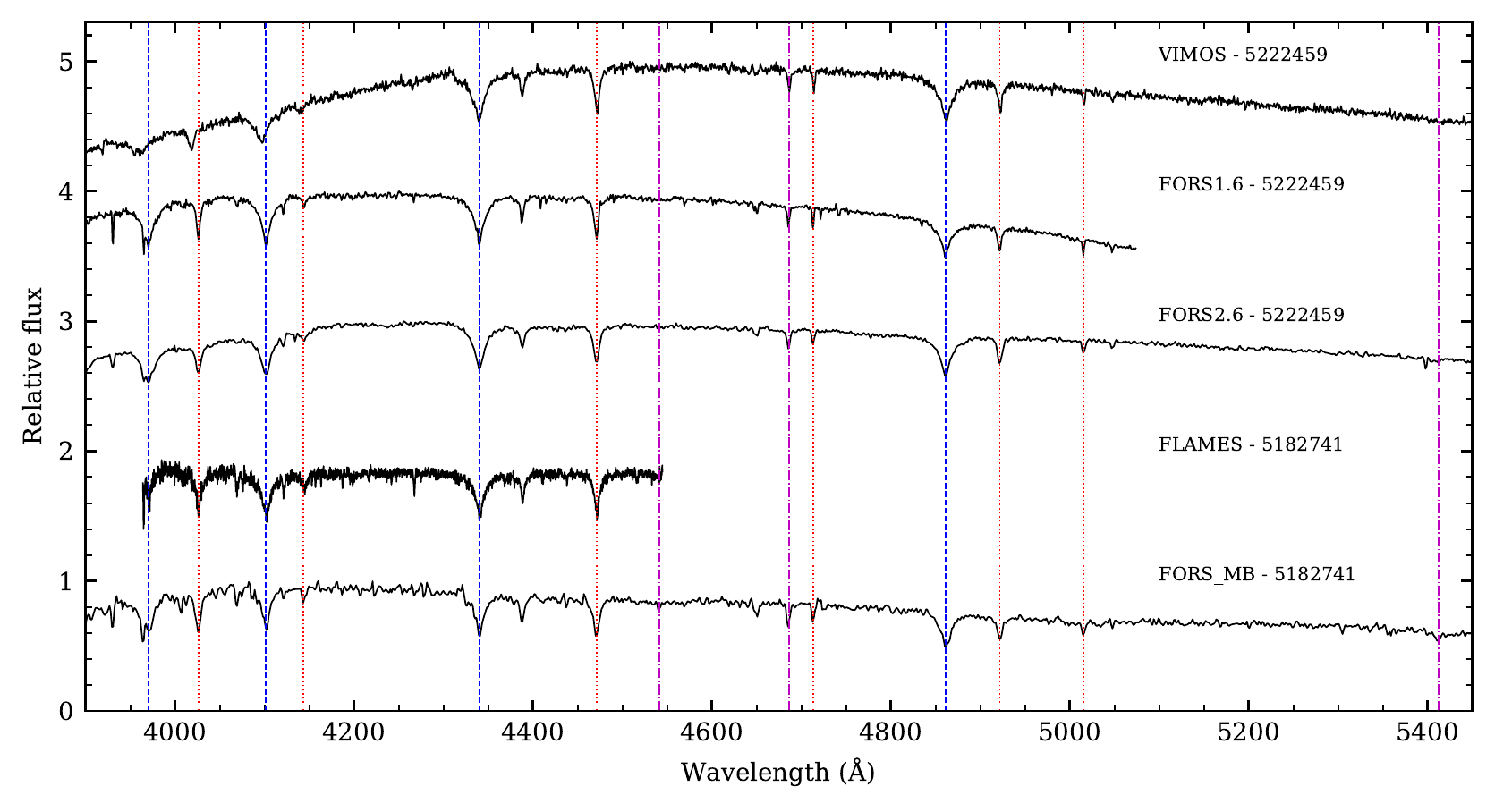}
\caption{Observed spectra from the different samples for two stars : 522459 from the VIMOS, FORS1.6, and FORS2.6 samples, and 5182741 from the FLAMES and \forsmb\ samples. The spectra have been corrected for radial velocity and have been shifted to rest wavelengths. The position of the main Balmer (dashed), \ion{He}{i} (dotted), and \ion{He}{ii} (dash-dotted) lines are indicated. The VIMOS spectrum of 522459 suffers from an inaccurate wavelength calibration blueward of 4300 \AA. }
\label{pspec}
\end{center}
\end{figure*}

\subsection{The VIMOS data}
The candidate EHB sample for the VIMOS observations was selected in the same way as the original EFOSC2/FORS2 sample of \citet{ran16}. Based on the merged ACS/WFI catalog presented in \citet{cast07}, we selected EHB stars from ACS where available (using a cut in magnitude of 17.8 $< F435W <$ 19.8 mag and in color of $-$0.3 $< F435W - F625W <$ 0.2 mag), and from WFI in the outer regions of the cluster not covered by ACS (using a cut in magnitude of 17.8 $< B <$ 19.8 mag and a cut in color of -0.3 $< B-V <$ 0.2 mag). The resulting EHB candidates were overplotted on the VIMOS mask centered on \omcen. The pointing was adjusted so that the largest possible number of EHB targets fell into the four VIMOS quadrants and then the slit assignment was done automatically by the VMMPS software. Targets identified as being of special interest from the \citet{ran16} sample (the pulsators and those EHB stars with the highest helium abundances; 12 stars in total) were given a higher priority in the FIMS fitting algorithm than the remaining EHB candidates. After the automatic slit assignment each target was checked by eye for apparent crowding using the pre-imaging, and targets that were severely crowded especially by bright nearby stars were discarded. Additional slits were then added manually for other EHB candidates not selected by the algorithm, the sole criteria being an acceptable level of crowding and the geometrical restrictions posed by the VIMOS mask specifications. A total of 102 stars were selected this way.

Since the main aim of the VIMOS observations was to find close EHB binaries based on relative radial velocity shifts on a time-scale of hours, the individual exposure times were kept to 10 minutes, which we deemed a reasonable compromise between getting sufficient S/N and not smoothing any radial velocity shifts too much. A group of three exposures were consecutively executed in an hour-long Observing Block, each Observing Block being observable independently to facilitate scheduling at the VLT.
We obtained a total of 42 useful exposures at 14 epochs between June 2014 and February 2016, obtaining spectra of widely varying quality, depending on the observing conditions. Please see Table \ref{obsvimos} for a log of the observations. We used the VIMOS MOS setting with the HR Blue grism and a slit width of 0.8\arcsec throughout.

The observations were reduced with the VIMOS pipeline, using the default parameters except for using a linear fit instead of a quadratic fit to describe the slit curvature.
We encountered severe issues with the wavelength calibration, especially for the VIMOS quadrants 2 and 3, due to the dearth of lines toward the blue end of the arc lamp spectrum. This is a common problem for VIMOS spectra taken with the HR blue grism. The net result was that the pipeline did not manage to find a wavelength solution for a significant number of spectra,  which precluded them from being extracted. For some of the spectra that {\it were} successfully extracted, we found the wavelength calibration to be inaccurate blueward of 4300 \AA\ based on the wavelength solution for the arc spectra. Redward of that the wavelength solution looked fine, therefore we have reason to believe the wavelength calibration of the science spectra is also satisfactory in that regime.

Given the crowdedness of the field and the strong variations in atmospheric conditions from one epoch to the next all the science target spectra extracted by the pipeline had to be identified, inspected and evaluated individually. Only spectra that showed characteristics of an EHB star (broad Balmer lines, otherwise relatively featureless spectrum apart from some He lines, continuum distribution characteristic of a blue star) were kept for the radial velocity analysis (see Sect. 3.1). This visual inspection may have also excluded binaries with a very bright, early main sequence companion; however since such binary systems would be expected to have long orbital periods on the scale of months or years our observations would not have been sensitive to them. In total, our VIMOS RV sample encompasses 75 stars, for each of which anywhere between 3 and 40 usable spectra were obtained. 

For the analysis of the stellar atmospheric parameters the bar was set higher in terms of quality. Only individual spectra with a S/N $\gtrsim$ 20 and free of visible pollution or artifacts were combined using the IRAF scombine task. The combined spectra were discarded if they presented obvious signs of pollution by neighboring stars. This implies that EHB stars with a main sequence binary companion typical of the long-period binaries in the field would also have been excluded from the spectroscopic sample. This yielded a spectroscopic sample of 67 stars for which we obtained averaged spectra with typical S/N $\sim$40-60. The VIMOS spectra have a wavelength resolution of $\Delta\lambda$ $\sim$1.6 \AA\ and cover the 3500 - 5500 \AA\ range, but some of them are truncated due to the position of the star on the VIMOS quadrant. The full wavelength range allows fitting of the Balmer lines from H$_{11}$ to H$_{\beta}$ as well as some \ion{He}{i} and \textsc{ii} lines. For stars with wavelength calibration problems toward the blue end we fitted only the lines redward of 4300 \AA.
The 67 stars selected for atmospheric parameter analysis are referred to as the VIMOS sample in the remainder of this paper.


\subsection{The FORS data}

We obtained spectra of EHB stars in \omcen\ in service mode in April and May 2008 using the FORS1 spectrograph on the UT1 Telescope of the VLT. We used the 1200B grism in MOS mode with slit widths of 0.5\arcsec\ and exposure time of 2700 s. The stars were observed during seeing conditions of 0.8\arcsec\ or better. The observation log is presented in Table~\ref{obsfors}. The data reduction was performed using the FORS pipeline\footnote{version fors1/4.2.5} up to the stage of obtaining rectified, wavelength-calibrated and rebinned frames. The sky background was then manually fitted and removed and the subsequent spectral extraction was performed using standard MIDAS routines. The individual spectra have S/N $\sim$40-50.
Some stars were observed two or three times, in which case the individual spectra were combined together using the IRAF scombine task. The spectra have a resolution of $\Delta\lambda$ $\sim$1.6 \AA. Although the nominal wavelength coverage is $\approx$3650$-$5200 \AA, many of the spectra are truncated at the blue or red end depending on their position on the CCD. Nevertheless, the majority of the spectra could be fitted up to the Balmer line H10 (3797.9 \AA) in the blue and have sufficient wavelength coverage to include \ion{He}{i} lines and \ion{He}{ii} $\lambda$4686. We obtained spectra of 21 stars, however two of them could not be used due to poor wavelength calibration affecting most of the spectral range.
 
Because the main goal of these observations was to measure carbon and nitrogen abundances in blue hook stars, the targets were selected from the faintest part of the EHB. The selection was based on the position of the blue hook in the $U$ vs $U - V$ CMD using WFI photometry \citep{cast07}. The blue hook region was defined using the position of EHB stars in \omcen\ with a helium abundance known to be close to or above the solar value (from \citealt{moe07}). 
This corresponds roughly to 18.4 $<$ $V$ $<$ 19.3 and $-$1.8  $< U - V <$ $-$1.2 mag in our Fig. 7. 
The 19 stars from this sample are labeled FORS1.6 (where 1.6 refers to the spectral resolution) in the remainder of this paper.

\subsection{Previously published data}

We complemented the new spectroscopic observations described above with previously published spectra from the samples of \cite{moe11}, \cite{moni12} and \cite{ran16}. The observations are described in detail in the respective publications, but we briefly summarize the main characteristics of these three additional samples here. 

From the EHB spectra presented in \cite{ran16} we selected the 38 ``clean'' spectra that were analyzed in \cite{lat14}. These spectra do not show signs of pollution by a main sequence companion or nearby star. The spectra were collected with the FORS2 spectrograph using the multi-slit (MXU) mode and the 600B grating. They have a wavelength resolution of $\sim$2.6 \AA\ and nominally cover the 3400$-$6100 \AA\ range, however some spectra are shortened at one end due to their position on the CCD. For most of the stars in this sample, the lines of the Balmer series (from H$\beta$ up to H11) as well as the strong \ion{He}{i} and \textsc{ii} lines from \ion{He}{i} $\lambda$4026 to \ion{He}{ii} $\lambda$5412 are available for constraining the atmospheric parameters. Since this spectroscopic sample was obtained to provide a mapping of the instability strip, the target selection favored the hotter part of the EHB domain (\teff\ $\ga$ 30\,000 K). In the following, we refer to this sample as FORS2.6 (where 2.6 refers again to the spectral resolution). 

The spectra analyzed in \cite{moni12} were also collected with the FORS2 spectrograph in MXU mode with the 600B grating. Their resolution and wavelength coverage are essentially the same as that of the FORS2.6 sample. The authors obtained spectra of stars covering the whole blue part of the cluster's HB. From their whole sample we selected the 37 targets with \teff\ $>$ 20\,000 K as determined by \cite{moni12}. We further inspected the spectra and found that four of them (91573, 97034, 157531, 175847)\footnote{IDs used in \cite{moni12}} show conspicuous signs of pollution by a cooler object according to the criteria described in \cite{ran16}. Thus, we kept 33 spectra from that sample, which we will refer to as FORS\textunderscore MB. 

Finally, the spectra from \cite{moe11} were obtained using the multi-object fiber spectrograph FLAMES$+$GIRAFFE on the VLT. These spectra have a resolution of $\sim$0.7 \AA\ and a wavelength coverage of 3964$-$4567 \AA. This shorter spectral range offers a more limited set of spectral lines that can be used to derive atmospheric parameters; the two Balmer lines H$\gamma$ and H$\delta$ and three \ion{He}{i} lines, 4026 \AA, 4388 \AA, and 4471 \AA. 
\cite{moe11} observed stars along the blue HB of \omcen\ and we selected from that sample the 45 objects with \teff\ $>$ 20\,000 K. These spectra will be referred to as the FLAMES sample. 

\begin{figure}
\begin{center}
\resizebox{\hsize}{!}{\includegraphics{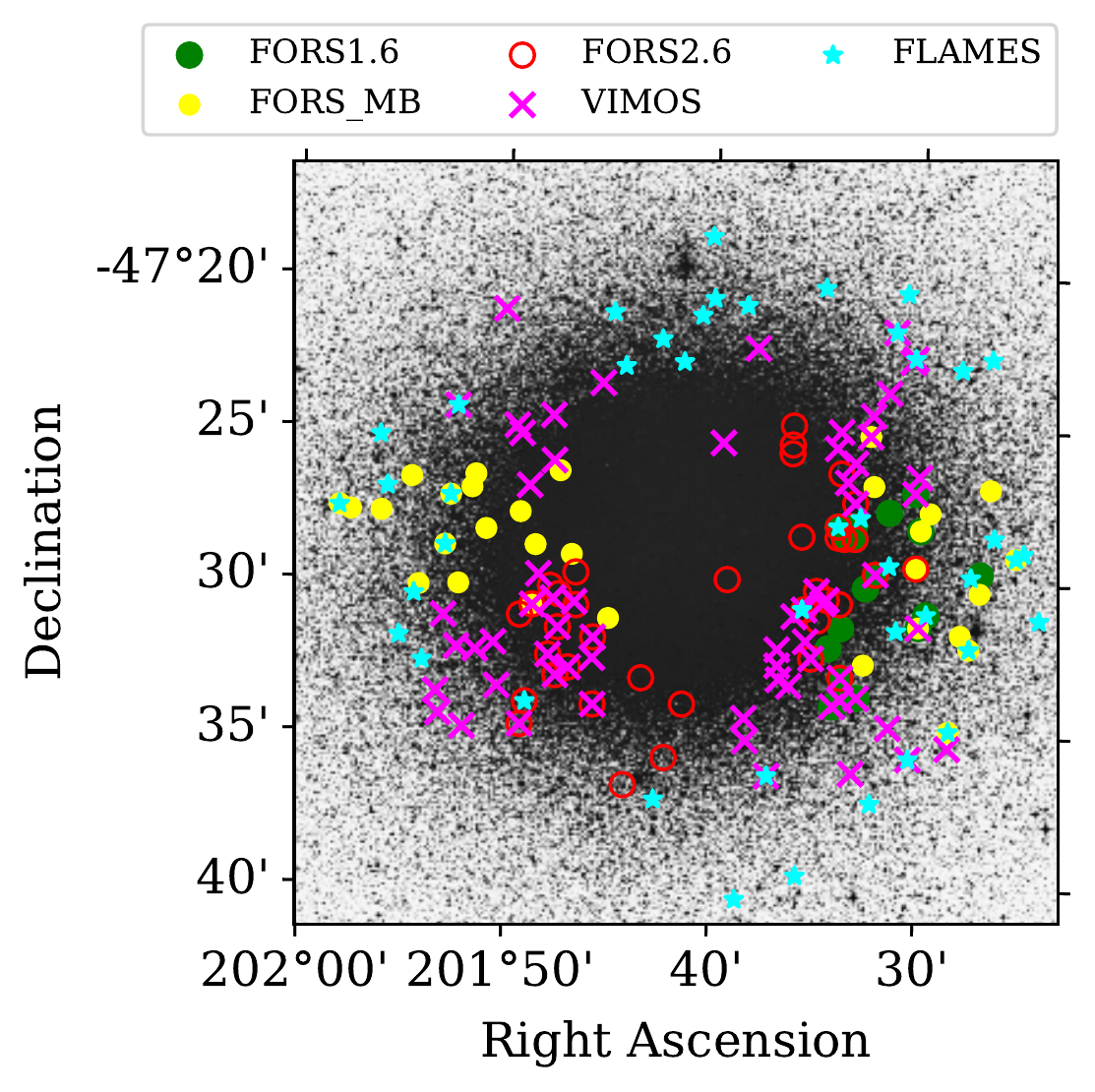}}
\caption{Spatial distribution of the stars included in the spectroscopic sample.}
\label{spdist}
\end{center}
\end{figure}

Fig. \ref{pspec} shows a representative spectrum from each of the five observed samples (VIMOS and FORS1.6 being the new observations, FORS2.6, FORS\textunderscore MB and FLAMES constituting the previously published data). The spectra were shifted to rest wavelengths, and the positions of the Balmer and helium lines are indicated. We note that the VIMOS spectrum shown in this figure suffers from the wavelength calibration problem mentioned in Sect. 2.1.
The spatial distribution across the cluster of the stars from all five samples can be seen in Fig. \ref{spdist}. The targets are well distributed across the cluster and one can notice that a significant number of stars were included in more than one sample. 

\section{Analysis method}

\subsection{Radial velocity determination}
We measured the radial velocities for the FORS2.6 and FORS1.6 spectra, as well as for the individual VIMOS spectra. The RVs of the \forsmb\ and FLAMES spectra were available from the literature, so we simply retrieved the published values.
For the other samples, the RVs were determined by fitting a set of mathematical functions (Gaussians, Lorentzians, and polynomials) to the spectral lines using the FITSB2 routine \cite{nap04}. Those three functions match the continuum, the line, and the line core, respectively and mimic the typical profile of spectral lines. The profiles are fitted to all lines simultaneously using $\chi^{2}$-minimization and the RV shift with the associated $1\sigma$ error is measured. Heliocentric corrections were applied to the RVs.

For the FORS2.6 and FORS1.6 spectra, we used the strongest Balmer lines (H$\beta$, H$\gamma$, H$\delta$) and, depending on the effective temperature and helium abundance, the \ion{He}{ii} 4686 \AA\ line, and the \ion{He}{i} lines at 4026 \AA, 4472 \AA, and 4922 \AA. We excluded H$\epsilon$ because of the blending with the interstellar \ion{Ca}{ii} H line. 

The VIMOS spectra were collected as part of a RV survey to look for close binary systems among the EHB stars of \omcen, therefore particular care was taken with the wavelength calibration procedure and arc lamp exposures were taken at the end of each one-hour observing block. Unfortunately, due to a lack of appropriate emission lines in the arc lamp spectrum toward the blue end of the VIMOS data, the wavelength calibration in this region is considered to be less reliable. To measure the RVs, we therefore only considered lines with rest wavelengths longer than $\sim4300\,{\rm \AA}$. Depending on the effective temperature and the helium abundance of the individual stars we used the H$\beta$ and H$\gamma$ lines, as well as the previously mentioned helium lines. Each fit was inspected visually and outliers caused by noisy spectra, cosmic rays and other artifacts were excluded. The average $1\sigma$ uncertainty per measurement is about $13\,{\rm km\,s^{-1}}$. Heliocentric corrections were applied to the RVs and mid-JDs. 

In total, we obtained RVs for $\sim$1800 individual VIMOS spectra for 75 different stars. 

One of the advantages of studying stars in a globular cluster is that the radial velocity of the cluster is well known. This information was used to correct for systematic shifts in the VIMOS spectra taken at different nights. We calculated the average RV of all stars observed in one night and compared it to the mean RV of \omcen\ (232.1 \kms, \citealt{har96}, 2010 edition).
Typical deviations from the mean RV of \omcen\ were on the order of $\pm10\,{\rm km\,s^{-1}}$ or less. Only for the last observing night the systematic shift was somewhat higher ($+22\,{\rm km\,s^{-1}}$), but still of the same order as the statistical uncertainties. We then corrected the individual RVs measured for each night by the systematic shift determined.

 According to the VIMOS manual, the internal accuracy of the wavelength calibration is about 0.3 pixels, not considering flexure effects. The HR blue grism we used has a dispersion of $0.51\,{\rm \AA \,px^{-1}}$, which translates to a systematic RV uncertainty of about  $10\,{\rm km\,s^{-1}}$. To also account for some flexure during the observations we added $15\,{\rm km\,s^{-1}}$ in quadrature to the statistical uncertainties. This systematic uncertainty is also consistent with the night-by-night shifts of the average RV we detected. The  uncertainties on the RVs of individual spectra are of importance for the computation of the false-detection probability (see Sect. 4.4).

\subsection{Models and fitting procedure}

We derived the atmospheric parameters of the stars by fitting the observed spectra with the grid of NLTE model atmospheres presented in \citet{lat14}. We recall that these models are computed using the public code TLUSTY and SYNSPEC \citep{lanz07,hub88} and include, besides H and He, a solar amount of C, N, and O as metallic elements. However two small changes were made with respect to the models used in \cite{lat14}.
Firstly, we extended the grid by computing a few additional models at lower temperature and lower log $g$ to provide a better parameter coverage, since some of our observed samples include stars cooler than those analyzed in \cite{lat14}. Our grid now covers a \teff\ range from 20\,000 to 58\,000 K (in steps of 2000 K) and a log $g$ range from 4.8 to 6.4 (in steps of 0.2 dex) for helium abundances log \nhe\ from $-$4.0 to 0.0 dex (in steps of 0.5 dex). Since the parameter ranges at which the helium-enriched stars are found were well covered by the original grid, we did not extend it for the models having log \nhe\ = 0.5, 1.0 and 1.5. These helium-rich models thus cover the original range from 26\,000 to 58\,000 K and log $g$ of 5.2 to 6.4. 
Secondly, we recomputed the synthetic spectra by including only hydrogen and helium lines. Some relatively strong lines of carbon and oxygen are blended with the Balmer (especially H$_\delta$) lines and affect the fitting procedure when they are absent from the observed spectra, especially at the higher resolution of the FORS1.6 and FLAMES spectra. The differences in the atmospheric parameters derived with both sets of synthetic spectra are however small when comparing the results of \cite{lat14} with our new parameters for the FORS2.6 sample. 
We fitted all the spectra included in the five spectroscopic samples with our model grid using the Balmer and helium lines available within the observed spectral range of each spectrum. The lines of both observed and model spectra are normalized and then simultaneously fit using a $\chi^2$ minimization technique similar to that described by \cite{saf94}.

All the individual fits were inspected and we discarded a few cases where the quality of the spectra ($S/N$, reduction artifacts) and the resulting fits were very poor. In the case of stars present in two or three samples, we paid particular attention to discrepant atmospheric parameters from one spectrum to the next, and when the fit to one spectrum was poor we discarded it. The atmospheric parameters of the stars observed in more than one sample were then averaged. After performing this selection, our sample includes 198 individual spectra of 152 distinct stars, of which 40 are included in two or three samples. 
The resulting atmospheric parameters of the stars are presented in Table \ref{bigtable}, and their positions in the various \teff, log $g$, helium planes are presented in Fig. \ref{app_gteff} and \ref{app_he}, color-coded by observed sample as well as in Figs. \ref{teffhe}, \ref{teffgravhe} and \ref{gravhe} color-coded by spectroscopic group (see Sect. 4.2).

\subsubsection{Error estimates on the atmospheric parameters }

Since many stars are present in more than one observed sample, we looked for differences and systematics in the atmospheric parameters derived from the different spectra. An extensive discussion of this is presented in Appendix~\ref{appA} and we report here only some results relevant to our uncertainty estimates for the atmospheric parameters. The formal errors returned by the fitting procedure only provide a lower limit for the uncertainties on the atmospheric parameters. To estimate the true uncertainties associated with the observational data (which are affected by the spectral resolution and wavelength coverage), we computed the ratio of the difference between each of the three atmospheric parameters derived for each pair of spectra of the same star and the corresponding formal error (see Eq. \ref{eq:A1}). If the uncertainties are realistic, this ratio should be normally distributed with a standard deviation of one. For each of the three atmospheric parameters we found a standard deviation larger than one, indicating that - as expected - our formal errors underestimate the true uncertainties. From the standard deviations we estimated correction factors of 2.5 for \teff, 1.6 for log $g$ and 1.8 for log \nhe, and applied these to the statistical errors in order to obtain more realistic uncertainties. These corrected errors are the ones provided in Table \ref{bigtable}. However, in some cases where the formal errors are already large (e.g., on the \teff\ of the hottest stars or on log \nhe\ of stars with a low helium abundance) such a correction might overestimate the uncertainties.

\subsection{Mass measurements}

Masses for the stars in our sample were computed by combining synthetic magnitudes with the observed WFI magnitudes. 
We first created grids of synthetic magnitudes ($m$) in the $V$ and $B$ band using spectra from our model grid described in Sect. 3.1. Magnitudes relative to the spectrum of Vega were computed using the Python package Pysynphot \citep{lim15} and zero point corrections were applied as described in appendix B of Lim et. al, but using the $V$-band correction of \citet{Boh07}.
Our synthetic fluxes are expressed in terms of the Eddington flux, $H_{\lambda}$, and since the synthetic flux is independent of the radius, our synthetic magnitudes $m$ are related to the absolute magnitudes ($M$) via the relation
   \begin{equation} \label{eq:magmod}
     M = m - 2.5 \log \left( \frac{4\pi R^2}{d^2} \right)
   \end{equation}
where $R$ is the radius of the star and $d$ the distance of 10 pc. 
We used the \teff, log $g$ and log $N$(He)/$N$(H) derived for every star to retrieve the appropriate synthetic magnitude from our grid using trilinear interpolation. 
 
In a second step, we rewrite Eq. \ref{eq:magmod} in terms of the stellar mass ($M_*$) using the relation 
   \begin{equation} \label{eq:r}
    R^2 = \frac{GM_*}{g}
   \end{equation}
and combine it with the equation of the true distance modulus ($\mu_0$)
   \begin{equation} \label{eq:dist}
    \mu_0 = V - M_V - A_V 
   \end{equation}
where $V$ is the observed WFI magnitude, and $A_V$ the visual extinction defined as 3.1$E(B-V)$. Finally, we derived the stellar masses (presented in Table \ref{bigtable}) using a distance modulus $\mu_0$ = 13.71$\pm$0.09 \citep{bra16} and a reddening index $E(B-V)$ = 0.11$\pm$0.01. The uncertainties on the masses were computed via error propagation and we considered the uncertainties on $\mu_0$, $E(B-V)$, log $g$ and the synthetic flux $m$. The uncertainty on $m$ was obtained by considering the uncertainty on \teff, which is the parameter dominating the emergent flux. 
We also estimated the reddening of each target by comparing the observed ($B-V$) colors with the theoretical colors computed from our synthetic $B$ and $V$ magnitudes, and report the values in the last column of Table~\ref{bigtable}.

\section{Results}

\subsection{RV distribution}

\begin{figure}
\begin{center}
\resizebox{\hsize}{!}{\includegraphics{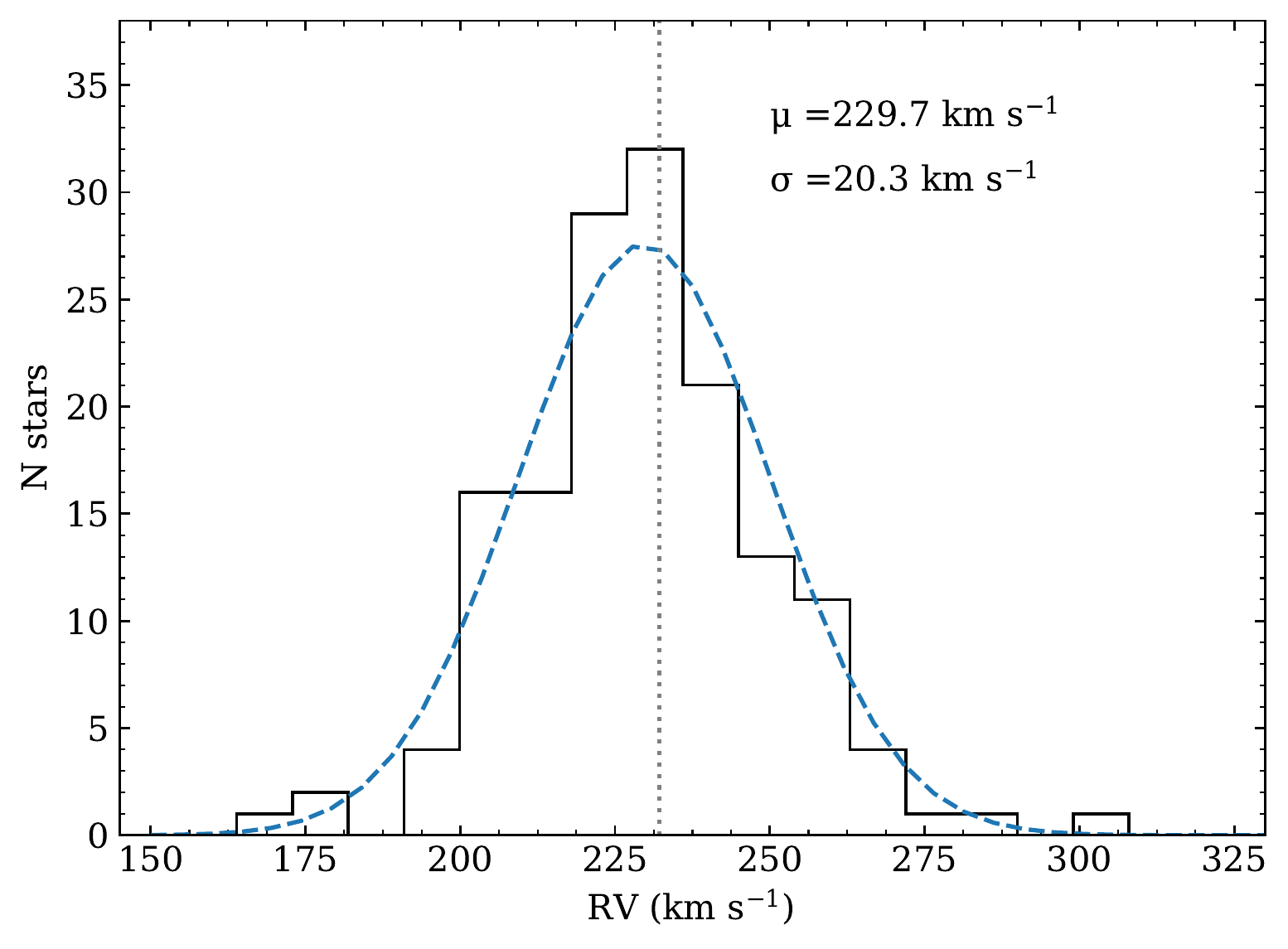}}
\caption{Distribution of heliocentric radial velocities for the 152 stars of our sample. The dashed curve shows the resulting best-fitting gaussian to our data and the dotted line indicates the RV of the cluster (232.1 \kms, \citealt{har96}, 2010 edition).  }
\label{histrv}
\end{center}
\end{figure}

In addition to the radial velocities measured for the VIMOS, FORS2.6 and FORS1.6 spectra, we collected the values presented in the literature for the \forsmb\ and FLAMES spectra \citep{moni12,moe11}. For the VIMOS spectra, we used the average RVs presented in Table \ref{tabrvvimos}.
For the stars found in more than one sample, we computed the average RV and used it as the final, representative RV. The resulting RVs for all stars are found in Table \ref{bigtable}. 
For stars with more than one RV value (from different spectra), we also provide the standard deviation ($\sigma$) of the individual measurements. This is meant to indicate how well, or not, the individual measurements agree with each other. 
The average RV of the whole sample is 229.7 \kms, the observed dispersion (one standard deviation) is $\sigma$ = 20.3 \kms, and the standard error on the average is 1.6 \kms. Figure~\ref{histrv} shows the RV distribution as well as a gaussian curve with the mean and dispersion indicated in the caption. The mean RV of our sample is quite close to that of the cluster (232.1 \kms, \citealt{har96}, 2010 edition). This is a good agreement, considering that our spectra have only moderate resolution, and that the RV measurements mostly rely on wide or weak spectral features.

A few stars in our sample have RVs rather far from the average value: seven stars lie outside the 2$\sigma$ interval, which is consistent with the expected value of 5\% of the sample for a normal distribution. However, two stars lie outside the 3$\sigma$ interval. As some of these stars might not be members of the cluster, we looked at their position in the $V$ vs $U - V$ CMD and in the $F435W$ vs $F435W - F625W$ CMD (for stars with ACS photometry), and at their derived masses (since a star at a different distance would present an anomalous mass). We did not find the ``outliers'' to have peculiar colors or masses in comparison with the rest of the sample, except for one star, 5062474, which has the lowest RV (164 \kms) as well as the reddest colors in the WFI and ACS CMD, and for which we derived a rather high mass (0.72 $\pm$ 0.38 \msun). The observed spectrum is not particularly good (see Fig.~1 in \citealt{lat14}), thus explaining the large uncertainty on the mass. Although this star has some peculiar properties, we nevertheless keep it in our sample.

\longtab{
\begin{landscape}
\scriptsize
\begin{longtable}{rcccccccccccccccc}
\caption{\label{bigtable}Spectroscopic sample of EHB stars in \omcen.}\\ 
\hline\hline
ACS ID & RA & DEC & Sample\tablefootmark{a} & \teff \tablefootmark{b} & log $g$\tablefootmark{b} & log \nhe \tablefootmark{b} & $U$ & $B$ & $V$ & $I$ & ID (1) & ID (2) & RV\tablefootmark{c} &  $\sigma_{RV}$ & Mass & E(B$-$V) \\ 
 &  &  &  & (K) & (cm s$^{-2}$) & (dex) & &  &  & & &  & (\kms) & (\kms)  & (\msun) & \\
\hline
\endfirsthead
\caption{continued.}\\
\hline\hline
ACS ID & RA & DEC & Sample\tablefootmark{a} & \teff\ \tablefootmark{b} & log $g$ \tablefootmark{b} & log \nhe\ \tablefootmark{b} & $U$ & $B$ & $V$ & $I$ & ID (1) & ID (2) & RV\tablefootmark{c}  &  $\sigma_{RV}$ & Mass & E(B$-$V)\\ 
 &  &  &  & (K) & (cm s$^{-2}$) & (dex) & &  &  & & &  & (\kms) & (\kms)  & (\msun) & \\
\hline
\endhead
\hline
\endfoot
    257150 &  201.763229 &  $-$47.542900 & 1 &  27482 $\pm$   927 &  5.49 $\pm$  0.07 & $-$1.40 $\pm$  0.08 &  16.959 &  18.121 &  18.238 &  18.369 & ... & ... & 224 $\pm$  20 & ... & 0.50 $\pm$ 0.11 &  0.11 \\
    264670 &  201.778290 &  $-$47.512329 & 1 &  33767 $\pm$  1420 &  6.15 $\pm$  0.16 & $-$0.16 $\pm$  0.11 &  17.474 &  18.780 &  18.781 &  18.666 & ... & ... & 228 $\pm$  17 & ... & 0.90 $\pm$ 0.34 &  0.24 \\
   5153131 &  201.807617 &  $-$47.497375 & 1 &  36076 $\pm$  1227 &  5.86 $\pm$  0.12 & $-$0.42 $\pm$  0.10 &  17.378 &  18.809 &  18.945 &  18.996 & ... & ... & 258 $\pm$  30 & ... & 0.35 $\pm$ 0.11 &  0.13 \\
   5137388 &  201.795517 &  $-$47.510567 & 1 &  33975 $\pm$  1075 &  6.20 $\pm$  0.12 & $-$0.57 $\pm$  0.10 &  17.675 &  19.128 &  19.203 &  19.115 & ... & ... & 233 $\pm$  18 & ... & 0.68 $\pm$ 0.22 &  0.17 \\
   5180639 &  201.873184 &  $-$47.537510 & 1 &  36497 $\pm$  1050 &  5.87 $\pm$  0.11 & $-$1.35 $\pm$  0.13 &  17.033 &  18.617 &  18.816 &  18.929 & ... & ... & 262 $\pm$  24 & ... & 0.41 $\pm$ 0.11 &  0.07 \\
   5196769 &  201.883987 &  $-$47.520283 & 1 &  56528 $\pm$  3807 &  5.94 $\pm$  0.16 & $-$0.77 $\pm$  0.14 &  16.600 &  18.168 &  18.344 &  18.373 & ... & ... & 250 $\pm$  28 & ... & 0.44 $\pm$ 0.17 &  0.13 \\
   5132323 &  201.887299 &  $-$47.573872 & 1 &  35372 $\pm$  1127 &  5.83 $\pm$  0.12 & $-$0.64 $\pm$  0.10 &  17.090 &  18.738 &  18.772 &  18.834 & ... & ... & 217 $\pm$  35 & ... & 0.40 $\pm$ 0.13 &  0.23 \\
   5179481 &  201.858093 &  $-$47.538803 & 1 &  29293 $\pm$   545 &  5.43 $\pm$  0.05 & $-$2.61 $\pm$  0.16 &  16.503 &  17.812 &  17.959 &  18.133 & ... & ... & 262 $\pm$  16 & ... & 0.50 $\pm$ 0.08 &  0.10 \\
   5183041 &  201.843475 &  $-$47.534912 & 1 &  39786 $\pm$  1475 &  6.02 $\pm$  0.19 &  0.55 $\pm$  0.13 &  17.395 &  18.949 &  19.074 &  19.068 & ... & ... & 222 $\pm$  22 & ... & 0.41 $\pm$ 0.18 &  0.15 \\
   5156440 &  201.840149 &  $-$47.558643 & 1 &  39110 $\pm$  2275 &  6.36 $\pm$  0.19 &  1.17 $\pm$  0.13 &  17.634 &  19.149 &  19.295 &  19.541 & ... & ... & 260 $\pm$  23 & ... & 0.81 $\pm$ 0.38 &  0.11 \\
   5121885 &  201.867691 &  $-$47.581158 & 1 &  43567 $\pm$  1612 &  5.86 $\pm$  0.11 & $-$1.70 $\pm$  0.16 &  16.542 &  17.944 &  18.189 &  18.383 & ... & ... & 242 $\pm$  21 & ... & 0.55 $\pm$ 0.15 &  0.05 \\
   5151410 &  201.889420 &  $-$47.561794 & 1 &  37180 $\pm$  1667 &  5.84 $\pm$  0.16 & $-$0.07 $\pm$  0.11 &  17.405 &  18.924 &  19.056 &  19.162 & ... & ... & 219 $\pm$  23 & ... & 0.29 $\pm$ 0.11 &  0.13 \\
    274052 &  201.797760 &  $-$47.503815 & 3 &  36018 $\pm$  1357 &  5.51 $\pm$  0.14 & $-$0.40 $\pm$  0.11 &  17.301 &  18.725 &  18.823 &  18.767 & ... & ... & 206 $\pm$  18 &  43.4 & 0.18 $\pm$ 0.06 &  0.17 \\
   5165122 &  201.792831 &  $-$47.552532 & 1,3 &  35034 $\pm$   780 &  5.74 $\pm$  0.08 & $-$0.70 $\pm$  0.07 &  17.160 &  18.601 &  18.775 &  19.000 & ... & ... & 234 $\pm$  15 &  16.3 & 0.33 $\pm$ 0.08 &  0.09 \\
   5170422 &  201.782700 &  $-$47.548168 & 1,3 &  38174 $\pm$   880 &  5.81 $\pm$  0.10 & $-$0.73 $\pm$  0.08 &  16.922 &  18.352 &  18.454 &  18.408 & ... & ... & 247 $\pm$  16 &   2.9 & 0.45 $\pm$ 0.11 &  0.17 \\
   5039935 &  201.762848 &  $-$47.531593 & 1,3 &  40481 $\pm$  1470 &  6.12 $\pm$  0.20 &  0.66 $\pm$  0.14 &  16.134 &  19.171 &  19.477 &  20.046 & ... & ... & 236 $\pm$  20 &   3.7 & 0.35 $\pm$ 0.17 & $-$0.03 \\
   5034421 &  201.799026 &  $-$47.541382 & 1,3(V1) &  49041 $\pm$  2400 &  5.98 $\pm$  0.12 & $-$1.70 $\pm$  0.16 &  16.813 &  18.269 &  18.459 &  18.519 & ... & ... & 234 $\pm$  24 &   2.3 & 0.49 $\pm$ 0.15 &  0.11 \\
    281063 &  201.812347 &  $-$47.514126 & 1,3,4(V5) &  51902 $\pm$  4120 &  5.91 $\pm$  0.14 & $-$1.97 $\pm$  0.24 &  16.957 &  18.349 &  18.386 &  18.230 & ... & 156638 & 261 $\pm$  22 &  24.7 & 0.42 $\pm$ 0.16 &  0.27 \\
   5119720 &  201.791504 &  $-$47.526360 & 1,3 &  33458 $\pm$   868 &  5.86 $\pm$  0.09 & $-$0.87 $\pm$  0.09 &  17.203 &  18.695 &  18.856 &  19.095 & ... & ... & 229 $\pm$  17 &  10.7 & 0.45 $\pm$ 0.11 &  0.09 \\
   5141232 &  201.762360 &  $-$47.568230 & 1,3 &  35077 $\pm$   880 &  5.77 $\pm$  0.09 & $-$0.62 $\pm$  0.08 &  17.134 &  18.578 &  18.739 &  18.968 & ... & ... & 212 $\pm$  15 &  27.6 & 0.37 $\pm$ 0.09 &  0.10 \\
   5124244 &  201.821182 &  $-$47.579502 & 1,3 &  38341 $\pm$  1557 &  5.98 $\pm$  0.15 &  0.12 $\pm$  0.10 &  17.250 &  18.892 &  19.033 &  19.193 & ... & ... & 232 $\pm$  16 &  21.8 & 0.40 $\pm$ 0.14 &  0.12 \\
    254318 &  201.757507 &  $-$47.392712 & 1 &  36595 $\pm$   672 &  5.62 $\pm$  0.07 & $-$1.20 $\pm$  0.07 &  17.395 &  18.648 &  18.665 &  18.460 & ... & ... & 221 $\pm$  15 & ... & 0.26 $\pm$ 0.05 &  0.26 \\
    273649 &  201.796875 &  $-$47.410667 & 1 &  35168 $\pm$   732 &  5.64 $\pm$  0.08 & $-$0.87 $\pm$  0.07 &  16.937 &  18.377 &  18.540 &  18.735 & ... & ... & 219 $\pm$  14 & ... & 0.32 $\pm$ 0.07 &  0.10 \\
   5226206 &  201.795822 &  $-$47.435371 & 1 &  33431 $\pm$  1002 &  5.89 $\pm$  0.11 & $-$1.36 $\pm$  0.12 &  17.390 &  18.703 &  18.945 &  19.366 & ... & ... & 203 $\pm$  21 & ... & 0.44 $\pm$ 0.13 &  0.02 \\
   5299498 &  201.826202 &  $-$47.416374 & 1 &  28062 $\pm$   697 &  5.58 $\pm$  0.06 & $-$3.61 $\pm$  0.81 &  17.360 &  18.776 &  18.880 &  19.097 & ... & ... & 269 $\pm$  20 & ... & 0.33 $\pm$ 0.06 &  0.13 \\
   5295674 &  201.822281 &  $-$47.420811 & 1 &  36765 $\pm$  1360 &  5.76 $\pm$  0.12 & $-$0.09 $\pm$  0.09 &  17.378 &  18.933 &  18.999 &  18.966 & ... & ... & 246 $\pm$  14 & ... & 0.26 $\pm$ 0.08 &  0.20 \\
   5370155 &  201.836258 &  $-$47.353096 & 1 &  35745 $\pm$   990 &  5.97 $\pm$  0.10 & $-$0.37 $\pm$  0.08 &  17.501 &  19.019 &  19.163 &  19.342 & ... & ... & 247 $\pm$  18 & ... & 0.38 $\pm$ 0.09 &  0.12 \\
   5094822 &  201.815491 &  $-$47.449116 & 1 &  36861 $\pm$   965 &  5.77 $\pm$  0.10 & $-$0.51 $\pm$  0.08 &  16.868 &  18.330 &  18.391 &  18.373 & ... & ... & 253 $\pm$  16 & ... & 0.46 $\pm$ 0.12 &  0.21 \\
   5307782 &  201.874130 &  $-$47.406052 & 1,5 &  42960 $\pm$   748 &  6.24 $\pm$  0.14 &  0.39 $\pm$  0.12 &  17.015 &  18.599 &  18.678 &  18.686 & 181678 & ... & 241 $\pm$  24 &   4.2 & 0.84 $\pm$ 0.28 &  0.20 \\
    157448 &  201.567551 &  $-$47.426430 & 1 &  31747 $\pm$  2322 &  5.46 $\pm$  0.22 & $-$3.41 $\pm$  0.78 &  16.929 &  18.122 &  18.192 &  18.120 & ... & ... & 229 $\pm$  31 & ... & 0.37 $\pm$ 0.22 &  0.19 \\
   5283552 &  201.552780 &  $-$47.434372 & 1 &  26302 $\pm$  2295 &  5.23 $\pm$  0.16 & $-$2.28 $\pm$  0.12 &  16.588 &  17.827 &  17.989 &  18.078 & ... & ... & 228 $\pm$  23 & ... & 0.37 $\pm$ 0.18 &  0.06 \\
   5338760 &  201.507812 &  $-$47.378567 & 1 &  27099 $\pm$  3950 &  5.50 $\pm$  0.27 & $-$2.01 $\pm$  0.17 &  17.543 &  18.885 &  19.029 &  19.157 & ... & ... & 199 $\pm$  31 & ... & 0.26 $\pm$ 0.21 &  0.08 \\
   5296709 &  201.540955 &  $-$47.419601 & 1,4 &  61960 $\pm$  9822 &  5.98 $\pm$  0.24 & $-$0.58 $\pm$  0.21 &  16.712 &  18.162 &  18.403 &  18.573 & 66703 & 113991 & 223 $\pm$  26 &  37.8 & 0.46 $\pm$ 0.27 &  0.07 \\
   5317711 &  201.526382 &  $-$47.395874 & 1 &  42330 $\pm$  2120 &  6.13 $\pm$  0.23 & $-$0.92 $\pm$  0.15 &  17.250 &  18.688 &  18.898 &  19.101 & ... & ... & 265 $\pm$  14 & ... & 0.55 $\pm$ 0.30 &  0.08 \\
   5097663 &  201.558792 &  $-$47.444893 & 1 &  37962 $\pm$  1875 &  5.68 $\pm$  0.24 & $-$0.17 $\pm$  0.17 &  16.933 &  18.357 &  18.511 &  18.529 & ... & ... & 238 $\pm$  28 & ... & 0.33 $\pm$ 0.19 &  0.12 \\
   5268317 &  201.504288 &  $-$47.450394 & 1,2 &  36018 $\pm$  1383 &  5.71 $\pm$  0.15 & $-$0.56 $\pm$  0.10 &  17.056 &  18.596 &  18.747 &  18.970 & 54377 & ... & 245 $\pm$  16 &  30.3 & 0.31 $\pm$ 0.12 &  0.12 \\
   5276767 &  201.501160 &  $-$47.441540 & 1 &  36537 $\pm$  1455 &  5.98 $\pm$  0.16 & $-$0.19 $\pm$  0.12 &  17.477 &  18.873 &  19.092 &  19.348 & ... & ... & 213 $\pm$  24 & ... & 0.40 $\pm$ 0.15 &  0.04 \\
   5306037 &  201.539047 &  $-$47.408272 & 1 &  36993 $\pm$  1115 &  5.86 $\pm$  0.13 & $-$0.37 $\pm$  0.09 &  17.258 &  18.712 &  18.869 &  18.945 & ... & ... & 212 $\pm$  22 & ... & 0.36 $\pm$ 0.11 &  0.11 \\
   5347296 &  201.633057 &  $-$47.372486 & 1 &  38583 $\pm$  1462 &  5.69 $\pm$  0.18 & $-$0.22 $\pm$  0.13 &  17.439 &  18.765 &  18.758 &  18.883 & ... & ... & 227 $\pm$  11 & ... & 0.26 $\pm$ 0.11 &  0.28 \\
   5262593 &  201.552444 &  $-$47.456291 & 1,3 &  30460 $\pm$  1230 &  5.48 $\pm$  0.11 & $-$2.98 $\pm$  0.30 &  16.546 &  17.810 &  17.938 &  18.028 & ... & ... & 226 $\pm$  21 &  46.4 & 0.53 $\pm$ 0.17 &  0.12 \\
    177238 &  201.603104 &  $-$47.414402 & 3(V3) &  47700 $\pm$  2027 &  6.01 $\pm$  0.11 & $-$1.85 $\pm$  0.20 &  16.825 &  18.328 &  18.536 &  18.739 & ... & ... & 223 $\pm$  17 &  35.5 & 0.50 $\pm$ 0.14 &  0.09 \\
    155799 &  201.564621 &  $-$47.417397 & 1 &  37104 $\pm$  1380 &  5.52 $\pm$  0.16 &  0.12 $\pm$  0.13 &  17.390 &  18.904 &  19.068 &  19.175 & 75364 & ... & 231 $\pm$  22 &  16.7 & 0.14 $\pm$ 0.05 &  0.10 \\
   5341196 &  201.505798 &  $-$47.376869 & 1,5 &  36554 $\pm$  1650 &  5.93 $\pm$  0.18 &  0.08 $\pm$  0.13 &  17.343 &  18.858 &  19.015 &  19.140 & 54733 & ... & 221 $\pm$  26 &   6.1 & 0.39 $\pm$ 0.16 &  0.10 \\
   5359493 &  201.521515 &  $-$47.362488 & 1,5 &  35081 $\pm$  1231 &  5.87 $\pm$  0.14 & $-$0.38 $\pm$  0.09 &  17.121 &  18.585 &  18.763 &  18.977 & 59786 & ... & 228 $\pm$  17 &   1.0 & 0.46 $\pm$ 0.15 &  0.08 \\
   5131824 &  201.596558 &  $-$47.529964 & 1 &  31104 $\pm$  1750 &  5.74 $\pm$  0.17 & $-$3.65 $\pm$  1.06 &  17.384 &  18.433 &  18.458 &  18.048 & ... & ... & 201 $\pm$  26 & ... & 0.57 $\pm$ 0.26 &  0.23 \\
   5136690 &  201.577499 &  $-$47.511143 & 1 &  25347 $\pm$  2937 &  5.29 $\pm$  0.23 & $-$3.10 $\pm$  0.33 &  17.586 &  18.680 &  18.636 &  18.326 & ... & ... & 231 $\pm$  23 & ... & 0.25 $\pm$ 0.17 &  0.26 \\
   5125408 &  201.523087 &  $-$47.578598 & 1 &  32418 $\pm$  1552 &  5.77 $\pm$  0.15 & $-$1.50 $\pm$  0.11 &  17.371 &  18.720 &  18.902 &  19.146 & ... & ... & 238 $\pm$  29 & ... & 0.37 $\pm$ 0.15 &  0.07 \\
    182549 &  201.612274 &  $-$47.553101 & 1 &  30353 $\pm$  1382 &  5.35 $\pm$  0.14 & $-$3.35 $\pm$  0.71 &  16.733 &  18.113 &  18.219 &  18.469 & ... & ... & 219 $\pm$  18 & ... & 0.30 $\pm$ 0.12 &  0.15 \\
    183403 &  201.613800 &  $-$47.536694 & 1 &  29630 $\pm$  1215 &  5.42 $\pm$  0.12 & $-$3.14 $\pm$  0.40 &  16.526 &  18.081 &  18.144 &  18.262 & ... & ... & 218 $\pm$  17 & ... & 0.41 $\pm$ 0.13 &  0.18 \\
    177825 &  201.603867 &  $-$47.556538 & 1 &  27718 $\pm$  2685 &  5.45 $\pm$  0.19 & $-$3.06 $\pm$  0.25 &  17.035 &  18.322 &  18.418 &  18.489 & ... & ... & 231 $\pm$  20 & ... & 0.38 $\pm$ 0.22 &  0.14 \\
   5111007 &  201.474808 &  $-$47.589355 & 1 &  33860 $\pm$   975 &  5.92 $\pm$  0.10 & $-$0.95 $\pm$  0.07 &  17.147 &  18.612 &  18.806 &  18.918 & 45734 & ... & 243 $\pm$  14 &  16.6 & 0.52 $\pm$ 0.14 &  0.06 \\
   5142759 &  201.568619 &  $-$47.567245 & 1,2 &  35710 $\pm$  1277 &  5.87 $\pm$  0.13 & $-$0.53 $\pm$  0.10 &  17.139 &  18.610 &  18.789 &  18.966 & 77044 & ... & 219 $\pm$  12 &  11.1 & 0.43 $\pm$ 0.15 &  0.08 \\
   5131557 &  201.639801 &  $-$47.574520 & 1 &  33997 $\pm$  1077 &  5.91 $\pm$  0.12 & $-$1.88 $\pm$  0.19 &  17.143 &  18.543 &  18.729 &  18.927 & ... & ... & 204 $\pm$  46 & ... & 0.55 $\pm$ 0.18 &  0.08 \\
   5114452 &  201.638382 &  $-$47.586830 & 1 &  49373 $\pm$  2727 &  6.11 $\pm$  0.15 & $-$1.61 $\pm$  0.18 &  16.937 &  18.351 &  18.548 &  18.826 & ... & ... & 207 $\pm$   8 & ... & 0.60 $\pm$ 0.23 &  0.10 \\
   5094098 &  201.552414 &  $-$47.603844 & 1 &  38506 $\pm$  1410 &  5.78 $\pm$  0.17 &  0.65 $\pm$  0.15 &  17.742 &  19.107 &  19.303 &  19.534 & 71099 & ... & 250 $\pm$  24 &  19.4 & 0.21 $\pm$ 0.08 &  0.07 \\
   5032350 &  201.613647 &  $-$47.545490 & 1 &  37532 $\pm$  1975 &  5.84 $\pm$  0.17 &  0.16 $\pm$  0.12 &  17.955 &  19.071 &  19.477 &  19.649 & ... & ... & 177 $\pm$  22 & ... & 0.20 $\pm$ 0.08 & $-$0.14 \\
    170679 &  201.591019 &  $-$47.532055 & 1 &  35085 $\pm$  1462 &  5.82 $\pm$  0.16 & $-$0.68 $\pm$  0.11 &  17.444 &  18.624 &  18.578 &  18.252 & ... & ... & 226 $\pm$  19 & ... & 0.48 $\pm$ 0.19 &  0.31 \\
    176008 &  201.600601 &  $-$47.518524 & 1 &  33861 $\pm$  2185 &  5.97 $\pm$  0.23 & $-$0.29 $\pm$  0.16 &  17.554 &  19.037 &  19.142 &  19.096 & ... & ... & 264 $\pm$  24 & ... & 0.43 $\pm$ 0.23 &  0.14 \\
   5150273 &  201.549225 &  $-$47.562553 & 1,2 &  38066 $\pm$  1927 &  5.75 $\pm$  0.21 &  0.92 $\pm$  0.16 &  17.609 &  19.065 &  19.236 &  19.433 & 69899 & ... & 229 $\pm$  26 & ... & 0.21 $\pm$ 0.10 &  0.09 \\
   5148322 &  201.562164 &  $-$47.563770 & 1,2 &  33435 $\pm$  1147 &  5.77 $\pm$  0.12 & $-$0.89 $\pm$  0.10 &  17.096 &  18.464 &  18.619 &  18.847 & 74566 & ... & 239 $\pm$  15 &   3.4 & 0.45 $\pm$ 0.14 &  0.10 \\
   5166220 &  201.562378 &  $-$47.551682 & 1,2,3(V4) &  50395 $\pm$  2333 &  5.87 $\pm$  0.13 & $-$1.25 $\pm$  0.12 &  16.721 &  18.205 &  18.415 &  18.568 & 74659 & ... & 249 $\pm$  16 &   9.4 & 0.39 $\pm$ 0.13 &  0.09 \\
   5193651 &  201.500046 &  $-$47.523582 & 1,2,4 &  28147 $\pm$  1350 &  5.53 $\pm$  0.11 & $-$1.94 $\pm$  0.12 &  16.894 &  18.248 &  18.404 &  18.569 & 53153 & 92018 & 217 $\pm$  19 &  10.7 & 0.45 $\pm$ 0.14 &  0.08 \\
   5214452 &  201.543167 &  $-$47.502560 & 2 &  35259 $\pm$   777 &  5.66 $\pm$  0.08 & $-$0.64 $\pm$  0.06 &  17.028 &  18.460 &  18.643 &  18.780 & 67585 & ... & 249 $\pm$  14 &  11.9 & 0.31 $\pm$ 0.07 &  0.08 \\
   5222459 &  201.535385 &  $-$47.494991 & 1,2,3 &  34049 $\pm$  1068 &  5.80 $\pm$  0.11 & $-$0.73 $\pm$  0.08 &  17.381 &  18.825 &  18.952 &  18.995 & 64729 & ... & 221 $\pm$  14 &   8.9 & 0.35 $\pm$ 0.10 &  0.13 \\
   5238307 &  201.592453 &  $-$47.514717 & 1,3 &  25187 $\pm$  1371 &  5.31 $\pm$  0.13 & $-$2.28 $\pm$  0.14 &  17.734 &  18.656 &  18.606 &  18.583 & ... & ... & 243 $\pm$  17 &  12.6 & 0.28 $\pm$ 0.10 &  0.26 \\
    168035 &  201.586395 &  $-$47.541367 & 1,3 &  28671 $\pm$  2295 &  5.33 $\pm$  0.17 & $-$3.63 $\pm$  0.68 &  16.593 &  17.931 &  18.098 &  18.263 & ... & ... & 202 $\pm$  20 &  16.5 & 0.36 $\pm$ 0.18 &  0.08 \\
   5091999 &  201.620468 &  $-$47.605801 & 1,5 &  31027 $\pm$  1540 &  5.37 $\pm$  0.15 & $-$3.06 $\pm$  0.38 &  17.161 &  18.536 &  18.670 &  18.882 & 95401 & ... & 218 $\pm$  23 &   8.9 & 0.20 $\pm$ 0.08 &  0.12 \\
    165943 &  201.582718 &  $-$47.504475 & 1,3 &  36361 $\pm$  1091 &  5.88 $\pm$  0.12 & $-$0.69 $\pm$  0.08 &  17.349 &  18.770 &  18.916 &  19.040 & ... & ... & 238 $\pm$  16 &   2.8 & 0.38 $\pm$ 0.12 &  0.12 \\
   5103569 &  201.506409 &  $-$47.595497 & 1,5 &  35856 $\pm$  1183 &  5.75 $\pm$  0.12 & $-$0.47 $\pm$  0.08 &  16.928 &  18.396 &  18.591 &  18.749 & 55158 & ... & 242 $\pm$  16 &  12.6 & 0.39 $\pm$ 0.11 &  0.07 \\
   5123061 &  201.474258 &  $-$47.580219 & 5,4 &  36099 $\pm$  1357 &  5.71 $\pm$  0.13 & $-$1.31 $\pm$  0.10 &  16.936 &  18.398 &  18.616 &  18.808 & 45556 & 81531 & 226 $\pm$  25 &   8.5 & 0.34 $\pm$ 0.12 &  0.06 \\
   5138707 &  201.574615 &  $-$47.509209 & 1,3 &  38075 $\pm$  1820 &  5.90 $\pm$  0.19 &  0.67 $\pm$  0.14 &  17.806 &  19.202 &  19.407 &  19.992 & ... & ... & 207 $\pm$  17 &   8.9 & 0.25 $\pm$ 0.11 &  0.06 \\
    165237 &  201.581421 &  $-$47.520538 & 3 &  43793 $\pm$   952 &  5.99 $\pm$  0.19 &  0.80 $\pm$  0.21 &  16.889 &  18.416 &  18.535 &  18.744 & ... & ... & 233 $\pm$  18 &  22.2 & 0.54 $\pm$ 0.23 &  0.17 \\
   5200280 &  201.493988 &  $-$47.516590 & 5,2 &  33938 $\pm$   926 &  5.63 $\pm$  0.09 & $-$1.10 $\pm$  0.07 &  17.389 &  18.595 &  18.533 &  18.305 & 51359 & ... & 210 $\pm$  10 &  11.7 & 0.34 $\pm$ 0.08 &  0.33 \\
   5249172 &  201.566086 &  $-$47.469170 & 5,2,3 &  34967 $\pm$  1303 &  5.61 $\pm$  0.12 & $-$1.10 $\pm$  0.09 &  16.899 &  18.313 &  18.513 &  18.692 & 75981 & ... & 225 $\pm$   8 &  20.1 & 0.32 $\pm$ 0.10 &  0.07 \\
   5226217 &  201.502853 &  $-$47.491199 & 2,3,4 &  34787 $\pm$  1026 &  5.83 $\pm$  0.12 & $-$0.65 $\pm$  0.09 &  17.313 &  18.724 &  18.906 &  19.130 & 53945 & 98857 & 209 $\pm$  21 &  12.8 & 0.37 $\pm$ 0.11 &  0.08 \\
    204071 &  201.654999 &  $-$47.498909 & 3 &  28351 $\pm$  2127 &  5.48 $\pm$  0.16 & $-$3.03 $\pm$  0.28 &  16.861 &  18.055 &  18.173 &  18.281 & ... & ... & 233 $\pm$  17 & ... & 0.49 $\pm$ 0.23 &  0.12 \\
   5139614 &  201.790863 &  $-$47.508301 & 3 &  27898 $\pm$  1177 &  5.45 $\pm$  0.10 & $-$4.01 $\pm$  2.39 &  17.613 &  18.868 &  19.039 &  19.207 & ... & ... & 222 $\pm$  18 & ... & 0.21 $\pm$ 0.06 &  0.06 \\
   5243164 &  201.566101 &  $-$47.475098 & 3 &  31322 $\pm$   547 &  5.36 $\pm$  0.06 & $-$2.63 $\pm$  0.21 &  16.847 &  18.206 &  18.359 &  18.444 & 75993 & ... & 240 $\pm$  10 &  28.2 & 0.25 $\pm$ 0.04 &  0.11 \\
   5180753 &  201.595352 &  $-$47.475014 & 3 &  33548 $\pm$   667 &  5.78 $\pm$  0.08 & $-$1.45 $\pm$  0.09 &  17.012 &  18.437 &  18.746 &  18.942 & ... & ... & 202 $\pm$  10 & ... & 0.41 $\pm$ 0.09 & $-$0.05 \\
   5142999 &  201.817123 &  $-$47.567085 & 3,5 &  33343 $\pm$  1066 &  5.71 $\pm$  0.11 & $-$1.10 $\pm$  0.09 &  17.217 &  18.549 &  18.735 &  19.008 & 164808 & ... & 225 $\pm$  11 &   5.1 & 0.35 $\pm$ 0.10 &  0.07 \\
   5164025 &  201.723480 &  $-$47.553406 & 3 &  34964 $\pm$  1027 &  5.78 $\pm$  0.11 & $-$0.60 $\pm$  0.09 &  17.466 &  18.878 &  19.037 &  19.241 & ... & ... & 260 $\pm$  13 & ... & 0.28 $\pm$ 0.08 &  0.10 \\
   5205350 &  201.563934 &  $-$47.511494 & 3 &  34966 $\pm$   877 &  5.52 $\pm$  0.09 & $-$0.65 $\pm$  0.08 &  17.012 &  18.462 &  18.616 &  18.804 & ... & ... & 252 $\pm$  11 & ... & 0.23 $\pm$ 0.06 &  0.11 \\
   5242504 &  201.560898 &  $-$47.475719 & 3,2 &  35667 $\pm$  1216 &  5.78 $\pm$  0.13 & $-$0.53 $\pm$  0.09 &  17.215 &  18.657 &  18.824 &  19.032 & 74057 & ... & 246 $\pm$  10 &   3.6 & 0.34 $\pm$ 0.11 &  0.10 \\
    264057 &  201.776947 &  $-$47.514145 & 3 &  36092 $\pm$  1042 &  5.69 $\pm$  0.11 & $-$0.79 $\pm$  0.09 &  17.280 &  18.721 &  18.871 &  19.239 & ... & ... & 220 $\pm$  13 & ... & 0.26 $\pm$ 0.07 &  0.12 \\
   5142638 &  201.690079 &  $-$47.567398 & 3 &  35880 $\pm$  1045 &  5.68 $\pm$  0.10 & $-$0.44 $\pm$  0.08 &  17.342 &  18.811 &  18.919 &  19.001 & ... & ... & 271 $\pm$  12 & ... & 0.24 $\pm$ 0.06 &  0.16 \\
   5102280 &  201.704132 &  $-$47.596725 & 3 &  36037 $\pm$   852 &  5.66 $\pm$  0.09 & $-$0.95 $\pm$  0.09 &  16.856 &  18.357 &  18.441 &  18.432 & ... & ... & 219 $\pm$  13 & ... & 0.36 $\pm$ 0.08 &  0.19 \\
    177711 &  201.603806 &  $-$47.429600 & 3 &  36185 $\pm$  1070 &  5.72 $\pm$  0.10 & $-$0.49 $\pm$  0.08 &  17.299 &  18.427 &  18.351 &  18.049 & ... & ... & 284 $\pm$  16 & ... & 0.44 $\pm$ 0.11 &  0.34 \\
   5220684 &  201.564041 &  $-$47.440182 & 3 &  36657 $\pm$   830 &  5.87 $\pm$  0.09 & $-$0.83 $\pm$  0.08 &  17.093 &  18.573 &  18.748 &  19.045 & ... & ... & 262 $\pm$  12 & ... & 0.42 $\pm$ 0.10 &  0.10 \\
   5062474 &  201.777328 &  $-$47.496262 & 3 &  36025 $\pm$  2352 &  5.94 $\pm$  0.23 & $-$0.12 $\pm$  0.16 &  17.672 &  18.570 &  18.385 &  17.886 & ... & ... & 163 $\pm$  17 & ... & 0.72 $\pm$ 0.38 &  0.44 \\
   5047695 &  201.822342 &  $-$47.519772 & 3 &  38323 $\pm$  1392 &  5.69 $\pm$  0.19 & $-$0.22 $\pm$  0.13 &  17.514 &  18.932 &  19.075 &  18.358 & ... & ... & 247 $\pm$  11 & ... & 0.19 $\pm$ 0.09 &  0.13 \\
   5085696 &  201.736816 &  $-$47.611923 & 3 &  38655 $\pm$   927 &  5.59 $\pm$  0.13 & $-$0.08 $\pm$  0.09 &  17.316 &  18.747 &  18.744 &  18.603 & ... & ... & 227 $\pm$  12 & ... & 0.21 $\pm$ 0.07 &  0.28 \\
   5242616 &  201.552444 &  $-$47.475616 & 3,2 &  43823 $\pm$  1520 &  5.90 $\pm$  0.12 & $-$1.41 $\pm$  0.12 &  16.921 &  18.427 &  18.551 &  18.497 & 70950 & ... & 205 $\pm$  12 &  22.0 & 0.42 $\pm$ 0.13 &  0.17 \\
    177614 &  201.603729 &  $-$47.424793 & 3 &  54676 $\pm$  3125 &  6.05 $\pm$  0.11 & $-$1.36 $\pm$  0.14 &  16.928 &  18.446 &  18.615 &  18.610 & ... & ... & 272 $\pm$  18 & ... & 0.45 $\pm$ 0.13 &  0.14 \\
   5368114 &  201.678329 &  $-$47.355011 & 5 &  30234 $\pm$   965 &  5.50 $\pm$  0.09 & $-$2.47 $\pm$  0.09 &  16.855 &  18.117 &  18.164 &  18.106 & 115194 & ... & 229  & ... & 0.46 $\pm$ 0.12 &  0.20 \\
   5374166 &  201.641983 &  $-$47.349293 & 5 &  29158 $\pm$  1247 &  5.26 $\pm$  0.10 & $-$3.24 $\pm$  0.18 &  16.488 &  17.723 &  17.858 &  17.993 & 102600 & ... & 213  & ... & 0.37 $\pm$ 0.11 &  0.11 \\
   5040549 &  201.644745 &  $-$47.673519 & 5 &  26689 $\pm$  1502 &  5.12 $\pm$  0.10 & $-$2.04 $\pm$  0.07 &  16.801 &  18.028 &  18.151 &  18.242 & 103563 & ... & 205  & ... & 0.24 $\pm$ 0.08 &  0.11 \\
   5406338 &  201.670563 &  $-$47.311852 & 5 &  31273 $\pm$  1450 &  5.69 $\pm$  0.13 & $-$1.39 $\pm$  0.09 &  17.084 &  18.504 &  18.621 &  18.831 & 112475 & ... & 232  & ... & 0.43 $\pm$ 0.16 &  0.13 \\
   5352738 &  201.710190 &  $-$47.368271 & 5 &  25950 $\pm$  1317 &  5.09 $\pm$  0.09 & $-$1.68 $\pm$  0.06 &  16.295 &  17.231 &  17.239 &  17.237 & 126350 & ... & 227  & ... & 0.55 $\pm$ 0.16 &  0.21 \\
   5332719 &  201.739243 &  $-$47.383190 & 5 &  31565 $\pm$   887 &  5.33 $\pm$  0.08 & $-$4.09 $\pm$  0.63 &  16.667 &  17.949 &  18.055 &  18.115 & 137299 & ... & 217  & ... & 0.31 $\pm$ 0.08 &  0.16 \\
   5369336 &  201.749069 &  $-$47.353909 & 5 &  36288 $\pm$  1252 &  5.78 $\pm$  0.11 & $-$1.05 $\pm$  0.08 &  17.186 &  18.646 &  18.794 &  19.037 & 141008 & ... & 226  & ... & 0.34 $\pm$ 0.10 &  0.12 \\
   5235851 &  201.883286 &  $-$47.481960 & 5,4 &  29196 $\pm$  1162 &  5.46 $\pm$  0.10 & $-$2.27 $\pm$  0.12 &  16.907 &  18.195 &  18.267 &  18.216 & 183592 & 229084 & 239 $\pm$  30 &   8.3 & 0.40 $\pm$ 0.11 &  0.17 \\
   5173726 &  201.900894 &  $-$47.544926 & 5 &  30916 $\pm$  1405 &  5.65 $\pm$  0.14 & $-$3.43 $\pm$  0.52 &  17.080 &  18.545 &  18.688 &  18.807 & 186476 & ... & 236  & ... & 0.38 $\pm$ 0.15 &  0.11 \\
   5208245 &  201.908035 &  $-$47.508541 & 5 &  47966 $\pm$  3372 &  5.80 $\pm$  0.15 & $-$2.80 $\pm$  0.98 &  16.890 &  18.260 &  18.705 &  18.704 & 187534 & ... & 217  & ... & 0.26 $\pm$ 0.10 & $-$0.15 \\
   5294343 &  201.936386 &  $-$47.422180 & 5 &  29981 $\pm$  2452 &  5.54 $\pm$  0.22 & $-$2.82 $\pm$  0.30 &  17.075 &  18.565 &  18.717 &  18.801 & 191111 & ... & 225  & ... & 0.30 $\pm$ 0.19 &  0.10 \\
   5264144 &  201.879150 &  $-$47.454666 & 5,4 &  36514 $\pm$  1212 &  5.27 $\pm$  0.09 & $-$2.38 $\pm$  0.16 &  16.288 &  17.764 &  17.929 &  18.112 & 182772 & 234333 & 239 $\pm$  30 &   2.6 & 0.23 $\pm$ 0.06 &  0.12 \\
   5257714 &  201.969009 &  $-$47.460850 & 5,4 &  25424 $\pm$  1121 &  5.11 $\pm$  0.09 & $-$2.31 $\pm$  0.12 &  16.208 &  17.430 &  17.531 &  17.430 & 194383 & 233133 & 213 $\pm$  30 &   5.9 & 0.46 $\pm$ 0.13 &  0.12 \\
   5235130 &  201.415817 &  $-$47.482506 & 5 &  33681 $\pm$  1322 &  5.82 $\pm$  0.13 & $-$1.26 $\pm$  0.09 &  16.995 &  18.411 &  18.602 &  18.775 & 29850 & ... & 240  & ... & 0.50 $\pm$ 0.17 &  0.07 \\
   5243680 &  201.439774 &  $-$47.474403 & 5 &  30225 $\pm$  1412 &  5.45 $\pm$  0.13 & $-$3.84 $\pm$  0.82 &  16.890 &  18.229 &  18.405 &  18.526 & 35828 & ... & 199  & ... & 0.32 $\pm$ 0.12 &  0.08 \\
   5341114 &  201.443649 &  $-$47.376816 & 5 &  44619 $\pm$  3952 &  5.49 $\pm$  0.18 & $-$2.85 $\pm$  0.83 &  16.760 &  18.127 &  18.378 &  18.633 & 36669 & ... & 205  & ... & 0.19 $\pm$ 0.09 &  0.05 \\
   5221335 &  201.458237 &  $-$47.495911 & 5 &  33020 $\pm$  2002 &  5.45 $\pm$  0.17 & $-$1.15 $\pm$  0.12 &  17.024 &  18.417 &  18.600 &  18.739 & 40846 & ... & 255  & ... & 0.22 $\pm$ 0.10 &  0.08 \\
   5191522 &  201.517944 &  $-$47.525745 & 5 &  26463 $\pm$  2065 &  5.34 $\pm$  0.15 & $-$3.10 $\pm$  0.20 &  17.199 &  18.296 &  18.363 &  18.507 & 58774 & 91573 & 213 $\pm$  30 &   4.9 & 0.34 $\pm$ 0.15 &  0.16 \\
   5227178 &  201.524353 &  $-$47.490322 & 5 &  43454 $\pm$  2110 &  5.45 $\pm$  0.10 & $-$3.41 $\pm$  1.17 &  16.564 &  18.029 &  18.236 &  18.392 & 60820 & ... & 226  & ... & 0.20 $\pm$ 0.06 &  0.09 \\
   5200778 &  201.536774 &  $-$47.620155 & 5 &  32935 $\pm$  1092 &  5.64 $\pm$  0.10 & $-$1.05 $\pm$  0.07 &  16.941 &  18.355 &  18.556 &  18.749 & 65373 & ... & 232  & ... & 0.37 $\pm$ 0.10 &  0.06 \\
   5254427 &  201.548111 &  $-$47.464264 & 5 &  30078 $\pm$  1015 &  5.33 $\pm$  0.09 & $-$3.90 $\pm$  0.55 &  16.590 &  17.870 &  18.039 &  18.165 & 69373 & ... & 219  & ... & 0.35 $\pm$ 0.09 &  0.08 \\
    172304 &  201.594086 &  $-$47.514648 & 5 &  28623 $\pm$  2035 &  4.95 $\pm$  0.15 & $-$3.28 $\pm$  0.28 &  16.255 &  17.271 &  17.251 &  17.102 & 86429 & ... & 206  & ... & 0.33 $\pm$ 0.15 &  0.27 \\
   5078284 &  201.711899 &  $-$47.619541 & 5 &  33828 $\pm$  1907 &  5.48 $\pm$  0.15 & $-$0.86 $\pm$  0.11 &  16.979 &  18.389 &  18.541 &  18.752 & 126892 & ... & 229  & ... & 0.24 $\pm$ 0.10 &  0.11 \\
   5336346 &  201.692383 &  $-$47.380478 & 5 &  35366 $\pm$  1210 &  5.94 $\pm$  0.11 & $-$0.22 $\pm$  0.08 &  17.660 &  18.852 &  19.000 &  19.392 & 120119 & ... & 230  & ... & 0.42 $\pm$ 0.12 &  0.11 \\
   5186047 &  201.919846 &  $-$47.531506 & 5 &  35253 $\pm$  1255 &  5.86 $\pm$  0.11 & $-$0.45 $\pm$  0.08 &  17.023 &  18.511 &  18.577 &  18.390 & 189080 & ... & 235  & ... & 0.52 $\pm$ 0.15 &  0.19 \\
   5377817 &  201.668640 &  $-$47.345703 & 5 &  41994 $\pm$   715 &  6.32 $\pm$  0.12 &  0.77 $\pm$  0.10 &  17.730 &  19.141 &  19.219 &  19.370 & 111785 & ... & 239  & ... & 0.66 $\pm$ 0.19 &  0.20 \\
   5268490 &  201.930328 &  $-$47.450127 & 5 &  38087 $\pm$  1585 &  5.72 $\pm$  0.14 & $-$0.70 $\pm$  0.10 &  17.034 &  18.656 &  18.800 &  18.858 & 190398 & ... & 233  & ... & 0.27 $\pm$ 0.09 &  0.13 \\
   5197853 &  201.402451 &  $-$47.518871 & 5 &  31770 $\pm$  1720 &  5.96 $\pm$  0.18 & $-$0.65 $\pm$  0.12 &  17.182 &  18.645 &  18.838 &  19.001 & 26774 & ... & 213  & ... & 0.62 $\pm$ 0.29 &  0.05 \\
   5232426 &  201.422150 &  $-$47.485092 & 5,4 &  34615 $\pm$  1615 &  5.81 $\pm$  0.16 & $-$0.38 $\pm$  0.12 &  17.350 &  18.803 &  18.987 &  19.186 & 31400 & 100171 & 228 $\pm$  30 &   1.7 & 0.33 $\pm$ 0.13 &  0.07 \\
   5182741 &  201.458923 &  $-$47.535099 & 5,4 &  34236 $\pm$  1347 &  5.63 $\pm$  0.15 & $-$0.45 $\pm$  0.10 &  17.058 &  18.521 &  18.678 &  18.807 & 41074 & 89638 & 229 $\pm$  30 &   2.3 & 0.30 $\pm$ 0.11 &  0.10 \\
   5333076 &  201.468033 &  $-$47.382702 & 5 &  35212 $\pm$  1167 &  5.80 $\pm$  0.10 & $-$0.65 $\pm$  0.08 &  17.204 &  18.580 &  18.791 &  18.998 & 43520 & ... & 227  & ... & 0.37 $\pm$ 0.10 &  0.05 \\
   5381809 &  201.512680 &  $-$47.341492 & 5 &  32839 $\pm$  1042 &  5.83 $\pm$  0.10 & $-$0.76 $\pm$  0.07 &  16.789 &  18.258 &  18.436 &  18.515 & 56896 & ... & 236  & ... & 0.64 $\pm$ 0.18 &  0.07 \\
   5047902 &  201.595901 &  $-$47.660110 & 5 &  33876 $\pm$  1120 &  5.88 $\pm$  0.10 & $-$0.87 $\pm$  0.07 &  17.109 &  18.543 &  18.716 &  18.966 & 87161 & ... & 236  & ... & 0.51 $\pm$ 0.15 &  0.08 \\
   5384240 &  201.579010 &  $-$47.338970 & 5 &  34194 $\pm$  1265 &  5.83 $\pm$  0.12 & $-$0.91 $\pm$  0.08 &  17.039 &  18.453 &  18.623 &  18.839 & 80690 & ... & 217  & ... & 0.50 $\pm$ 0.16 &  0.09 \\
   5249607 &  201.552963 &  $-$47.468742 & 2 &  35722 $\pm$   915 &  5.81 $\pm$  0.10 & $-$0.69 $\pm$  0.08 &  17.134 &  18.619 &  18.756 &  18.888 & 71204 & ... & 196 $\pm$   7 & ... & 0.38 $\pm$ 0.10 &  0.13 \\
   5247607 &  201.499390 &  $-$47.470615 & 2,4 &  34410 $\pm$  1053 &  5.70 $\pm$  0.12 & $-$0.88 $\pm$  0.10 &  17.065 &  18.467 &  18.646 &  18.796 & 52905 & 103232 & 219 $\pm$  19 &   7.1 & 0.36 $\pm$ 0.11 &  0.08 \\
   5257735 &  201.524918 &  $-$47.461021 & 2 &  33466 $\pm$   752 &  5.68 $\pm$  0.08 & $-$0.99 $\pm$  0.08 &  17.049 &  18.472 &  18.649 &  18.810 & 60969 & ... & 228 $\pm$   7 & ... & 0.35 $\pm$ 0.08 &  0.08 \\
   5223089 &  201.450623 &  $-$47.494194 & 2 &  37054 $\pm$  1360 &  5.87 $\pm$  0.15 & $-$0.84 $\pm$  0.12 &  16.921 &  18.319 &  18.514 &  18.665 & 38670 & 98189 & 227 $\pm$  19 &  25.7 & 0.51 $\pm$ 0.20 &  0.08 \\
   5182383 &  201.572433 &  $-$47.535667 & 2 &  36576 $\pm$  1162 &  5.82 $\pm$  0.12 & $-$0.48 $\pm$  0.09 &  17.781 &  18.964 &  18.894 &  18.524 & 78393 & ... & 224 $\pm$   6 & ... & 0.33 $\pm$ 0.10 &  0.34 \\
   5192767 &  201.562485 &  $-$47.524570 & 2 &  37728 $\pm$   995 &  5.84 $\pm$  0.11 & $-$0.27 $\pm$  0.08 &  17.317 &  18.756 &  18.853 &  18.855 & 74644 & ... & 255 $\pm$   8 & ... & 0.35 $\pm$ 0.09 &  0.17 \\
   5257577 &  201.491806 &  $-$47.461132 & 4 &  35797 $\pm$  3095 &  5.43 $\pm$  0.28 &  0.53 $\pm$  0.17 &  17.760 &  19.151 &  19.258 &  19.041 & ... & 105290 & 237 $\pm$  30 & ... & 0.11 $\pm$ 0.07 &  0.15 \\
   5270645 &  201.443527 &  $-$47.447819 & 4 &  38094 $\pm$  3022 &  5.85 $\pm$  0.33 &  0.23 $\pm$  0.21 &  17.497 &  18.967 &  19.108 &  19.181 & 36725 & 108102 & 237 $\pm$  30 &   5.7 & 0.28 $\pm$ 0.22 &  0.12 \\
   5271711 &  201.537659 &  $-$47.446865 & 4 &  37548 $\pm$  1315 &  5.84 $\pm$  0.15 & $-$0.70 $\pm$  0.12 &  17.095 &  18.975 &  18.651 &  18.630 & ... & 108309 & 228 $\pm$  30 & ... & 0.41 $\pm$ 0.15 &  0.60 \\
    251042 &  201.750732 &  $-$47.521118 & 4 &  35796 $\pm$  1015 &  5.39 $\pm$  0.10 & $-$1.17 $\pm$  0.09 &  16.137 &  17.541 &  17.714 &  18.000 & ... & 154412 & 232 $\pm$  30 & ... & 0.38 $\pm$ 0.10 &  0.10 \\
   5069716 &  201.780869 &  $-$47.486443 & 4 &  34681 $\pm$  1462 &  5.68 $\pm$  0.16 & $-$0.87 $\pm$  0.14 &  16.395 &  19.233 &  19.170 &  19.611 & ... & 166106 & 223 $\pm$  30 & ... & 0.21 $\pm$ 0.08 &  0.33 \\
    280071 &  201.810272 &  $-$47.481472 & 4 &  36438 $\pm$  1010 &  5.68 $\pm$  0.11 & $-$1.00 $\pm$  0.09 &  17.007 &  18.460 &  18.508 &  18.509 & ... & 167821 & 247 $\pm$  30 & ... & 0.35 $\pm$ 0.10 &  0.23 \\
   5255164 &  201.822739 &  $-$47.463608 & 4 &  23829 $\pm$   755 &  5.11 $\pm$  0.06 & $-$2.41 $\pm$  0.14 &  ... &  17.600 &  17.589 &  17.474 & ... & 173876 & 262 $\pm$  30 & ... & 0.49 $\pm$ 0.10 &  0.22 \\
    270816 &  201.791092 &  $-$47.440907 & 4 &  42686 $\pm$  2657 &  5.86 $\pm$  0.17 & $-$1.98 $\pm$  0.32 &  17.071 &  18.576 &  18.758 &  18.907 & ... & 181428 & 308 $\pm$  30 & ... & 0.33 $\pm$ 0.14 &  0.11 \\
   5213165 &  201.904282 &  $-$47.503742 & 4 &  36220 $\pm$  1002 &  5.68 $\pm$  0.11 & $-$0.45 $\pm$  0.08 &  16.901 &  18.402 &  18.595 &  18.719 & ... & 224916 & 223 $\pm$  30 & ... & 0.32 $\pm$ 0.09 &  0.07 \\
   5213936 &  201.872253 &  $-$47.503036 & 4 &  36095 $\pm$   950 &  5.66 $\pm$  0.10 & $-$0.62 $\pm$  0.08 &  17.112 &  18.619 &  18.754 &  18.807 & ... & 225063 & 258 $\pm$  30 & ... & 0.26 $\pm$ 0.07 &  0.13 \\
   5245132 &  201.850220 &  $-$47.473099 & 4 &  32768 $\pm$   790 &  5.60 $\pm$  0.09 & $-$1.42 $\pm$  0.09 &  16.848 &  18.011 &  18.000 &  17.888 & ... & 230786 & 203 $\pm$  30 & ... & 0.56 $\pm$ 0.14 &  0.27 \\
   5255018 &  201.935120 &  $-$47.463570 & 4 &  35780 $\pm$  1272 &  5.46 $\pm$  0.13 & $-$0.52 $\pm$  0.11 &  16.859 &  18.342 &  18.478 &  18.481 & ... & 232593 & 214 $\pm$  30 & ... & 0.22 $\pm$ 0.08 &  0.13 \\
   5255436 &  201.959717 &  $-$47.463112 & 4 &  23463 $\pm$   842 &  4.89 $\pm$  0.07 & $-$1.94 $\pm$  0.10 &  16.170 &  17.279 &  17.359 &  17.305 & ... & 232682 & 205 $\pm$  30 & ... & 0.37 $\pm$ 0.08 &  0.12 \\
   5268276 &  201.862000 &  $-$47.450474 & 4 &  24015 $\pm$  1337 &  4.75 $\pm$  0.12 & $-$1.74 $\pm$  0.15 &  16.403 &  17.495 &  17.549 &  17.598 & ... & 235103 & 236 $\pm$  30 & ... & 0.22 $\pm$ 0.08 &  0.16 \\
   5273495 &  201.910889 &  $-$47.444885 & 4 &  27006 $\pm$   725 &  5.16 $\pm$  0.07 & $-$4.00 $\pm$  0.50 &  16.495 &  17.814 &  17.931 &  17.970 & ... & 236142 & 212 $\pm$  30 & ... & 0.32 $\pm$ 0.07 &  0.12 \\
   5275033 &  201.859039 &  $-$47.443348 & 4 &  24494 $\pm$   795 &  5.09 $\pm$  0.07 & $-$3.42 $\pm$  0.53 &  16.626 &  17.592 &  17.613 &  17.528 & ... & 236428 & 199 $\pm$  30 & ... & 0.43 $\pm$ 0.09 &  0.19 \\
   5174292 &  201.544006 &  $-$47.544468 & 4 &  23752 $\pm$  1830 &  5.09 $\pm$  0.14 & $-$1.87 $\pm$  0.09 &  16.575 &  17.772 &  17.913 &  18.052 & ... & 87776 & 205 $\pm$  30 & ... & 0.34 $\pm$ 0.14 &  0.06 \\
   5212417 &  201.450882 &  $-$47.504456 & 4 &  30507 $\pm$  1522 &  5.39 $\pm$  0.16 & $-$2.24 $\pm$  0.34 &  17.115 &  18.351 &  18.493 &  18.596 & ... & 95987 & 180 $\pm$  30 & ... & 0.25 $\pm$ 0.11 &  0.11 \\
   5189642 &  201.466141 &  $-$47.527626 & 4 &  33326 $\pm$  1172 &  5.72 $\pm$  0.13 & $-$1.24 $\pm$  0.12 &  17.087 &  18.468 &  18.680 &  18.866 & 43148 & 91164 & 208 $\pm$  30 &  26.9 & 0.38 $\pm$ 0.13 &  0.05 \\
\hline   
\end{longtable}
\tablefoot{
\tablefoottext{a}{(1) VIMOS; (2) FORS1.6; (3) FORS2.6; (4) \forsmb; (5) FLAMES. The 4 pulsators are identified by their name according to \citet{ran16}. }
\tablefoottext{b}{The errors include the correction factor determined in Appendix A.}
\tablefoottext{c}{\citet{moe11} do not provide uncertainties on their RVs, thus no errors are provided for the stars exclusive to the FLAMES sample.}
}
\tablebib{
(1)~\citet{moe11}; (2)~\citet{moni12}
}
\end{landscape}   
}

\subsection{Atmospheric parameters}

\begin{figure}
\begin{center}
\resizebox{\hsize}{!}{\includegraphics{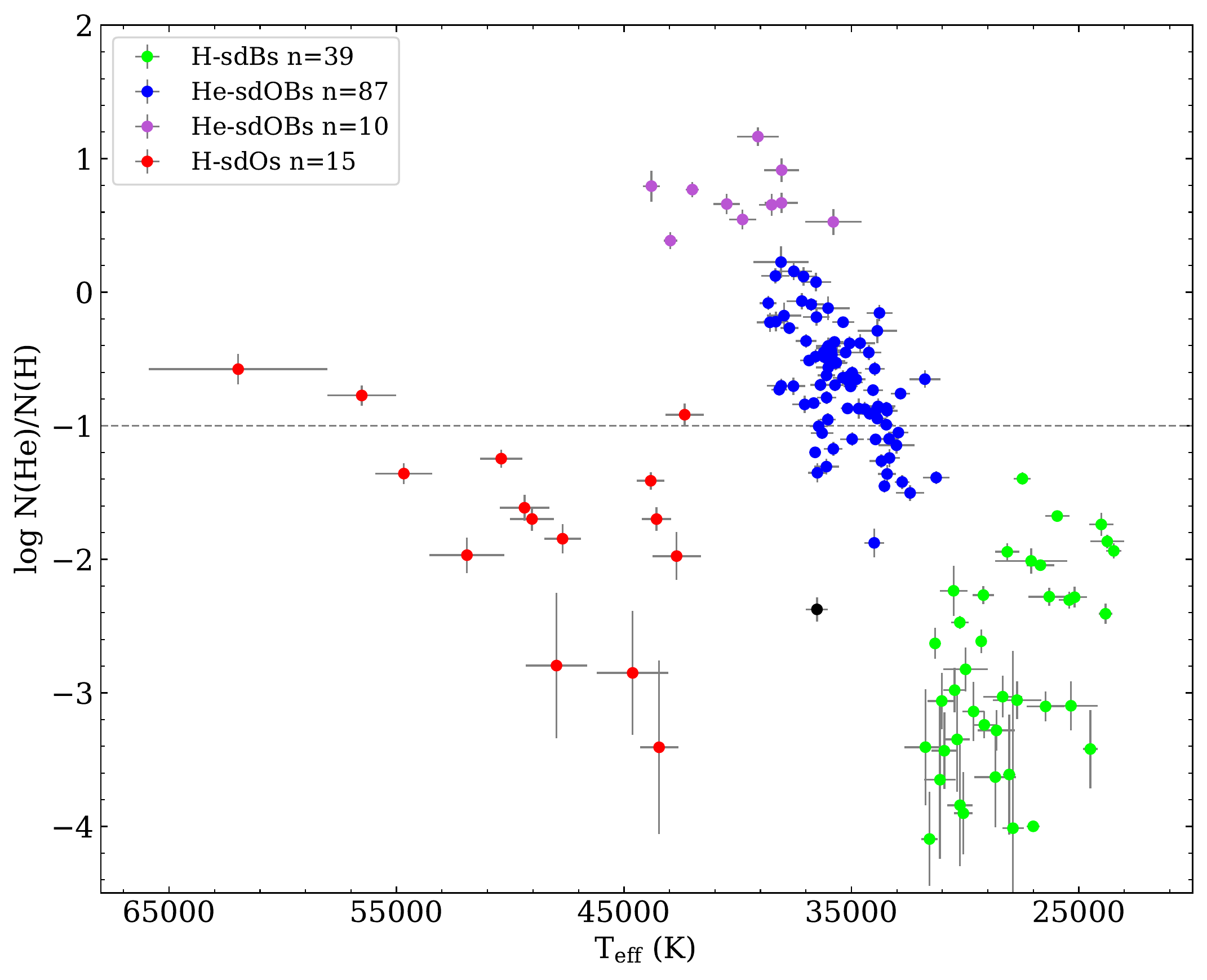}}
\caption{Helium abundance as a function of effective temperature for the 152 stars in our sample. The spectroscopic groups mentioned in the text are plotted in different colors. The dashed line represents the solar helium abundance. The error bars used for individual stars are the statistical uncertainties returned by the fitting procedure. 
}
\label{teffhe}
\end{center}
\end{figure}

It was previously reported in \citet{lat14} that the EHB stars in \omcen\ can be divided into three different groups according to their atmospheric parameters. That result was based on a sample of 38 stars. Looking at the distribution of our extended sample in the \teff\ $-$ log \nhe\ plane (Fig. \ref{teffhe}) we distinguish the same pattern with three prominent spectroscopic groups. There is a first group of helium-poor stars found at \teff\ $\la$ 30\,000 K, corresponding to an sdB spectral type. We will refer to these as the H-sdBs, to highlight that their atmosphere is enriched in hydrogen. A second group of stars with higher helium abundances, mainly super-solar (log \nhe $\ga$ $-1.0$), can be found at \teff\ between $\sim$33\,000$-$43\,000 K. The stars in this group show a clear trend of increasing helium abundance with effective temperature. In the smaller sample of \citet{lat14}, three stars out of the 25 forming that group had a helium abundance below solar. In our extended sample, it becomes clear that this group of ``helium-enriched'' stars is not strictly helium-rich but also extends to helium abundances slightly below solar. Nevertheless, we refer to this group of stars as the He-sdOBs. 
Interestingly, the ten most helium-rich objects of this group seem to distinguish themselves with the peculiarity that the helium abundance versus temperature correlation vanishes.
Although we plot these most helium-rich objects with a distinct color (purple) in some of the following figures, we consider them part of the He-sdOB group.
While the bulk of the He-sdOBs shows a rather tight \teff\ $-$ helium correlation, the situation is different among the H-sdBs, which have a much larger scatter in their helium abundances. This large scatter in helium abundance is also observed among hydrogen-rich sdBs of other clusters (see Fig. 8 of \citealt{moni12}).

The third group comprises the hottest stars, corresponding to an sdO spectral type. These stars are mostly helium-poor and also seem to show a correlation between helium abundance and temperature. However, one must be careful in interpreting this feature since the three most helium-poor objects are from the FLAMES sample (see also Fig. \ref{app_he}), where the spectral range does not cover any \ion{He}{ii} lines and only two Balmer lines were used to derive the atmospheric parameters (which also explains the large uncertainties). We identify the stars from this group as the H-sdOs, since the majority of them have an atmosphere enriched in hydrogen. Finally, there is one star (plotted in black in Fig. \ref{teffhe}) that could not be associated with any of the group described above. This particular object is part of two samples (\forsmb\ and FLAMES) and the parameters derived from both spectra are in good agreement. 

\begin{figure*}
\begin{center}
\includegraphics[width=0.48\linewidth]{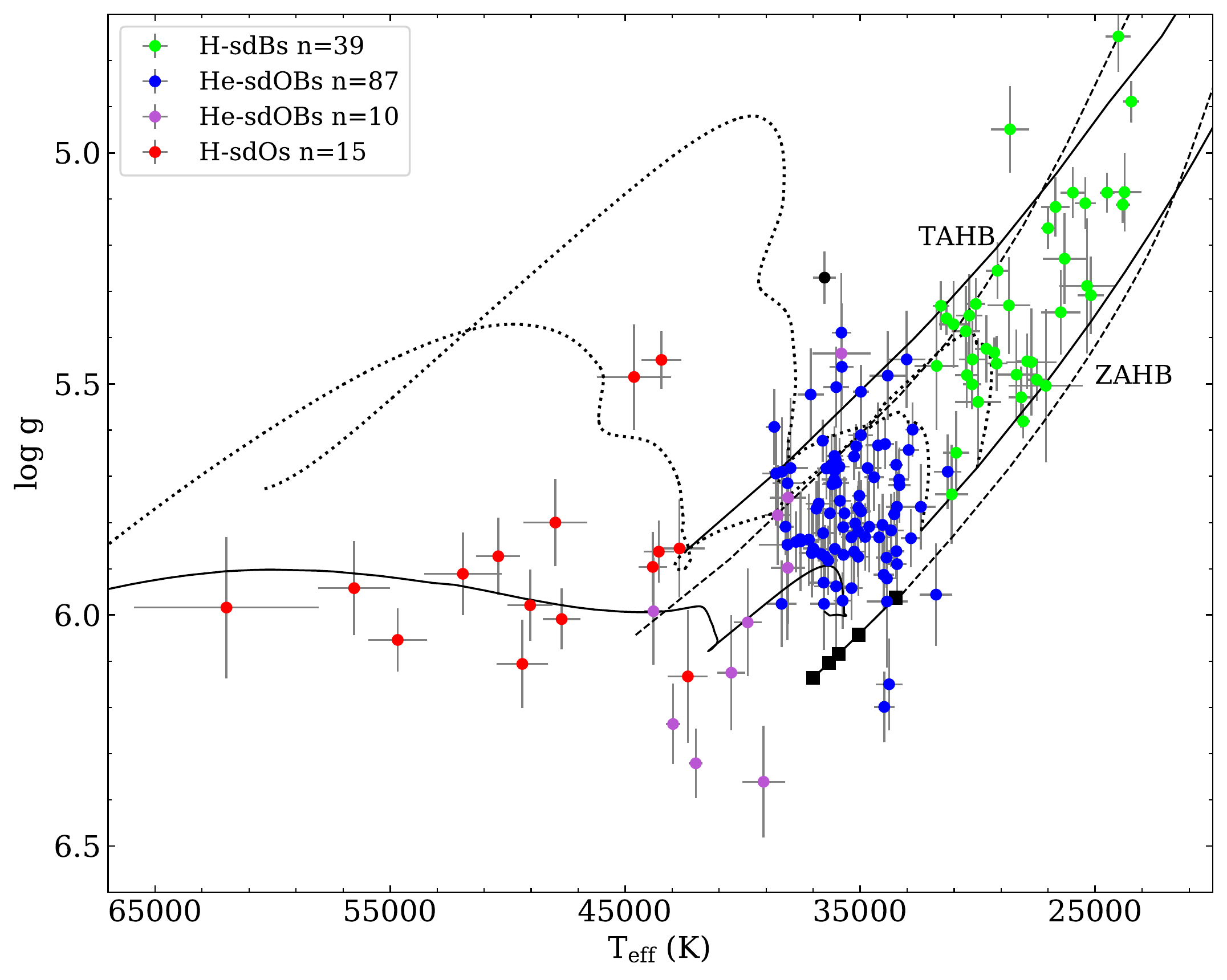}
\includegraphics[width=0.48\linewidth]{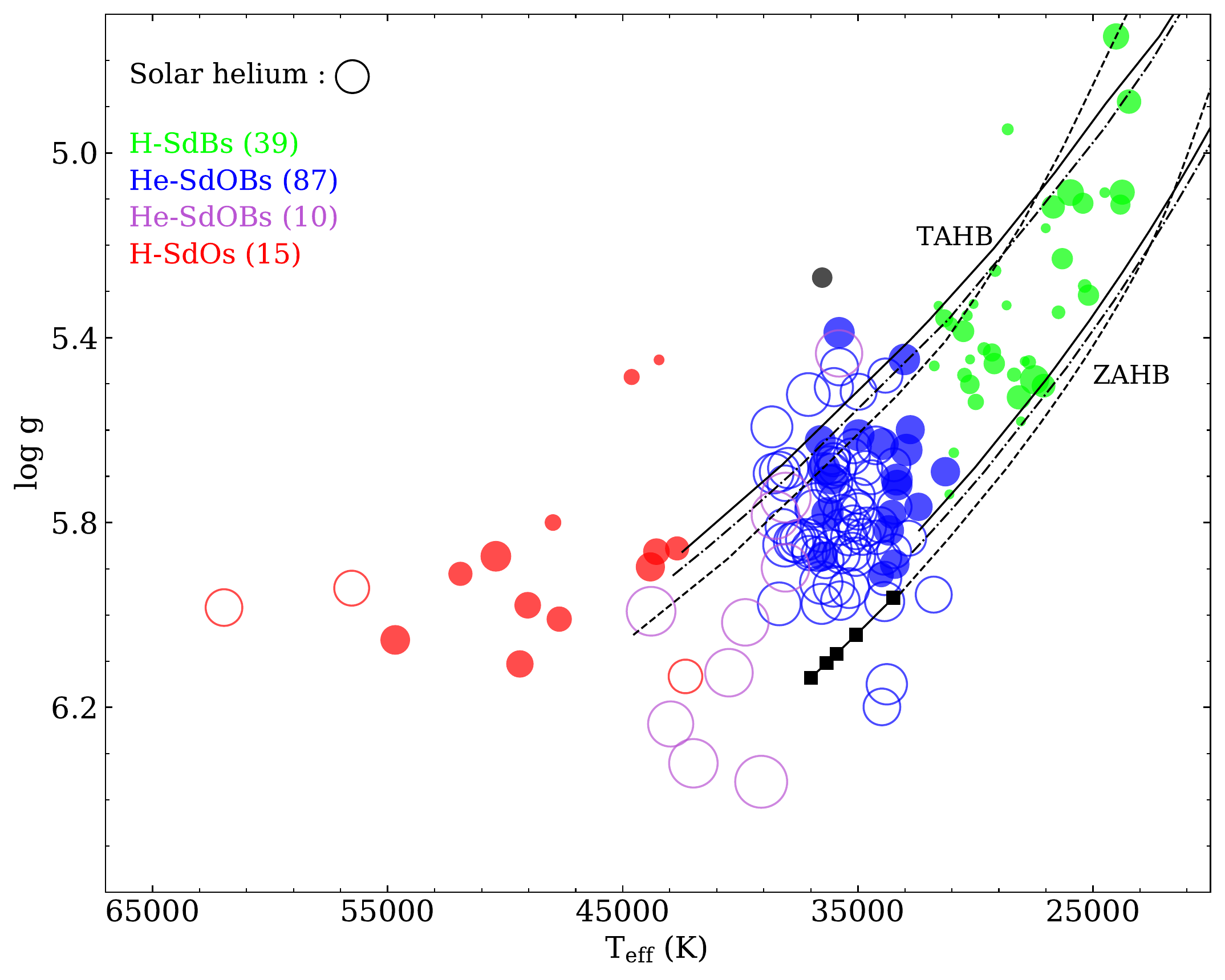}
\caption{$Left:$ Position of the stars in the log $g$ $-$ \teff\ diagram. The spectroscopic groups are plotted as in Fig.~\ref{teffhe}. The error bars used for individual stars are the statistical uncertainties returned by the fitting procedure. 
The ZAHB and TAHB sequences are plotted for two different helium contents $Y=0.24$ (solid lines) and $Y=0.40$ (dashed lines) and a metallicity of $[M/H]$ = -1.8. Canonical evolutionary tracks with $Y=0.24$ are shown for models having 0.498 and 0.5 \msun\ (dotted curves). The connected squares represent a series of ZAHB models computed by adding a hydrogen-rich layer to the surface of a late-flasher ZAHB model. The squares indicate hydrogen layer masses (with decreasing temperature) of 0, 10$^{-7}$, 10$^{-6}$, 10$^{-5}$, and 10$^{-4}$ \msun\ \citep{moe02}.
An additional late-flasher evolutionary track is plotted (solid curve), corresponding to the $Z = 0.001$ and $M = 0.491$ \msun\ shallow-mixing case of \citet{mil08}. 
$Right:$ Similar to the left panel, but here the size of the circles is proportional to the logarithmic helium abundance, super-solar and sub-solar abundances being represented by open and filled circles respectively. The circle size for the solar helium abundance is shown for reference. The ZAHBs and TAHBs are plotted as in the left panel, with an additional sequence at a metallicity of $[M/H]$ = -0.96 and $Y = 0.24$ added (dash-dotted lines). }
\label{teffgravhe}
\end{center}
\end{figure*}

Figure \ref{teffgravhe} shows the distribution of our stars in the log $g$ $-$ \teff\ diagram with theoretical models from the BASTI database\footnote{http://albione.oa-teramo.inaf.it/} \citep{pie06} overplotted. We selected a Zero-Age Horizontal Branch (ZAHB) and a Terminal Age Horizontal Branch (TAHB) sequence for a normal helium content, $Y = 0.24$, and a metallicity representative of the \omcen\ stellar population ($Z=0.0003$, $[M/H]$ = -1.8, solid lines), as well as a ZAHB and TAHB for a helium-enhanced stellar mixture ($Y=0.40$ and $Z=0.0002$ ($[M/H]$ = -1.8), dashed lines). The right panel of Fig. \ref{teffgravhe} shows an additional ZAHB and TAHB for a different metallicity, $Z=0.002$ ($[M/H]$ = -0.96, dashed-dotted lines), and normal helium content ($Y=0.24$). The ZAHB is the starting point of the He-core burning, while the TAHB represents the end of helium burning in the center of the star. These two sequences define the He-core burning region that is considered the evolutionary EHB region. After leaving the EHB, the star starts to burn helium in the outer shell, with the post-EHB evolution proceeding around 10 times faster than the EHB phase. The EHB models represent stars with masses in the canonical range 0.488$-$0.510 \msun.
While the position of HB stars predicted by the helium-enhanced models is quite different to that of the normal helium models for stars with \teff\ $<$ 20\,000 K, the difference is not very pronounced for the EHB domain shown in the plot. These models can be refered to as canonical in the sense that they do not result from a delayed helium-flash. Stars experiencing a late flash and the resulting mixing of the helium-rich core material with the hydrogen envelope will end up having higher effective temperatures. 

The majority of the stars (from the H-sdB and the He-sdOB groups) are sitting on the EHB, as would be expected for He-core burning objects. 
Although both groups are found along the EHB, there is a clear gap between them, which could be explained by the gap in \teff\ and log $g$ predicted between the hottest canonical EHB models and the models undergoing a delayed He-flash \citep{moe11,bro01}. Following the delayed He-flash, the hydrogen that survived the process is expected to diffuse toward the surface, causing the star to become colder as a hydrogen layer builds up on the surface \citep{mil08}. To illustrate this, we plotted a sequence of ZAHB models (black squares) computed by adding a hydrogen-rich layer (of varying mass: 0, 10$^{-7}$, 10$^{-6}$, 10$^{-5}$, and 10$^{-4}$ \msun) to the surface of a late-flasher model that consists almost entirely of helium (96 \%) and carbon (3$-$4 \%; see Sect. 5 of \citealt{moe02} for a detailed description of the models). The hydrogen layer has the effect of reducing the effective temperature of the star by increasing the shielding of the hot core and changing the atmospheric opacity. According to this scenario, our group of stars with the highest helium abundances (plotted in purple) with on average higher \teff\ than the other He-sdOBs could be newly born late-flashers, while the other He-sdOBs would have already undergone some level of diffusion.

\begin{figure}
\begin{center}
\resizebox{\hsize}{!}{\includegraphics{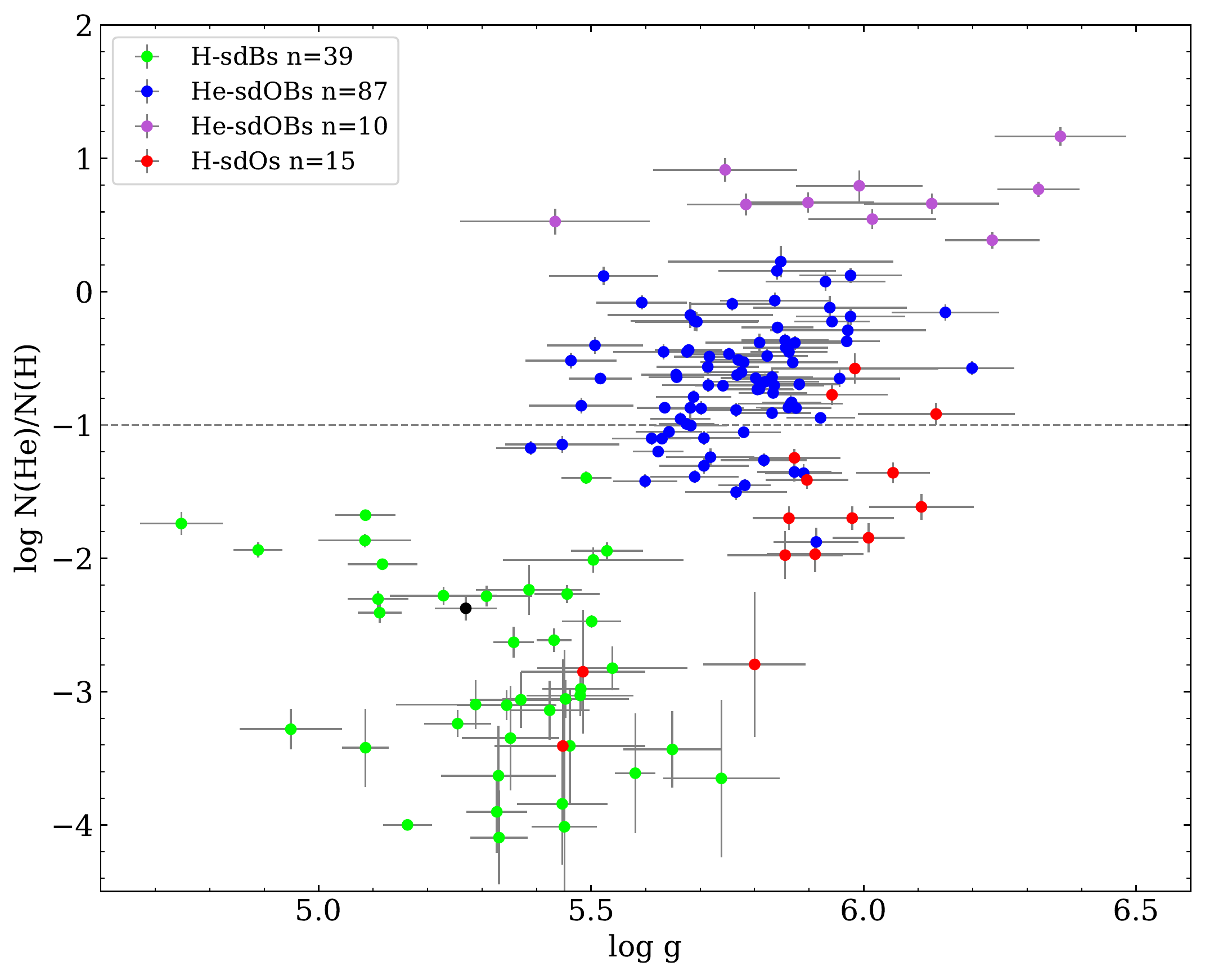}}
\caption{
Helium abundance as function of the surface gravity derived for the 152 stars of the sample. The spectroscopic groups are plotted as in Fig~\ref{teffhe}.
The error bars used for individual stars are the statistical uncertainties returned by the fitting procedure. }

\label{gravhe}
\end{center}
\end{figure}

In the left-hand panel of Fig. \ref{teffgravhe} we display evolutionary tracks for canonical models with a normal helium abundance and masses of 0.498 and 0.5 \msun\ (dotted curves; taken from the BASTI database), as well as a late hot flasher evolutionary track (solid curve, taken from \citealt{mil08}, the shallow mixing case for $Z=0.001$ and M = 0.491 \msun). An interesting difference between the canonical and late-flasher evolution is in the post-EHB region: while canonical models predict a rise in luminosity after core-helium exhaustion, the post-EHB evolution of the late flasher proceeds at a relatively constant surface gravity due to the thinner hydrogen envelope. 
The hydrogen-rich stars lying above the TAHB (including the one star that could not be associated with a specific spectroscopic group) are likely in the He-shell burning post-EHB phase, their low surface gravity matching the predictions from the canonical post-EHB tracks. The hottest H-sdOs, which are found around log $g$ = 6.0, could be the progeny of the He-sdOBs, given that diffusion leads to a decrease in atmospheric helium abundance over time. This idea was already suggested by \citet{lat14}, who found that the He-sdOBs and the H-sdOs show the same correlation between the helium and carbon abundances (see their Fig. 7).  
Figure \ref{gravhe} presents our stars in the helium $-$ surface gravity plane, where the H-sdBs and He-sdOBs are also well separated, the latter being found at higher gravities. Looking at the number of stars included in the three spectroscopic groups, we find that the He-sdOB stars account for 64~\% of the sample, the H-sdBs for 26~\% and the H-sdOs for the remaining 10~\%.
Our full spectroscopic sample combines data from five observed samples that were subject to different selection criteria, as described in Sect. 2. For instance, the FORS1.6 and FORS2.6 samples are biased toward hotter stars, and indeed, these two samples show the lowest fractions of H-sdBs. The sample least likely to be affected by selection effects is \forsmb, as it targeted the HB of \omcen\ all the way from the blue edge of the RR Lyrae gap to the hot end of the EHB (see \citealt{moni12}). This sample indeed includes a larger fraction (33~\%) of H-sdBs and slightly fewer He-sdOBs ($\sim$57~\%). 

\subsubsection{The blue hook} 
\begin{figure}
\begin{center}
\resizebox{\hsize}{!}{\includegraphics{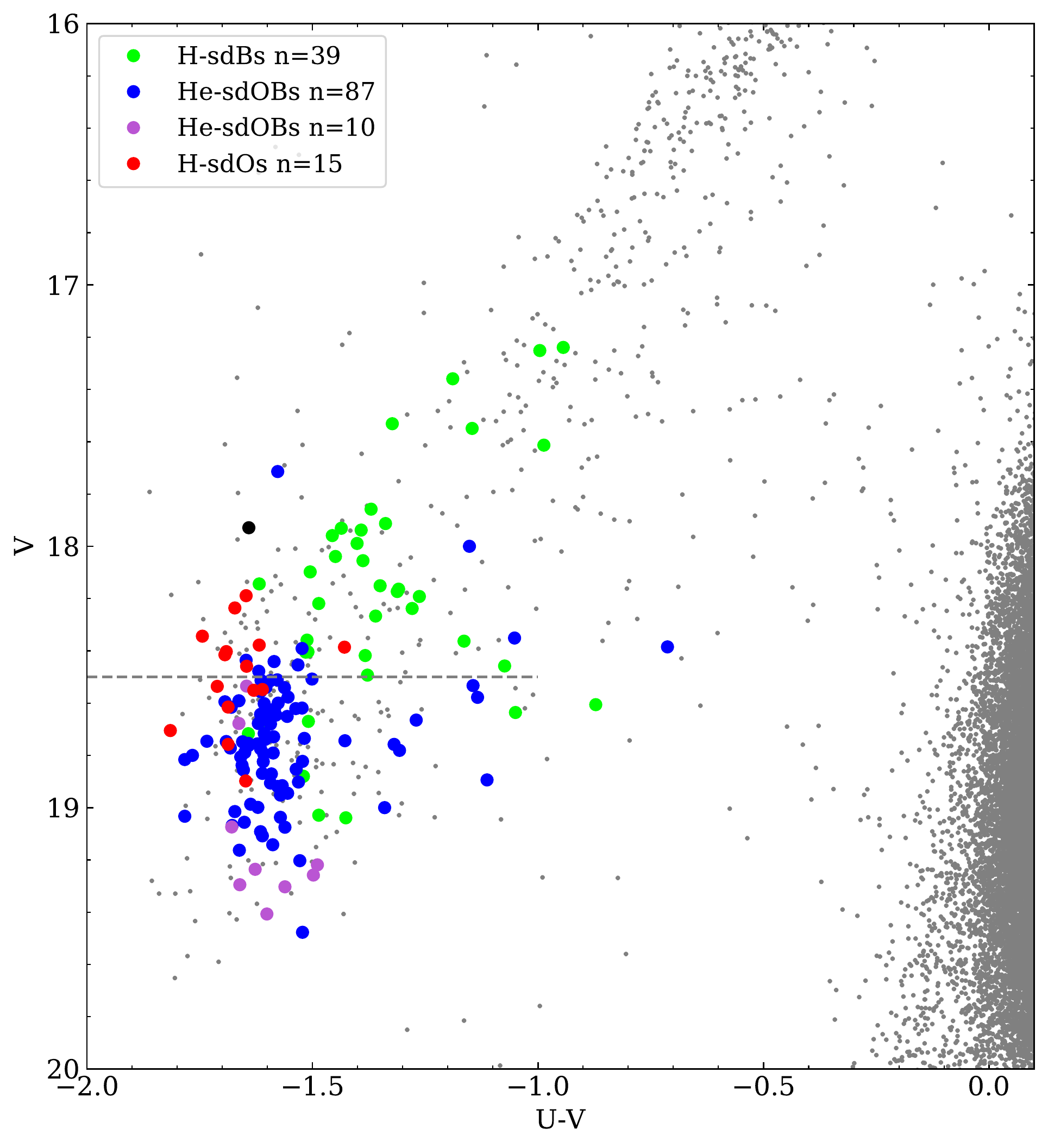}}
\caption{Position of our stars colored according to spectroscopic group in the $V$ vs $U-V$ CMD. The dashed line represents the separation between the EHB and the blue hook region \citep{moe02}. }
\label{cmd}
\end{center}
\end{figure}

An important aspect of the HB morphology of \omcen\ is the presence of its prominent blue hook that is most conspicuous in the UV CMD \citep{dcruz00, bro16}. This particular feature has been studied mostly in \omcen\ and NGC~2808, where it was attributed to the presence of helium enriched stars \citep{dcruz96,bro10,dan10}. However, very few of these blue hook objects have been observed spectroscopically \citep{moe02,moe04,bro12}. The main reason for this is that they are best identified in UV CMDs which rely on space observations (nowadays mainly with the Hubble Space Telescope). These observations usually target the dense central region of globular clusters that cannot be easily resolved using ground-based observations. On the other hand, spectroscopic data are mostly obtained with ground-based telescopes for stars found in the outskirts of the clusters where crowding is less severe; this is also the case for the EHB stars in \omcen, as illustrated in Fig. \ref{spdist}. Given the large size of our sample, we thought it worthwhile to plot the position of the targets according to spectroscopic group in the optical CMD (Fig. \ref{cmd}). A few stars have redder colors than the bulk of the EHB in the $V$, $U-V$ CMD, which is likely due to inaccurate WFI photometry since these stars are found among the bulk of EHB stars when using the ACS photometry.
As expected, the faintest part of the EHB consists predominantly of He-rich objects. \citet{moe02} identified the blue hook region in the optical CMD of \omcen\ as the region with $V \ga$ 18.5, based on the distance-corrected magnitude at which the blue tail of the HB of the globular cluster NGC~6752 ends. NGC~6752 is well known to host a fair number of EHB stars, but these all have \teff\ $\la$ 30\,000 K and are helium-poor, thus it can be inferred that EHB stars fainter than those of NGC~6752 should constitute the blue hook population of \omcen. According to Fig. \ref{cmd}, the majority of our targets with $V \ga$ 18.5 indeed belong to the He-sdOB group and this magnitude limit provides a good (although not perfect) separation between H-sdB and He-sdOB stars. We thus confirm that the He-sdOBs account for the majority of the blue hook region.
As for the H-sdOs (red circles), although they cluster on the blue side of the EHB (at low $U-V$), they cannot be easily isolated from the other spectroscopic groups by their position in the CMD. 
On Fig.~\ref{cmd} we also see that the most He-rich stars (purple circles) are among the faintest objects, as expected from \cite{bro01}. 
As discussed above, our spectroscopic samples are subject to selection effects and the He-sdOBs are likely overrepresented. Another estimate of the fraction of He-sdOBs present in \omcen\ can be made using the position of the stars in the CMD. We first calculated the number of stars in the WFI catalog (gray dots in Fig. \ref{cmd}) that are found in the blue hook region ($V > 18.5$ and $U-V < -0.5$), 216 stars, and in the EHB region ( $17.2 < V < 18.5$), 210 stars. Considering the fraction of He-sdOBs spectroscopically observed in each region (i.e., 85\% in the blue hook region and 19\% in the EHB region) we estimated a fraction of He-sdOBs of 52\%. This number, although a rough estimate, should be less affected by selection effects than the higher fraction (57\% to 64 \%) obtained from our spectroscopic sample only.

\subsection{Mass distribution}

\begin{figure}[t]
\begin{center}
\resizebox{\hsize}{!}{\includegraphics{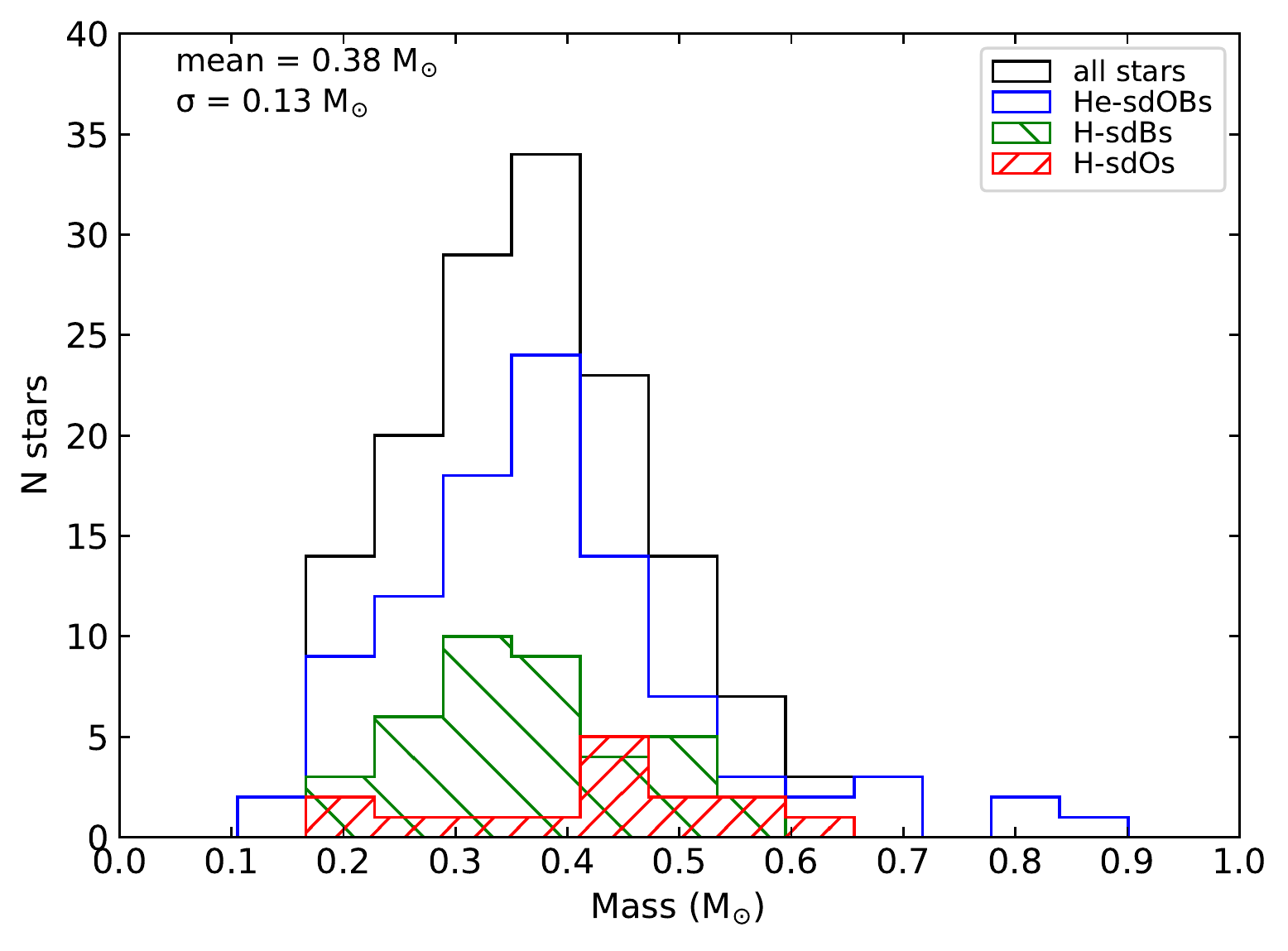}}
\caption{Distribution of spectroscopic masses for all stars in the sample as well as separated for the three spectroscopic groups. }
\label{dmass}
\end{center}
\end{figure}

Using the atmospheric parameters, $V$ magnitudes and the distance to \omcen, we derived spectroscopic masses following the method described in Sect. 3.3. The resulting masses and their uncertainties are presented in Table~\ref{bigtable}, and the total mass distribution is shown by the black histogram in Fig.~\ref{dmass}.  The distribution is characterized by a mean mass of 0.38 \msun\ and a standard deviation ($\sigma$) of 0.13 \msun.  This derived average mass is uncomfortably low given that the canonical value required to ignite helium in the core is $\sim$0.45 \msun. However, low masses have been reported previously for HB and EHB stars in \omcen\ \citep{moni2011,moni12,lat17}. 

Figure~\ref{dmass} also shows the individual mass distributions of the three spectroscopic groups, namely the H-sdBs, He-sdOBs and H-sdOs.  The most populous groups (He-sdOBs and H-sdBs) display a similar shape in their distribution, with the exception that no high-mass objects ($>$0.6~\msun) are found among the H-sdBs. This contrasts with the presence of some EHB stars with high spectroscopic masses reported by Moni Bidin et al. (\citeyear{moni07,moni09}) in NGC~6752 and M~80.
The average masses of the stars in both spectroscopic groups are also similar, with a mean mass of 0.38~\msun\ for the He-sdOBs and 0.36~\msun\ for the H-sdBs. As for H-sdOs, their average mass is 0.41 \msun. This is slightly larger (by 0.03 \msun) than the mean mass of the rest of the sample, but this difference is not as large as the one reported by \citet{lat17} (0.13~\msun) based on the FORS2.6 sample only.
The mass of an EHB star consists almost entirely of its He-core, and the hydrogen envelope contributes to at most 0.02 \msun. 
Given the conditions required to ignite helium under degenerate conditions, the possible range of core masses is not very extended and we do not expect the stars of different spectral groups to show statistically significant differences in masses. Moreover, given that the H-sdOs are thought to be the direct progeny of the cooler EHB stars, they are expected to have similar masses. In this regard at least our results are self-consistent\footnote{We performed $t$-Tests comparing the masses of each spectroscopic groups with the masses of all the other stars to check that the mass distributions are not statistically different.}.

Our method for deriving the spectroscopic masses is based on the emergent flux predicted by our own grid of model atmospheres given the specific atmospheric parameters (\teff, log $g$, and \nhe) of each star. Some previous investigations used a bolometric correction to estimate the theoretical stellar flux required to derive the stellar mass (e.g., \citealt{moni12}). Since our own method led to stellar masses significantly lower than predicted from evolutionary models, we decided to re-compute our masses using the method presented in \cite{moe17}, using the bolometric corrections of \citet{flo96}. For consistency we used the same values of reddening and distance modulus as in Sect. 3.3.   

\begin{figure}[t]
\begin{center}
\resizebox{\hsize}{!}{\includegraphics{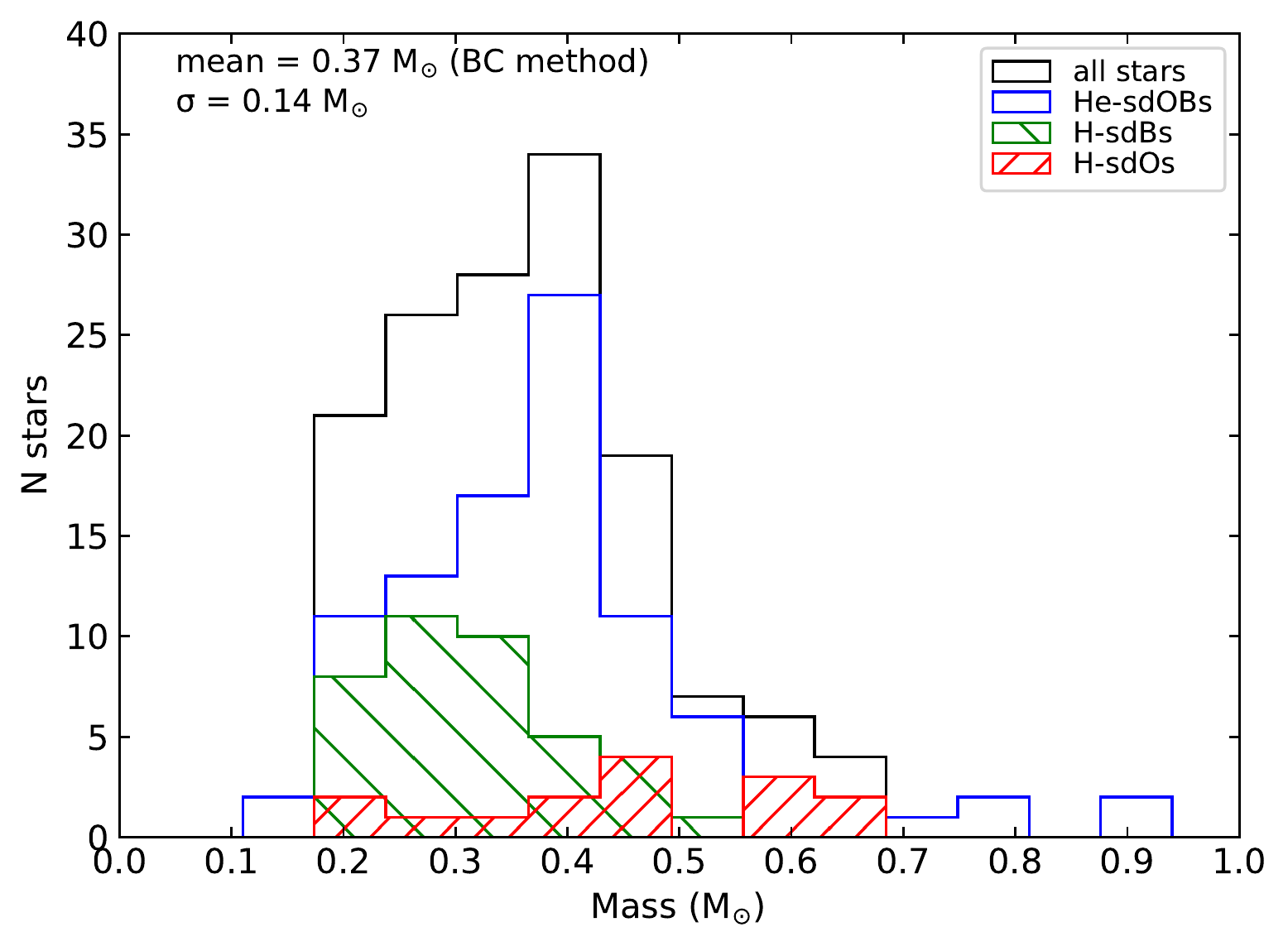}}
\caption{Same as Fig.~\ref{dmass} but for masses computed using the bolometric correction method.}
\label{massbc}
\end{center}
\end{figure}

The total mass distributions derived using the two different methods have a very similar shape (see Fig.~\ref{dmasscomp}) and the average mass obtained with the bolometric correction (BC) method, 0.37 \msun, is even slightly lower than the average mass of 0.38 \msun\ obtained previously with the spectroscopic method. The use of an empirical BC for computing the masses has one obvious caveat; the correction (applied to the \teff\ range of our EHB targets) is derived using hot main-sequence (MS) stars that have a solar helium abundance, but the stars in our sample have helium abundances varying from one thousandth to a hundred time solar. As this could induce some systematic effects on the derived masses, we also looked at the mass distribution and mean mass per spectroscopic group (Fig. \ref{massbc}). This is in fact more revealing than the average distribution as it highlights differences between the spectroscopic groups. The H-sdBs are found to have, on average, significantly smaller masses (0.31 \msun) than the He-sdOBs (0.39 \msun), while the H-sdOs appear to be even more massive (0.45 \msun).  Such large differences in mass between the three spectroscopic groups are not expected and undermine the reliability of the BC method when applied to stars with different helium abundances.
In addition, the abundance patterns of hot subdwarfs are in general quite different from the solar abundances of the MS stars used for calibrating the bolometric corrections \citep{geier13}.

Notwithstanding the differences in the individual masses obtained with the two methods, we are in both cases left with the puzzling conclusion that the masses derived for our EHB stars are on average smaller than those predicted from evolutionary models. 

\begin{table}[ht]
\center
\caption{Radial velocity statistics for 75 EHB stars in \omcen\ from the VIMOS data.}\label{tabrvvimos}
\scriptsize
\begin{tabular}{llclll}
\hline
\hline
ID & Spectral type & No. of spectra & RV$_{\rm average}$ & $\Delta$ RV$_{\rm stdev}$ & ln $p$ \\
     &     &    & ${\rm [km\,s^{-1}]}$ &  ${\rm [km\,s^{-1}]}$  &    \\
\hline
 5125408 & He-sdOB & 28 & 238.5 & 29.1 & -11.68\\
 5131557 & He-sdOB & 10 & 204.1 & 46.2 & -10.22\\
\hline
  5034421 & sdO & 24 & 237.9 & 32.1 & -9.07\\
  5132323 & He-sdOB & 24 & 217.1 & 35.5 & -7.31\\
  5150273 & He-sdOB & 25 & 229.3 & 26.2 & -6.25\\
  157448 & sdB & 20 & 229.2 & 31.2 & -5.75\\
  5097663 & He-sdOB & 13 & 238.6 & 28.1 & -5.62\\
  5262593 & sdB & 24 & 193.7 & 26.7 & -4.22\\
  5153131 & He-sdOB & 18 & 258.3 & 30.2 & -4.17\\
  5196769 & sdO & 30 & 250.1 & 28.6 & -4.15\\
  5128088 & sdO & 19 & 230.5 & 27.6 & -4.14\\
  5307782 & He-sdOB & 22 & 238.3 & 24.9 & -4.05\\
  5341196 & He-sdOB & 22 & 226.0 & 26.4 & -3.96\\
  5214452 & He-sdOB & 34 & 258.3 & 22.1 & -3.85\\
  5276767 & He-sdOB & 20 & 213.3 & 24.1 & -3.56\\
  5094098 & He-sdOB & 26 & 236.8 & 24.4 & -3.55\\
  5338760 & sdB & 19 & 199.8 & 31.8 & -3.46\\
  5268317 & He-sdOB & 23 & 223.6 & 24.1 & -3.46\\
  5131824 & sdB & 25 & 201.5 & 26.5 & -3.3\\
  5039935 & He-sdOB & 9 & 239.3 & 23.8 & -2.98\\
  5180639 & He-sdOB & 26 & 263.0 & 24.6 & -2.97\\
  155799 & He-sdOB & 21 & 243.8 & 22.1 & -2.95\\
  5222459 & He-sdOB & 24 & 218.2 & 22.6 & -2.88\\
  5148322 & He-sdOB & 32 & 241.5 & 21.6 & -2.69\\
  5207762 & He-sdOB & 17 & 250.3 & 30.1 & -2.59\\
  5156440 & He-sdOB & 18 & 260.3 & 23.8 & -2.5\\
  5283552 & sdB & 25 & 228.8 & 23.4 & -2.46\\
  274052 & He-sdOB & 18 & 237.0 & 22.8 & -2.37\\
  5121885 & sdO & 40 & 242.0 & 21.4 & -2.36\\
  5306037 & He-sdOB & 27 & 212.5 & 22.9 & -2.18\\
  5123061 & He-sdOB & 32 & 220.6 & 21.5 & -1.97\\
  5165122 & He-sdOB & 37 & 246.2 & 20.8 & -1.79\\
  5091999 & sdB & 9 & 224.8 & 23.7 & -1.74\\
  5032350 & He-sdOB & 10 & 177.7 & 22.9 & -1.73\\
  5166220 & sdO & 32 & 256.6 & 21.3 & -1.71\\
  5119720 & He-sdOB & 29 & 237.5 & 21.2 & -1.67\\
  168035 & sdB & 40 & 214.5 & 23.4 & -1.6\\
  176008 & He-sdOB & 8 & 264.5 & 24.7 & -1.55\\
  5193651 & sdB & 35 & 219.1 & 22.8 & -1.5\\
  257150 & sdB & 32 & 224.4 & 20.1 & -1.21\\
  5138707 & He-sdOB & 21 & 214.0 & 22.0 & -1.18\\
  165943 & He-sdOB & 31 & 236.9 & 19.8 & -1.14\\
  5296709 & sdO & 22 & 219.6 & 22.0 & -1.13\\
  5136690 & sdB & 34 & 231.6 & 23.4 & -1.08\\
  5170422 & He-sdOB & 26 & 249.5 & 21.5 & -1.06\\
  170679 & He-sdOB & 31 & 226.4 & 20.0 & -1.0\\
  5299498 & sdB & 7 & 269.8 & 20.8 & -0.89\\
  5226206 & He-sdOB & 16 & 203.0 & 21.7 & -0.8\\
  5137388 & He-sdOB & 19 & 233.5 & 18.3 & -0.7\\
  281063 & sdO & 32 & 271.3 & 21.7 & -0.68\\
  5183041 & He-sdOB & 19 & 222.2 & 23.0 & -0.64\\
  5370155 & He-sdOB & 28 & 247.9 & 18.8 & -0.64\\
  182549 & sdB & 6 & 219.9 & 18.8 & -0.63\\
  5142759 & He-sdOB & 29 & 211.7 & 17.7 & -0.61\\
  183403 & sdB & 11 & 218.2 & 17.5 & -0.58\\
  177825 & sdB & 31 & 231.6 & 20.2 & -0.56\\
  177238 & sdO & 3 & 198.6 & 16.4 & -0.5\\
  5094822 & He-sdOB & 32 & 253.5 & 16.3 & -0.46\\
  5238307 & sdB & 33 & 234.2 & 24.7 & -0.44\\
  5359493 & He-sdOB & 26 & 227.8 & 17.2 & -0.43\\
  5124244 & He-sdOB & 21 & 248.1 & 19.5 & -0.41\\
  5151410 & He-sdOB & 20 & 219.0 & 23.5 & -0.4\\
  165237 & He-sdOB & 31 & 249.0 & 18.7 & -0.38\\
  5102767 & He-sdOB+MS & 12 & 233.1 & 15.2 & -0.29\\
  264670 & He-sdOB & 11 & 228.1 & 17.9 & -0.27\\
  254318 & He-sdOB & 38 & 221.3 & 15.2 & -0.18\\
  5347296 & He-sdOB & 4 & 227.0 & 11.7 & -0.18\\
  5317711 & sdO & 12 & 265.5 & 14.3 & -0.15\\
  5141232 & He-sdOB & 37 & 232.5 & 15.5 & -0.14\\
  5103569 & He-sdOB & 37 & 233.9 & 16.1 & -0.11\\
  273649 & He-sdOB & 40 & 219.6 & 14.8 & -0.08\\
  5111007 & He-sdOB & 33 & 232.0 & 14.8 & -0.04\\
  5295674 & He-sdOB & 34 & 246.5 & 14.7 & -0.04\\
  5179481 & sdB & 40 & 262.9 & 16.5 & -0.01\\
  5114452 & sdO & 9 & 207.4 & 8.3 & 0.0\\
\hline
\end{tabular}
\end{table}

\subsection{Search for binaries in the VIMOS sample}

In addition to the atmospheric parameters of our EHB stars, we are also interested in their radial velocity properties. Of particular interest is the apparent general lack of close binaries among EHB stars in globular clusters (e.g., NGC~6752, M~80, and NGC~2808, Moni Bidin et al. \citeyear{moni06,moni09,moni11binary}), compared to a fraction of about 50\% among the field sdBs \citep{max01,nap04}. 
The closest binary systems observed in the field have periods ($P$) $\sim$0.05$-$0.3 d and semi-amplitudes ($K$) of typically 50$-$200 \kms\, while the longer period systems ($P$ up to 10 days) have values of $K$ down to 40$-$60 \kms\ \citep{kup15}. Given the short exposure time (600 s), the multiple epochs (14) obtained over 32 months, and the typical uncertainties on the individual RVs ($\pm$ 20 \kms), \footnote{this includes both statistical and systematic uncertainties, see Sect. 3.1.} our VIMOS observations are tailored to detect close binaries with properties similar to those in the field. Our VIMOS dataset is not only the largest ever obtained for this purpose in any globular cluster, but also the first to include blue hook stars. 

To estimate the fraction of false detection produced by statistical fluctuations and the significance of the measured RV variations, we used the method described in \citet{max01}. For each star we calculate the average velocity from all measured epochs and assuming this velocity to be constant, we calculate the $\chi^{2}$. Comparing this value with the $\chi^{2}$-distribution for the appropriate number of degrees of freedom we calculate the probability ($p$) of obtaining the observed value of $\chi^{2}$ or higher from random fluctuations around a constant value. The natural logarithm of the false-detection probability (ln $p$) is given in Table \ref{tabrvvimos} for each of the 75 stars that are part of the RV survey. Table \ref{tabrvvimos} also provides the average RV for each star, the standard deviation $\sigma$ of the individual RVs as well as a spectral classification, which is especially useful for the stars that did not have their atmospheric parameters derived.
We consider the detection of RV variability to be significant if the false-detection probability is smaller than $0.01\%$ (ln $p$ $<-9.2$). 

\begin{figure*}
\begin{center}
\includegraphics[width=0.48\linewidth]{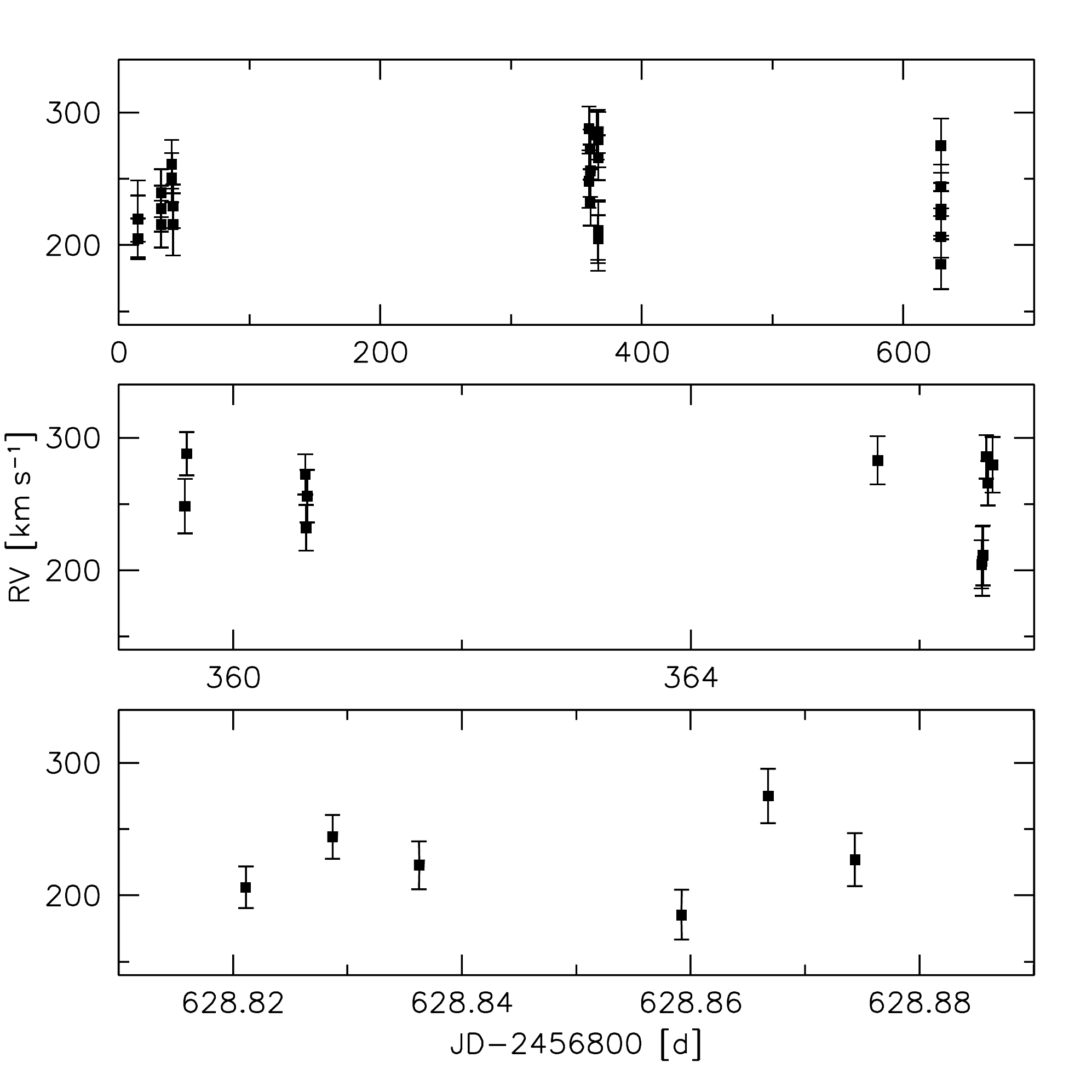}
\includegraphics[width=0.48\linewidth]{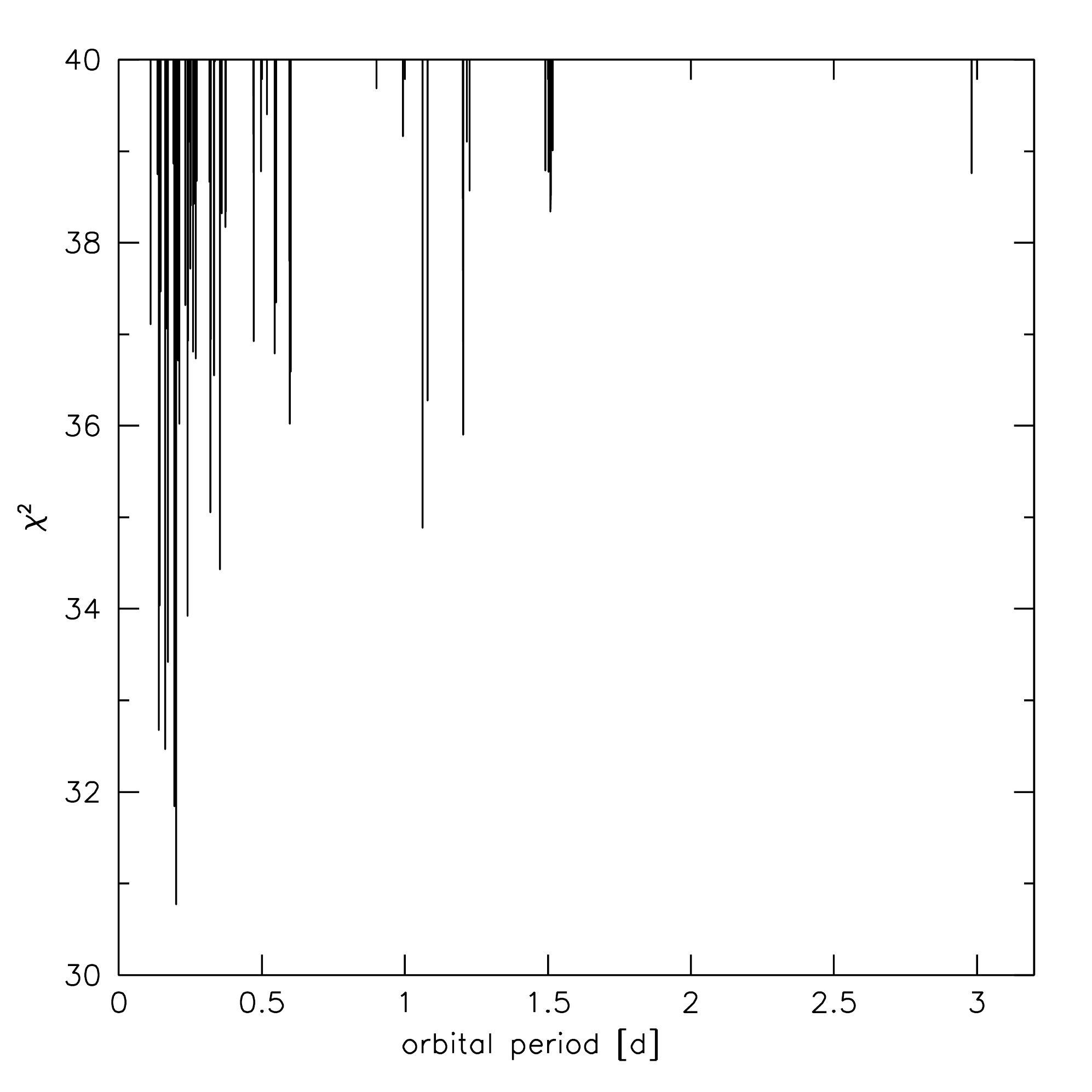}
\caption{$Left:$ Radial velocities measured for one of the close-binary candidates, 5125408. The different panels show different time ranges and as such illustrate the RV variations detected on different time scales. $Right:$ $\chi^2$ from the period fitting routine plotted against binary period. No unique orbital solution could be found. }
\label{rvcurve}
\end{center}
\end{figure*}

We only found two stars among the full sample of 75 stars that show statistically significant RV variations according to this criterion. Star 5125408 shows a maximum RV shift 
of about 100 \kms\ and star 5131557 appears to have a maximum RV shift of about 140 \kms. 
Although both stars are part of the He-sdOB spectroscopic group, they are among the coolest and least He-rich stars of that group.
Assuming circular orbits, sine curves were fitted to the RV data points of the two binary candidates in fine
steps over a range of test periods. For each period the $\chi^2$ of the best fitting sine curve was determined. The result is similar to a power spectrum with the lowest $\chi^2$ indicating the most likely
period \citep{geier11}. No unique solution could be found.
The RVs measured for 5125408 as well as the $\chi^2$ distribution are shown in Fig.~\ref{rvcurve}.
Since based on the maximum RV shifts measured the RV semi-amplitudes of both systems should be about 60 \kms, the individual uncertainty of the RVs (about 20 \kms) might simply be too high to find a significant solution. Alternatively, the sampling of the RV curves might be insufficient to solve the orbit. 

Assuming that these two targets are indeed binaries and the companions are white dwarfs with 0.5 \msun\ (which is the case for many of the close companions to field sdBs, see \citealt{kup15}), the orbital periods would very likely exceed several days. If the companions are M dwarfs with 0.1 \msun, which is also quite typical of field sdB binaries, the periods should be on the order of 0.1-0.2 days. In this case, characteristic sinusoidal variations caused by reflection effects and/or eclipses should be visible in the light curves of the binaries. The lack of such variations would be an indication for compact companions like WDs.

Here we consider these two stars to be close-binary candidates. Follow-up observations are needed to confirm their binary nature and put constraints on the orbital parameters. For the remaining 73 stars of our VIMOS sample, we can adopt the average RV uncertainty of 20 \kms\ as an upper limit for any RV variations. The upper limit for the RV semi amplitudes of hidden close binaries should then be about twice this number. As can be seen in Fig. 6 of \citet{kup15}, this excludes most known types of sdO/B close binaries. Any undetected binaries would have low-mass companions ($<0.2\,M_{\rm \odot}$) and/or orbital periods of several days quite different from the known field population.

\subsubsection{Binary fraction among the EHB stars of \omcen}

\begin{figure*}[t]
\begin{center}
\includegraphics[width=0.49\linewidth]{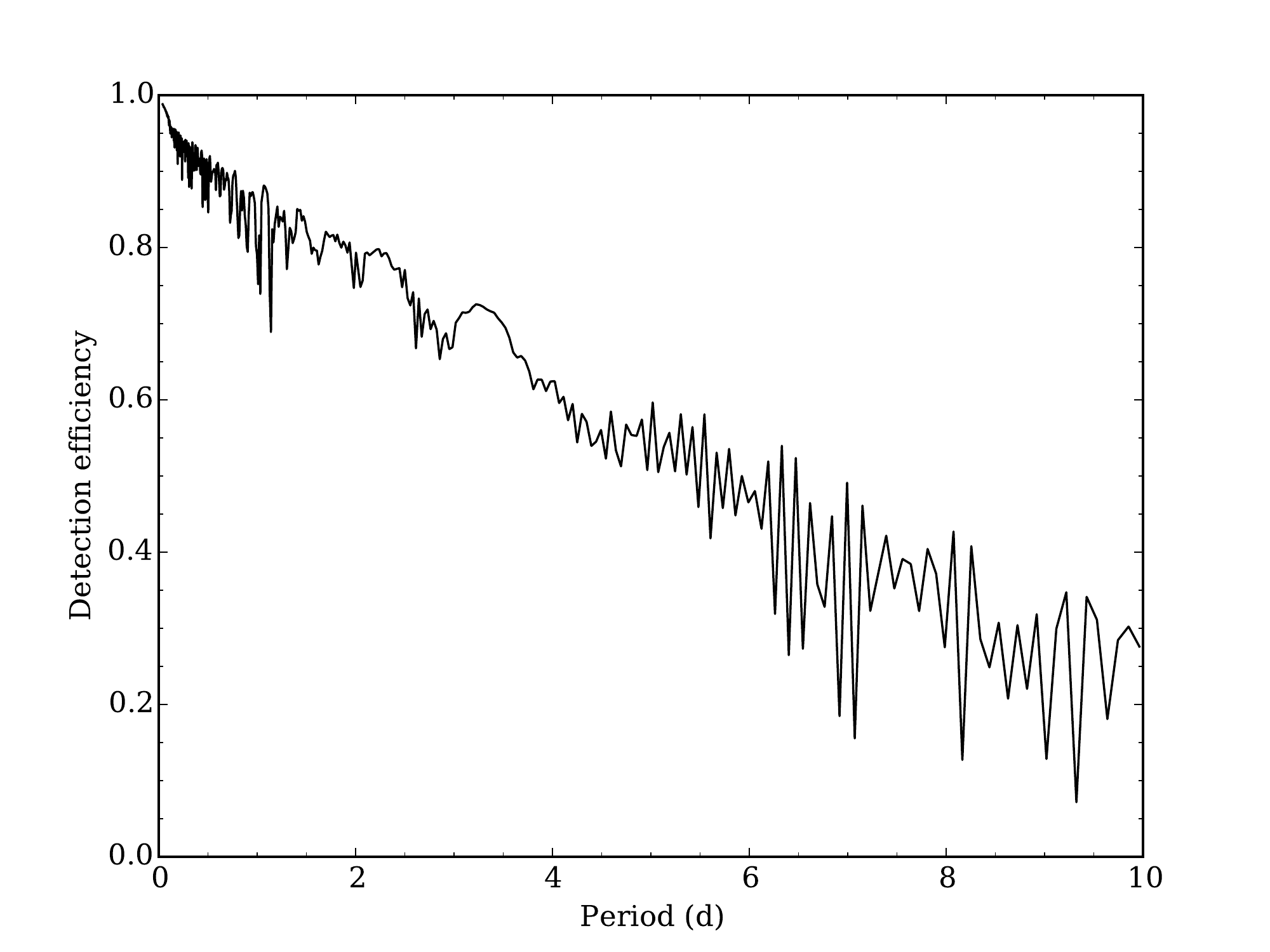}
\includegraphics[width=0.49\linewidth]{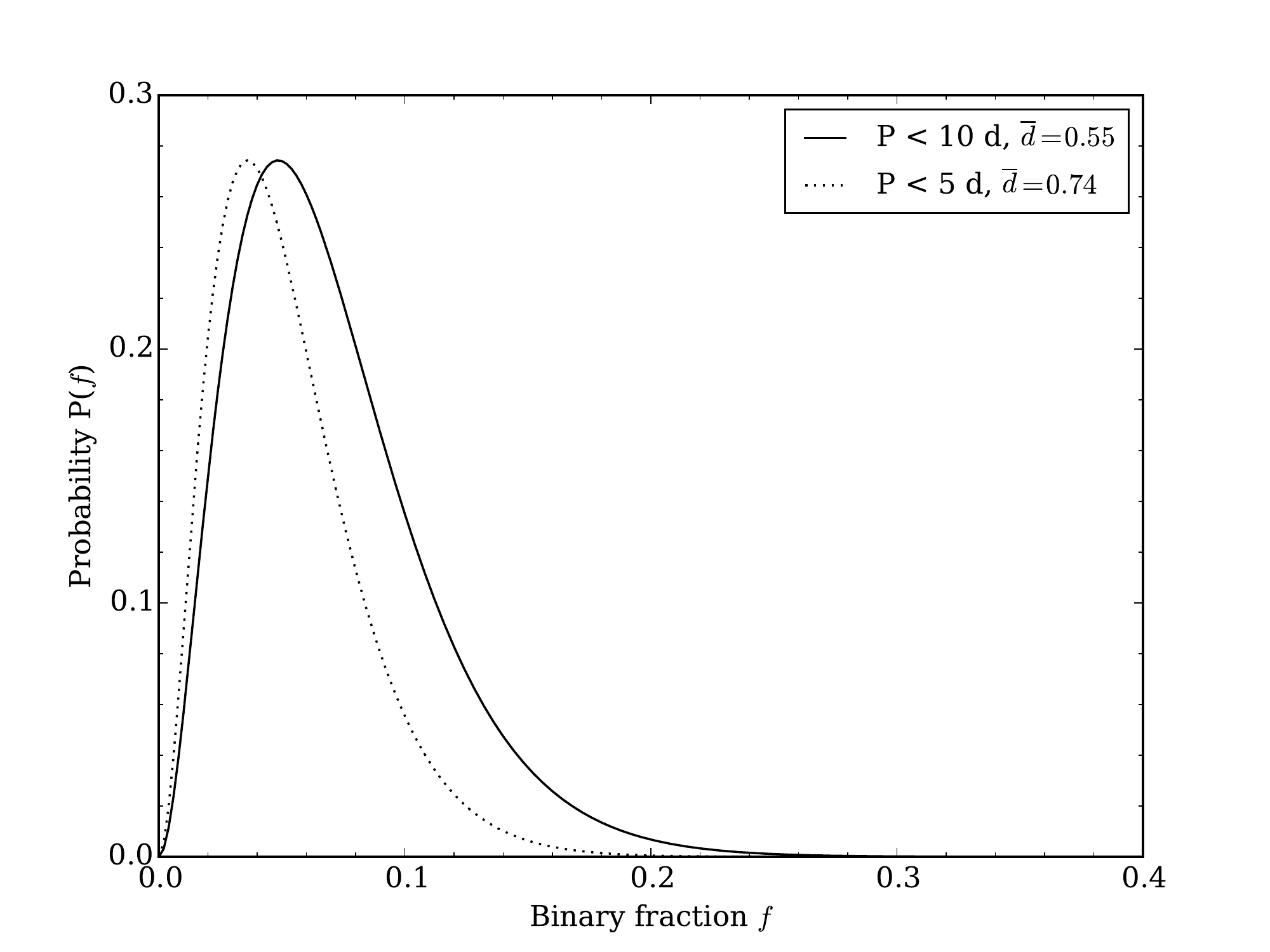}
\caption{$Left: $ Average detection efficiency versus period for the 75 stars of the VIMOS sample. $Right: $ Probability curve of having a binary fraction $f$ in our VIMOS sample for binaries with periods shorter than 10 days (solid) and shorter than 5 days (dotted).}
\label{bin_frac}
\end{center}
\end{figure*}

The two binary candidates that we discovered among our VIMOS sample suggests a  binary fraction $f$ of about 2.7\%, assuming that we detected all the binaries among our sample. However the detection efficiency ($\overline{d}$) can never be 100\% due to the possible inclinations of the systems, and it is also a strong function of the binary period. Given a fraction of binaries $f$ in a sample of $N$ stars and a detection efficiency $\overline{d}$, the probability of detecting $N_B$ binaries is 
\begin{equation}
P = \frac{N!}{(N-N_B)!N_B!}(\overline{d}f)^{N_B}(1 - \overline{d}f )^{N-N_B}.
\end{equation}
With a detection efficiency of $\overline{d} = 1$ and two detected binaries out of 75 stars, the probability function indeed peaks around 2.7\%. However, a better estimate can be obtained by evaluating the detection efficiency of the VIMOS survey. For that we closely followed the method used by \citet{max01} in their study of field EHB stars to compute the detection efficiency at a given period $P$ for each star in our sample. First, we assume masses of 0.5~\msun\ for both the EHB and companion star. Such a companion mass actually corresponds to one of the peak in the companion mass distribution for hot subdwarfs in the field and largely corresponds to WD companions \citep{kup15,kawka15}. We then compute the maximum value of the semi-amplitude $K_{max}$ (i.e., for an inclination of $i=90^{\circ}$) assuming a circular orbit. We used the observation times $T_{\rm obs}$ of the star to simulate a set of mock RVs of an hypothetical binary using the equation
\begin{equation}
v_{rad} = K_{\rm max} sin(\frac{2\pi}{P}(T_{\rm obs}-T_0)).
\end{equation}
The mock RVs are then used to compute the $\chi^2$ value of this hypothetical binary, $\chi^2_{\rm max}$, using the RV uncertainties of the actual observations. 
This calculation was repeated over 50 values of $T_{0}$ to cover all possible orbital phases and averaged to $\chi^2_{\rm max}$. We can then compare the value of $\chi^2_{\rm max}$ to the value required, $\chi^2_{\rm crit}$, to satisfy our detection criterion (ln $p$ $<-9.2$; see previous section). If $\chi^2_{\rm max} < \chi^2_{\rm crit}$ then no binaries with that orbital period can be detected and $d$ is zero. Otherwise, we calculate the semi-amplitude for which $\chi^2_{\rm max} = \chi^2_{\rm crit}$, $K_{\rm crit}$. For randomly oriented orbits, $i$ is distributed as $\cos{i}$, so the detection efficiency for a given combination of observations, period and mass is $d = \sqrt{1-(K_{\rm crit}/K_{\rm max})^2}$. This way, we computed $d$ for periods up to 10 days for every star in the VIMOS survey and took the average values to produce the detection efficiency curve presented in the left panel of Fig. \ref{bin_frac}.

To calculate the binary fraction of our sample using Eq. 4, the detection efficiency must be averaged over a certain range of periods. Ideally, one would compute a weighted mean of $d$ over the period distribution of known binaries. Although the period distribution of EHB binaries in the field is rather well described, it is not at all clear that this can be blindly applied to globular clusters, especially considering the only EHB binary in a globular cluster that has a known orbital solution has quite peculiar characteristics (period and companion mass) when compared to the field population. We thus simply use the straight average of the detection efficiencies, which we calculated for periods up to 5 days ($\overline{d}= 0.74$) and 10 days ($\overline{d}= 0.55$). The resulting probability curves are illustrated in the right panel of Fig. \ref{bin_frac},  and peak at 3.6\% for binaries with $P < 5$ days and 4.8\% for binaries with $P < 10$ days. The probability curve reaches a 1\% probability for a binary fraction of 14\% and 18.5\% for periods up to 5 and 10 days respectively. These values can be seen as a conservative upper limit on the binary fraction of our sample. 

In addition to WD companions, low-mass M dwarfs stars are also common companions to hot subdwarfs in the field, where they are usually found in close orbits ($P < 1$ d; \citealt{kup15,kawka15}). We computed additional detection efficiency curves assuming this time a 0.13~\msun\ companion. A lower mass companion produces smaller RV variations, leading to a rather low detection efficiency of $\overline{d} = 0.29$ for $P < 1$ d. Such numbers would result in a most likely binary fraction of  9\%, with an upper limit (1\% probability) of 35\%.

\section{Discussion}

\subsection{Contrasting the properties of the field and \omcen\ EHB stars.}
\subsubsection{Atmospheric properties}
\begin{figure*}
\begin{center}
\resizebox{\hsize}{!}{\includegraphics{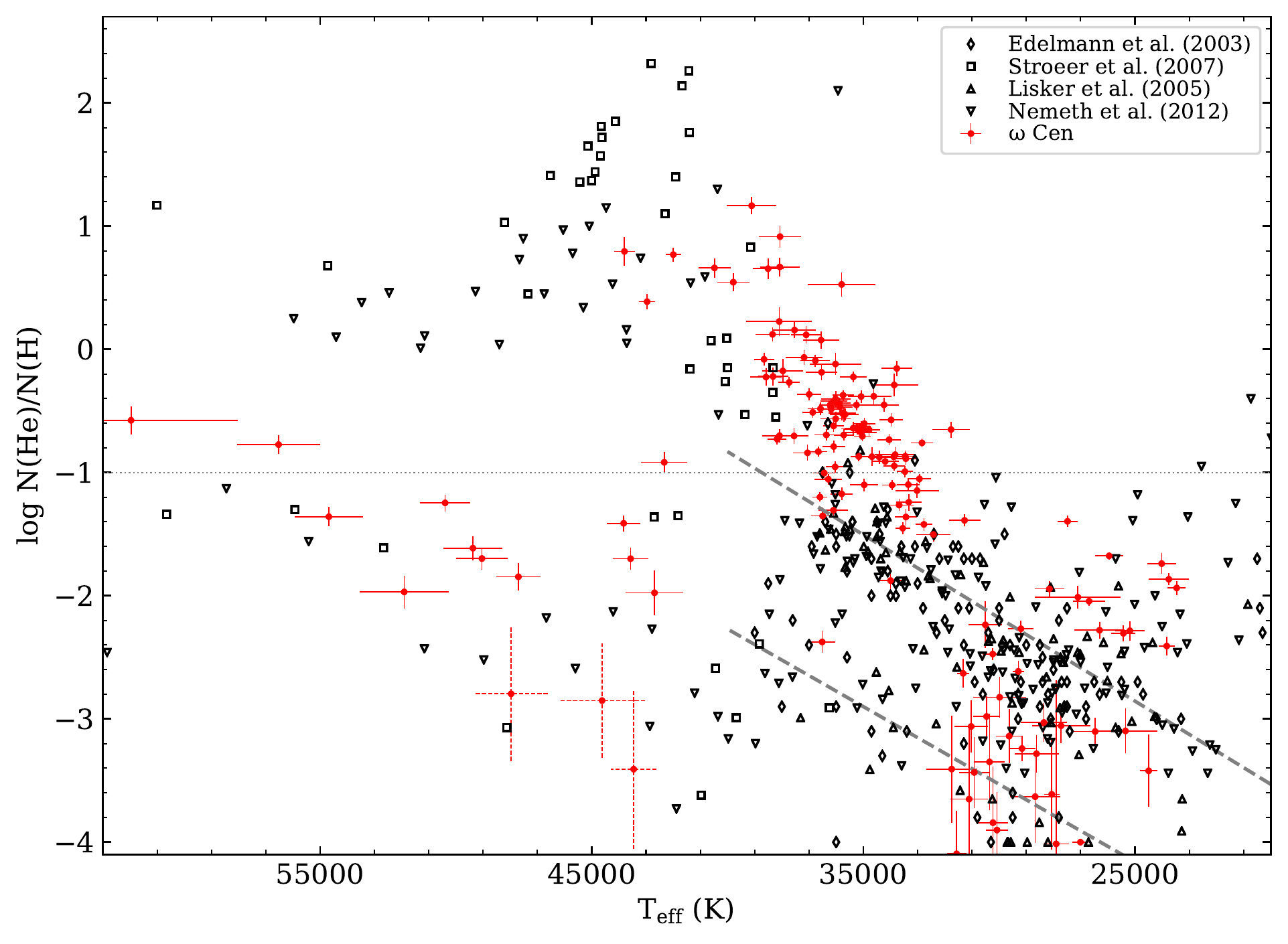}}
\caption{Helium abundance versus \teff\ for the EHB stars in \omcen\ (red circles) and for hot subdwarf stars among the galactic field population from the samples of \citet{ede03}, \citet{lisk05}, \citet{stro07}, and \citet{nem12}. The three H-sdOs with uncertain parameters are indicated with dashed errorbars (see Sect. 4.2). The two helium $-$ \teff\ sequences identified by \citet{ede03} are shown in dashed lines. }
\label{complitt}
\end{center}
\end{figure*}

The populations of EHB stars in globular clusters and the galactic field are different in terms of spectroscopic properties and binary fractions.
A good summary of the spectroscopic properties of hot subdwarfs from the galactic field population, based on the results of five major studies, is presented in Sect. 2.2 of \citet{heb16}. In short, the sdB stars (\teff~$\la$~40 000 K) outnumber the sdOs by a factor of $\sim$3, and the vast majority of sdB stars ($\sim$95 \%) have a hydrogen-rich atmosphere \citep{heb09,nem12}. The picture is however different among the sdO spectral type, where about two-thirds of the stars have an atmosphere that is strongly enriched in helium (log \nhe\ $\ga$ $-$1.5; \citealt{stro07,nem12}). Those stars are usually refereed to as He-sdOs.

Our sample of EHB stars in \omcen\ is the first in any globular cluster that is large enough to be comparable to surveys undertaken for the field EHB star population.
One of the main differences between the population observed in \omcen\ and that in the field lies in the helium composition of the stars with \teff\ below 38\,000 K. 
We illustrate this in Fig.~\ref{complitt}, where we compare the distribution of our \omcen\ stars in the log \nhe\ $-$ \teff\ diagram with that of four samples \citep{ede03,lisk05,stro07,nem12} representative of the field hot subdwarf population (see also Fig. 5 of \citealt{heb16}).
As mentioned above, in the galactic field these stars form a homogeneous group of helium-poor sdBs, while in \omcen\ the stars cooler than 38\,000 K form two very distinct groups based on their temperature and helium abundance.  
In fact, stars with \teff\ and helium abundance similar to the majority of the EHB stars in \omcen\ (the He-sdOBs) are rare among the field population (see also Fig. 23 of \citealt{heb16}), which consists mostly of hydrogen-poor sdBs. The distribution of the field sdBs follows a \teff\ $-$ \nhe\ relation along two ``sequences"  identified by \citet{ede03} that are also indicated in Fig.~\ref{complitt}. While our He-sdOBs also follow a clear \teff\ $-$ \nhe\ trend, the helium abundances of the H-sdBs in \omcen\ are much more scattered.
Among the hot subdwarfs in the galactic field, there is an important population of He-sdO stars showing an atmosphere even richer in helium than any of our \omcen\ stars. As can be inferred from Fig.~\ref{cmd}, a higher helium content in the atmosphere results in a fainter magnitude. It therefore seems possible that such objects might have simply been missed in our spectroscopic observations, since in \omcen\ we are generally more limited by the magnitude of our targets than is the case for field star studies. However, looking more closely at the CMD we find that there are few stars fainter than our targets in the EHB of \omcen, therefore it would seem that \omcen\ does not harbour a significant population of these very He-rich sdOs. 

The differences between the hot subdwarf populations of \omcen\ and the galactic field could be related to different formation mechanisms at play in both environments as well as to differences in the metallicity and age of the progenitor stars. The hot subdwarfs included in field surveys are mostly bright objects belonging to the galactic disk \citep{alt04,mar17}, thus they are likely to have younger and more metal rich progenitors than their counterparts in \omcen. \citet{lat14} suggested that the \omcen\ He-sdOB population may have a counterpart among the galactic halo stars, an idea that was initially supported by the presence of similar objects in the SDSS sample, which is thought to contain more halo targets than other studies in the field (\citealt{hirschthesis}, P. N\'{e}met, priv. comm, 2014).  However, the preliminary results of \citet{geier17} could not confirm this hypothesis. The authors kinematically identified halo sdBs from the subdwarf catalog of \citet{geiercat} and found that only 23 \% of their sample corresponded to our He-sdOB type\footnote{We note that their sample only included stars with \teff\ up to 40\,000~K.}.  That fraction is larger than that of their disk sample (5\%) but nowhere near as high as observed in \omcen.

In spite of the differences between the \omcen\ and the field populations, Fig.~\ref{complitt} highlights a common characteristic: the relationship between effective temperature and helium abundance in the sdB and sdOB stars. Although the majority of stars are found at different effective temperatures in the field and in \omcen, the stars with \teff\ $\la$ 40 000 K show a clear positive correlation between \teff\ and the helium abundance. This relation was first uncovered by \citet{ede03}, who also found that a small fraction of the sdBs ($\sim$10\%) had a helium abundance about 1$-$1.5 dex lower than the other stars. The two ``helium sequences'' identified by \citet{ede03} are indicated in Fig.~\ref{complitt}. Although this correlation between the effective temperature and helium abundance is well documented by the different surveys among field sdBs it is not yet fully understood \citep{oto08}.  
 In fact, the helium abundances observed in sdBs are strongly influenced by diffusion processes. The abundances are larger than can be accounted for considering the balance between radiative levitation and gravity, thus an additional mechanism is needed to counteract gravitational settling \citep{mich89}. For example, a weak stellar wind ($\dot{M}$ $\approx$ 10$^{-13}-10^{-14}$ \msun\ yr$^{-1}$) could prevent the helium from sinking \citep{font97,ung01}. Also, weak turbulent mixing in the upper atmosphere can be invoked to explain the observed helium abundances \citep{hu11,mich11}. \citet{ung05} performed diffusion calculations and found that mass loss rates of $\dot{M}$ $\la$ 10$^{-14}$ \msun\ yr$^{-1}$ can produce an atmospheric composition (of H and He) similar to that of the helium-enriched EHB stars in \omcen\ and NGC~2808 \citep{moe04} if the initial ratio of H/He is $\approx$0.02. This initial amount of hydrogen is larger than what is predicted by the late-flasher scenario \citet{cass03}, but could be explained by the shallow-mixing case of \citet{mil08} or by a lower mixing efficiency during the helium flash.  
However, the \teff\ $-$ helium correlation has not been specifically addressed in previous work and remains unexplained. Nevertheless, it seems clear that diffusion plays an important role in shaping this correlation that is definitely present, and similar, in both populations. The diffusion processes make it more difficult to directly connect the measured atmospheric abundances with the predictions from evolutionary models, since these do not usually include all the processes at play in the very thin photospheric layers. 
 
 \subsubsection{Binarity}

An important difference between the EHB populations of the field and \omcen\ concerns the fraction of stars in short binary systems (P $\la$ 10 d). As reported in Sect. 4.4, our search for such systems among 75 EHB stars in \omcen\ did not allow us to characterize any binary systems. We detected only two binary candidates that showed statistically significant RV variations, but for which no orbital solution was found. Our estimated binary fraction derived from the probability curve shown in Fig.~\ref{bin_frac} is about 5\% with an upper limit of 18.5 \% at a confidence level of 99\%. This applies to binaries with $P < 10$ d and companion's mass of 0.5~\msun and indicates a rather low fraction of short period binaries among the EHB stars of \omcen\ \footnote{Our survey would not have detected EHB stars in wide binaries; the amplitudes of such binaries are too low and they would in any case have been excluded by our selection criteria}.
Our findings are in line with the lack of such binaries in other globular clusters, namely NGC~6752, NGC~5986, M~80, and NGC~2808 (Moni Bidin et al. \citeyear{moni06,moni09,moni11binary}), compared to a fraction of about 50\% among the field sdBs \citep{max01,nap04,cop10}. 
Besides the two binary candidates we found in our sample, only five other EHB star binary candidates are known in other GCs. \citet{moni09} reported the discovery of significant RV shifts with a maximum variation of 26.9$\pm$7 \kms\ for the H-sdB 4175 in their study of NGC~5986. In NGC~2808, \citet{moni11binary} found the three RV variable candidates 9519, 7700, and 55759 with maximum RV shifts of 70 \kms, 40 \kms and 60 \kms, respectively. The spectral type of those stars was not determined, but the temperatures derived from photometry indicate cool EHB objects (\teff\ $\sim$20\,000 K). Compared to the RV shifts detected for our two candidates, those shifts are somewhat lower. 

The most interesting EHB binary candidate to date was found in NGC 6752 by \citet{moni08,moni15}. In addition to a significant RV shift of 16.1$\pm$1.6 \kms measured from high resolution spectra obtained with FLAMES-GIRAFFE, this sdB star shows the signature of a cool MS companion in its spectrum. Follow-up spectroscopy has been obtained and preliminary results indicate an orbital period of several days (Moni Bidin priv. comm.). This binary is therefore not only unique in a GC, but there is also no counterpart in the field population, because all solved sdB+MS binaries have periods of several hundred days (e.g., \citealt{vos17}).

Including the results presented here, a total of seven RV variable EHB stars are now known in three GCs, but none of them could so far be fit with a credible orbital solution. Although this has rarely been attempted before this work, a likely reason for this lack of orbital solution is the combination of rather small RV variations with rather long orbital periods on the order of several days. To solve such binary systems, more epochs of accurate RV measurements are needed.

The difference in age between the field and globular cluster populations has been invoked as an explanation for the striking difference in the close binary fractions, older systems favoring evolutionary channels forming EHB stars in wide binaries and/or single stars \citep{moni08}. \citet{han08} predicted that for a stellar population older than 10 Gyr, the fraction of EHB stars in close binary systems (P $<$ 5 d, formed via a common-envelope channel) is below 3\% and instead the dominant formation channel is the merger of two He-core white dwarfs (WD). 
The binary population synthesis models of \citet{han08} do not make any prediction of the atmospheric properties of the stars produced, but the progeny of white dwarf mergers (He-WDs and hybrid CO(He) WDs) has been extensively studied in the context of the field He-sdOs, which are also mostly single stars (e.g., \citealt{web84,just11,zhang12}). The models predict sdO stars (\teff\ $\la$ 40\,000 K) that have an atmosphere enriched in helium and CNO-processed material \citep{saio00}.
Although He-sdOs are not common in \omcen, the work of \citet{clau11}, \citet{hall16} and \citet{schwab18} suggests that mergers (of two He-WDs but also a He-WD + an M dwarf) could also produce cooler objects with a hydrogen-rich atmosphere. This could potentially explain the single H-sdBs in our sample. 

\subsubsection{Pulsation}
A final difference between the field and the \omcen\ EHB population concerns the pulsating hot subdwarfs.
For completeness, we briefly mention this topic but it is discussed at length in Randall et al. (\citeyear{ran12,ran16}). Among the field population, an estimated fraction of about 10\% of H-sdBs around 33,000 K show low-amplitude, multi-periodic luminosity variations with periods in the range 60 $-$ 600 s \citep{font08b,bil02,ost10}. These variations are explained by  pressure modes excited by the $\kappa-$mechanism driven by an increased opacity of iron, and iron-like elements, in the subphotospheric layers of the star \citep{char97}. No direct counterparts have so far been found in \omcen, however the cluster hosts a well-defined class of pulsating H-sdOs \citep{ran09,ran11,ran16}. Four of the five pulsators so far known are included in our sample (V1, V3, V4, and V5, indicated in Table~\ref{bigtable}; V2 was excluded due to its spectrum being polluted). These stars have \teff\ close to 50\,000~K according to their optical spectra. However, an analysis of low resolution UV Hubble Space Telescope Cosmic Origin Spectrograph\footnote{program GO-13707, PI: Randall} data of two of these pulsators indicated that their effective temperature may in fact be closer to 60\,000~K \citep{lat17}\footnote{A similar discrepancy between effective temperatures determined from the optical and UV spectra has also been reported recently in the case of a UV-bright star in M~4 \citep{dix17}.}. It is thus possible that the effective temperatures of all hot H-sdOs have been systematically underestimated (both in \omcen\ and the field). So far, these objects have no confirmed counterpart among the field sdO population \citep{john14}. 
It is not yet clear whether each of these types of pulsator in fact does not exist in the other environment, or whether we are simply limited by the observational data obtained so far. 
 
\subsection{The mass conundrum}

Despite being faint and challenging to observe in spectroscopy, studying the EHB stars in \omcen\ has a major advantage compared to the field population: the stars are all at the same, known distance.  
This important information (as well as the reddening) allows one to compute spectroscopic masses, essentially by comparing their observed magnitude with the flux predicted from model atmospheres. The results presented in Sect. 4.3 pose an obvious problem; our stars have masses that are on average too low to ignite helium in the core. 

Many parameters are involved in the computation of the mass and we examined them to find a possible explanation for the systematic underestimation of the masses. 
The first important parameters are the temperature and surface gravity of the stars. Figure~\ref{teffgravhe} shows that the majority of our stars (excluding the H-sdOs thought to be post-EHB objects) have fundamental parameters (\teff\ and log $g$) in good agreement with those expected for helium-core burning stars of 0.48$-$0.50 \msun. Our atmospheric parameters are also in good agreement with those derived from previous studies \citep{moe11,moni12}. We thus consider them to be quite reliable. 
Two additional important quantities are the distance to the cluster and the average reddening. Both of these quantities have been extensively studied in the case of \omcen\ and the values derived from numerous studies are in good agreement (see Table 9 of \citealt{bra16} for a summary). While \omcen\ is known to have differential reddening \citep{cal05}, the variations across the cluster are rather small ($\sim$10\%), and since our stars are well-distributed around the cluster, using the average reddening should be an appropriate assumption. 
The absolute magnitude of the stars is computed using our own model atmospheres, which has the advantage of taking into account the derived helium abundance for each individual star. Our models account for NLTE effects and the opacity of C, N, and O, but the opacity of heavier metals is not considered.
However, the second method we used for deriving the masses is based on an empirical bolometric correction and is independent of our models. The fact that both methods result in a very similar average mass indicates that the mass problem is not caused by the flux distribution of our models. 
As for the observed V magnitude, the offset of 0.2 mag that would be necessary to increase the average mass to 0.45~\msun\ is too large compared to any possible systematic calibration errors ($<$ 4\%) of the WFI $V$ magnitudes \citep{cast07}. As a check we
recomputed the masses using the observed and synthetic $B$ magnitudes and the corresponding extinction $A_{B}$, but this affected the average mass by only 2\%.

Even though none of the above parameters alone appears to be able to cause a systematic underestimation of the masses on the scale observed, (unknown) smaller effects acting on more than one parameter could still lead to systematically lower masses when combined together. 
The average uncertainty on the individual mass determination is 0.13 \msun, a value similar to the standard deviation of our mass distribution,
and an important contributor to this value is the uncertainty on log $g$, which is on average 0.13 dex. 
For example, a shift of $+$0.1 dex in log $g$ would result in a mean mass of 0.47 \msun. An underestimation of the surface gravity could be caused, for example, by missing opacities in our model atmospheres that would affect the Balmer line profiles. However these effects are not expected to be important below $\sim$35 kK \citep{lat14f48} and our derived masses do not show a trend with temperature. 

Low masses have been derived previously for EHB (as well as HB) stars in \omcen\ by \cite{moni2011}. Interestingly the authors reported that this low mass problem was encountered only for the stars in \omcen\ and not in the other three globular clusters for which stellar masses were measured in a similar way (NGC\,6752, M\,80 and NGC\,5986; \citealt{moni07,moni09}). This is rather intriguing and it was suggested that there could be an intrinsic difference between the blue HB stars of \omcen\ and those of the other clusters. 

Precise mass estimates of field hot subdwarfs are rather scarce but a small sample was put together by \citet{fon12}. The mass estimates for these stars were obtained via asteroseismic modeling for pulsating stars, and via light curve and RV analyses for close-binary systems (mostly eclipsing binaries). From this sample of 22 stars, the average mass was found to be 0.47 \msun, which is in excellent agreement with predictions from stellar evolution models. These two ways of deriving masses do not strongly rely on atmospheric modeling and are thus independent of our spectroscopic method. A mass distribution obtained using the methods applied to globular clusters for field EHB stars cannot be derived until we have accurate distances for a large sample of field hot subdwarfs. Luckily, such an analysis will soon be possible based on the second data release of Gaia and it will be very interesting to see whether the mass problem we have for \omcen\ is also encountered for the field population.

\section{Conclusion}

In this work we characterized the largest sample of EHB stars ever analyzed in a globular cluster in terms of spectroscopic properties. We derived atmospheric parameters for 152 individual stars, using new FORS and VIMOS observations, as well as previously published FORS and FLAMES spectra. This represents about 20\% of the EHB population of \omcen, which consists of $\approx$730 stars up to about one degree from the cluster's center\footnote{Using the ACS and DECam catalogs \citep{cast07,cal17}}.
We also searched for RV variations in 75 stars that were observed over multiple epochs as part of the VIMOS survey.
We summarize our results as follows:

\begin{itemize}

\item The EHB population of \omcen\ can be divided into three distinct spectroscopic groups that are best discernible in the \teff\ $-$ helium abundance plane. We divided our targets into H-sdBs (the coolest H-rich stars, 26\% of our sample), H-sdOs (the hottest H-rich stars, 10\% of the sample) and He-sdOBs, which are found at intermediate temperatures (33 $-$ 43 kK) and have an atmospheric helium abundance close to or above the solar value. The He-sdOBs can be further sub-divided into two sub-groups according to their helium abundance. The location of the spectroscopic groups on the CMD confirms that the He-sdOBs form the blue hook population of \omcen.

\item The He-sdOBs found in \omcen\ are not well represented in the field. Surveys among the galactic disk population found some stars with similar atmospheric parameters, but their fraction ($\sim$5\%; \citealt{geier17}) is much lower than in \omcen\ ($\sim$52\%). The galactic halo contains a larger fraction of these objects (23\%), but still nowhere near as high as in \omcen. This suggest that the formation of the blue hook objects in \omcen\ (and by deduction probably in other clusters showing a blue hook, such as NGC~2808) is favored by the globular cluster's particular populations and environment.

\item There is a clear positive correlation between the helium abundance and effective temperature among the He-sdOBs. Such a correlation is also seen among the field sdBs, thus suggesting a common mechanism responsible. Although the \teff\ $-$ He relation has not yet been explained at the quantitative level, it is very likely governed by the diffusion processes taking place in the atmosphere. The gravitational settling of helium must be counteracted by another phenomenon, such as a weak stellar wind or the presence of turbulence in the upper atmospheric layers\footnote{The radiative forces on helium are essentially negligible.}. Even though diffusion modifies the initial atmospheric composition (such as the He and C abundance, which is observed to be lower than predicted by the late-flasher models), the initial composition will influence the subsequent equilibrium abundances (see e.g., \citealt{ung05}). 

\item The mean mass of our EHB sample (0.38 \msun) is significantly lower than the mass predicted by evolutionary models ($\sim$0.5 \msun).  We find masses that are too low regardless of whether we use our own model atmospheres or a bolometric correction to estimate the absolute magnitudes. Interestingly, \citet{moni2011} derived similarly low masses for EHB and HB stars in \omcen, but not for targets in the other globular clusters included in their work. The mass conundrum remains unexplained, but so far it seems to be unique to \omcen.  

\item  We estimate a close binary fraction of about 5\% among the EHB stars of \omcen. Out of the 75 stars included in our VIMOS RV survey, two showed statistically significant RV variations, however no periodicity could be detected. This close binary fraction is lower than for field sdB stars, but in line with the low fractions found among the EHB and HB stars of other globular clusters. 
This indicates that the common envelope channel, which is responsible for the production of close-binaries among the field sdBs, does not significantly contribute to the formation of hot subdwarfs in \omcen.
\end{itemize}

The results presented in this paper clearly reveal that the population of EHB stars in \omcen\ in fact has little in common with the hot subdwarf population in the galactic field. This strongly suggests that a large fraction of the \omcen\ EHB population owes its existence to the cluster's particular environment. For example, an initial helium enrichment combined with rapid rotation might enhance the probability of delayed helium flashes and thus favor the formation of the He-sdOB (blue hook) stars, as suggested by \citet{tailo15}. If such progenitors are peculiar to \omcen\ (and possibly other globular clusters), this would explain the relative lack of corresponding He-sdOBs in the field. In addition, the age of the cluster's population likely affects the efficiency of the different EHB formation channels. As was shown by \citet{han08}, hot subdwarfs are preferentially formed via mergers (instead of common-envelope ejection and Roche-lobe overflow) in populations older than about 10 Gyr. However, late flasher and merger events have  also been invoked to explain the formation of field He-sdO stars, a population that has no direct counterpart in \omcen. Whether these two mechanisms indeed produce hot subdwarfs with quite different properties in the field compared to GGCs needs further ingestigation, both on the modeling and the observational front. On our side, we plan to continue the SHOTGLAS project with the aim of providing statistically significant spectroscopic constraints for several globular clusters with an observationally accessible EHB. This will form the observational basis for in-depth studies of different EHB formation scenarios and their relative importance in the different environments where these enigmatic stars are found.

\begin{acknowledgements}

We are most grateful to  C. Moni Bidin for providing us with his observed spectra, to W. V. Dixon for help with the Pysynphot package, and to U. Heber and S. Dreizler for supporting this work. 
This work was supported by a fellowship for postdoctoral
researchers from the Alexander von Humboldt Foundation awarded to M.L., who also 
acknowledges fundings by the Deutsches Zentrum f\"{u}r
Luft- und Raumfahrt (grant 50 OR 1315) and the Deutsche Forschungsgemeinschaft (grant DR 281/35-1).
S.G. acknowledges funding by Heisenberg program of the Deutsche Forschungsgemeinschaft under grant GE 2506/8-1.
This study was supported by NASA through grant GO-
13707 from the Space Telescope Science Institute, which is operated by AURA, Inc., under NASA contract NAS5-26666.
This research has made use of NASA's Astrophysics Data System.
We also thank the referee, C. Moni Bidin, for his helpful comments. 
\end{acknowledgements}

%
%

\bibliographystyle{aa}


\begin{appendix} 
\section{Comparisons of atmospheric parameters derived from the various samples}\label{appA}

\begin{figure}[h]
\begin{center}
\resizebox{\hsize}{!}{\includegraphics{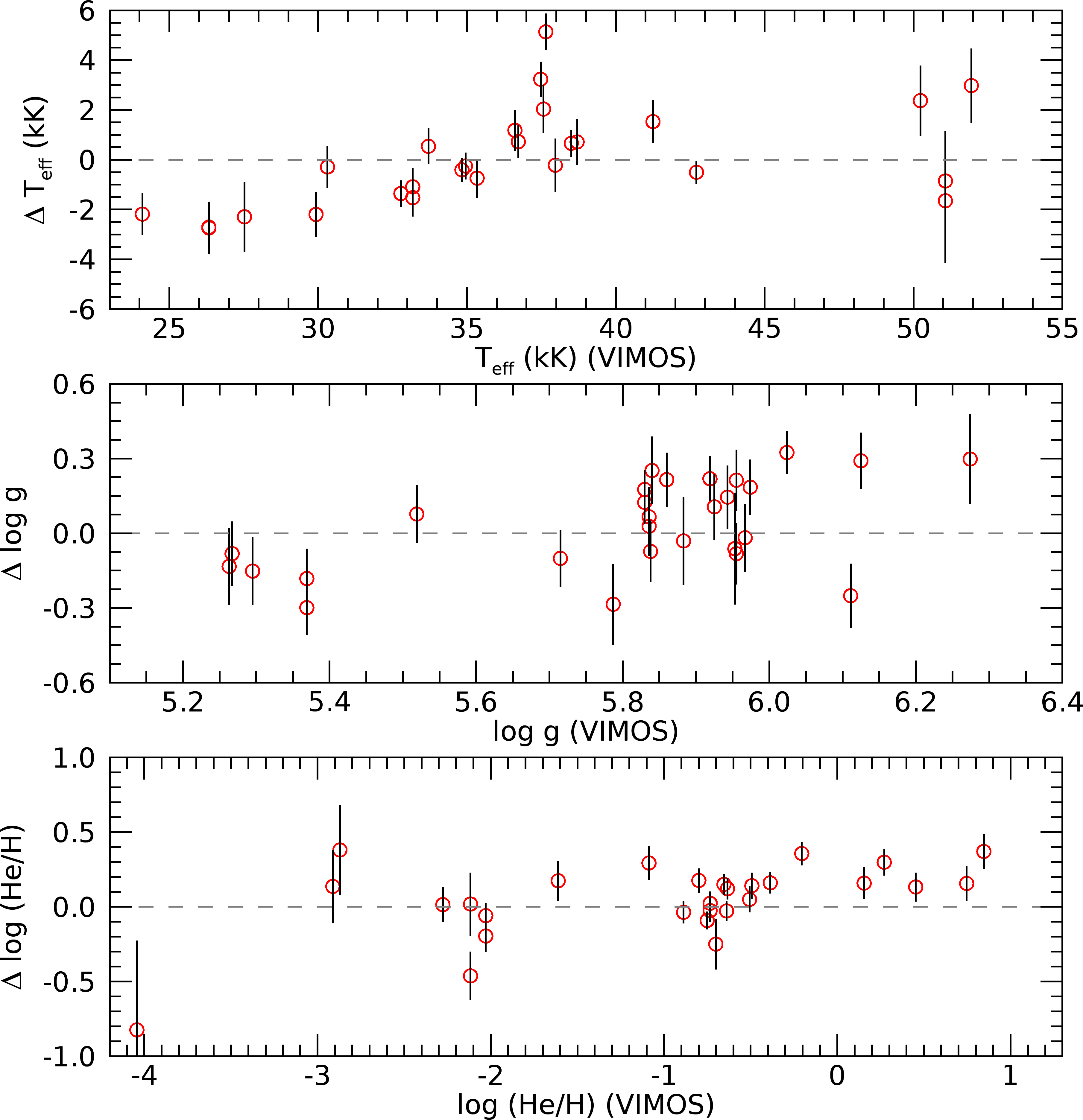}}
\caption{Comparison between the parameters (\teff, log $g$, and log \nhe) derived for the stars in common between the VIMOS sample and the other samples. The differences ($\Delta$) are expressed in terms of the value derived with the VIMOS spectrum minus the value derived from the other sample, e. g. $\Delta T_{\rm eff}$ = \teff\ (vimos) $-$ \teff\ (other).   }
\label{compvimos}
\end{center}
\end{figure}

\begin{figure}[h]
\begin{center}
\resizebox{\hsize}{!}{\includegraphics{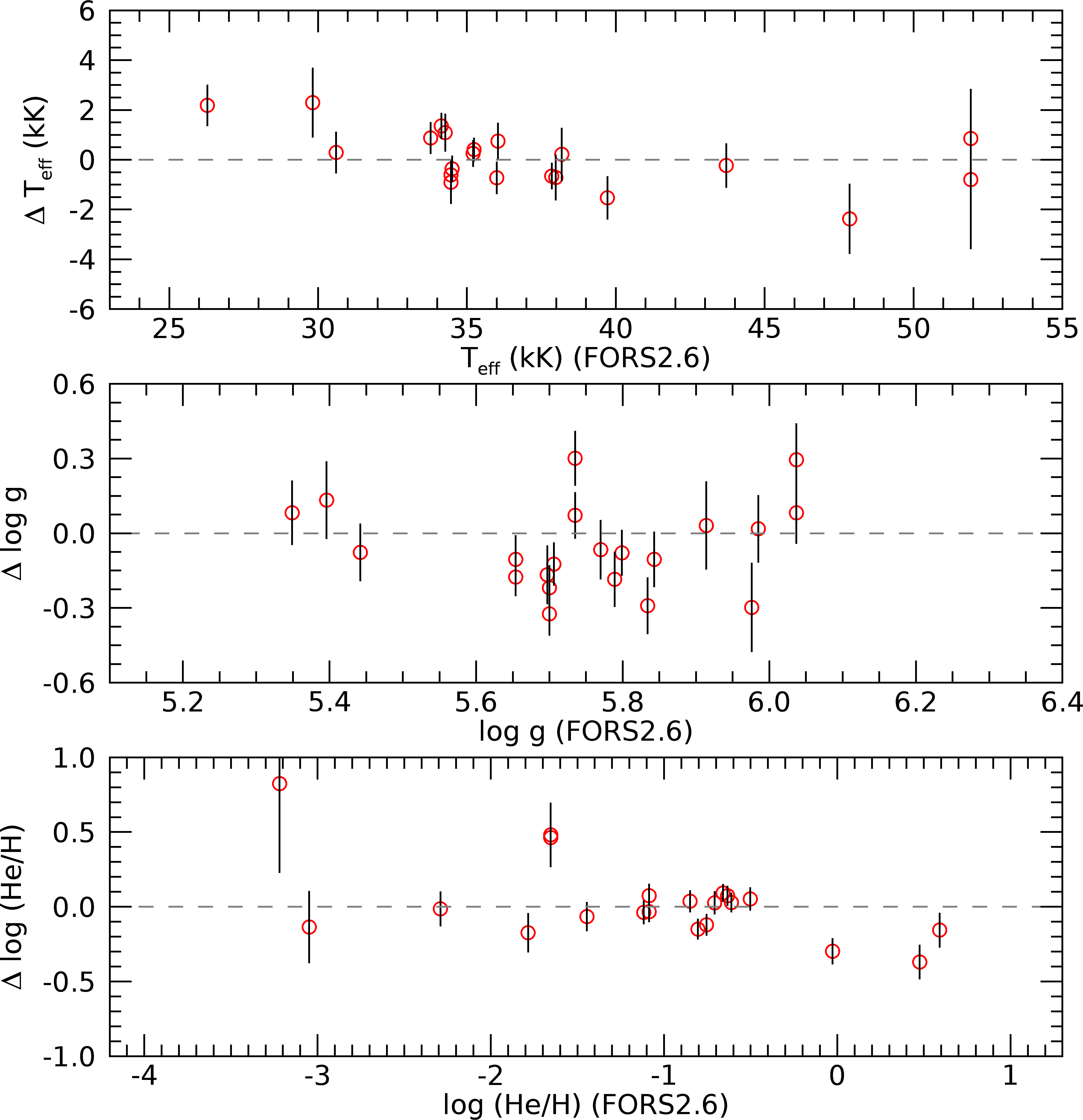}}
\caption{Same as Fig. \ref{compvimos} but for the stars in common between the FORS2.6 sample and the other samples. }
\label{compourfors}
\end{center}
\end{figure}

Given that 40 stars are present in more than one observed sample, we compared the atmospheric parameters derived using the spectra from different samples. Indications of systematic shifts between literature results have been reported in previous studies and our combined sample allows us to investigate this issue more thoroughly. \citet{lat14} reported that the He-rich stars in their sample were clustering at a higher temperature than those of \cite{moe11}, while \citet{moni12} found an offset in the helium abundances measured for the EHB stars in common between their sample and the FLAMES sample of \citet{moe11}. 
Because we analyzed all of our spectra in a homogeneous way, using the same model atmospheres and the same fitting procedure, a comparison of the atmospheric parameters obtained for a given star using spectra from different samples is sensitive to effects specifically associated with the observed spectra, such as wavelength coverage, resolution and possible background contamination. Among the 40 stars in common, 34 are found in two samples, and six are found in three samples. 

\begin{figure}
\begin{center}
\resizebox{\hsize}{!}{\includegraphics{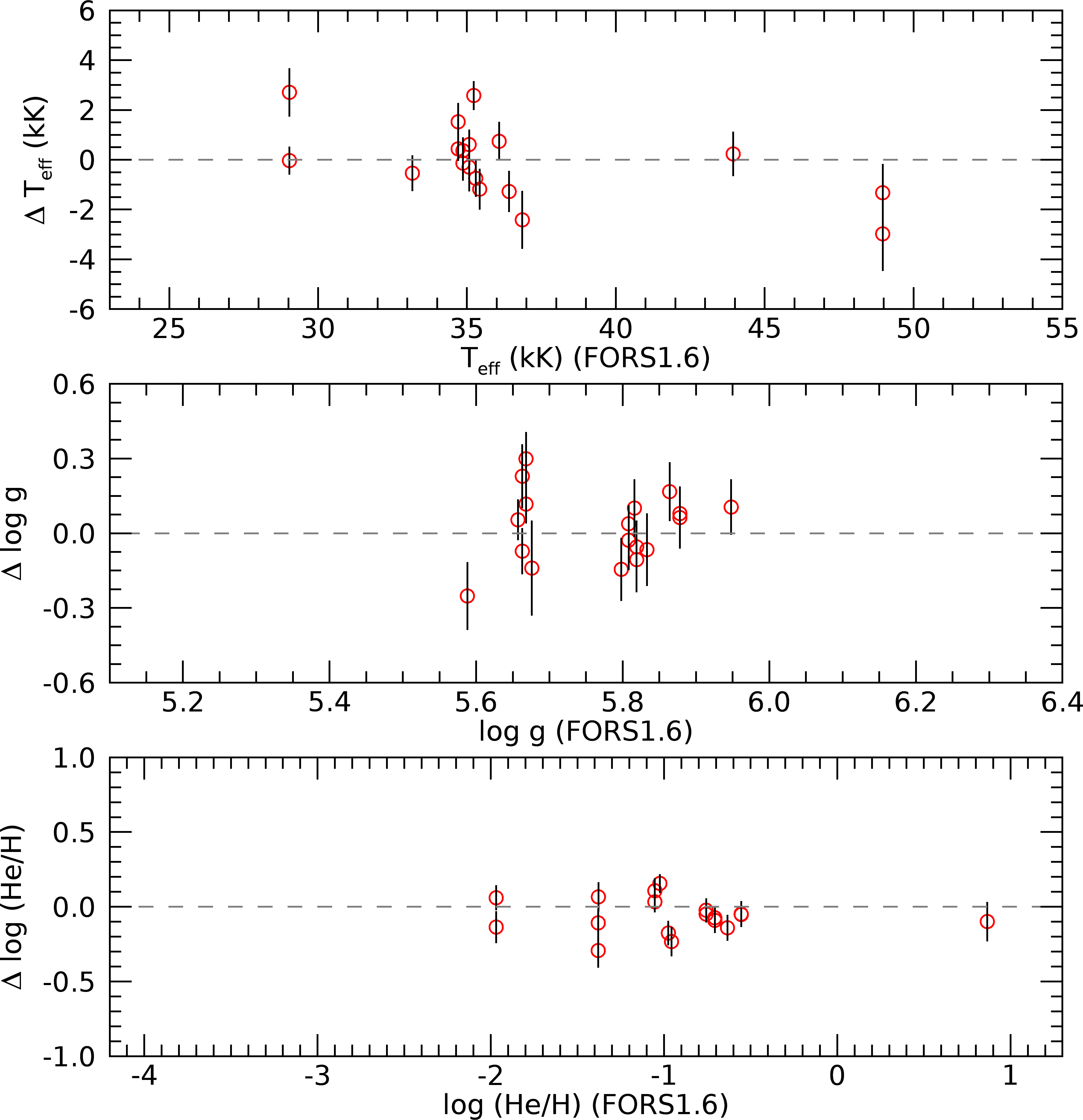}}
\caption{Same as Fig. \ref{compvimos} but for the stars in common between the FORS1.6 sample and the other samples. }
\label{compsabfors}
\end{center}
\end{figure}

\begin{figure}
\begin{center}
\resizebox{\hsize}{!}{\includegraphics{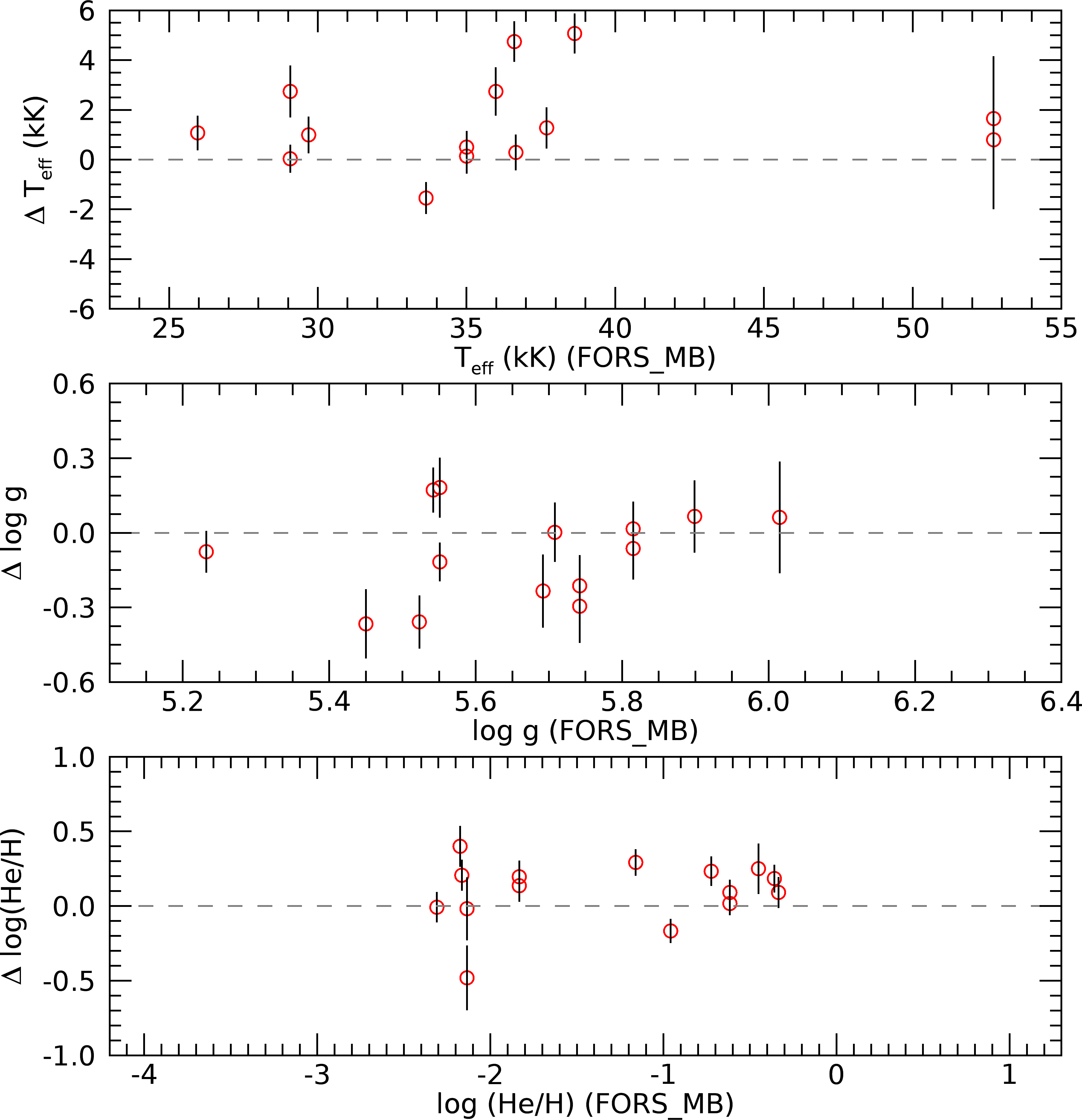}}
\caption{Same as Fig. \ref{compvimos} but for the stars in common between the \forsmb\ sample and the other samples. }
\label{compmb}
\end{center}
\end{figure}

\begin{figure}
\begin{center}
\resizebox{\hsize}{!}{\includegraphics{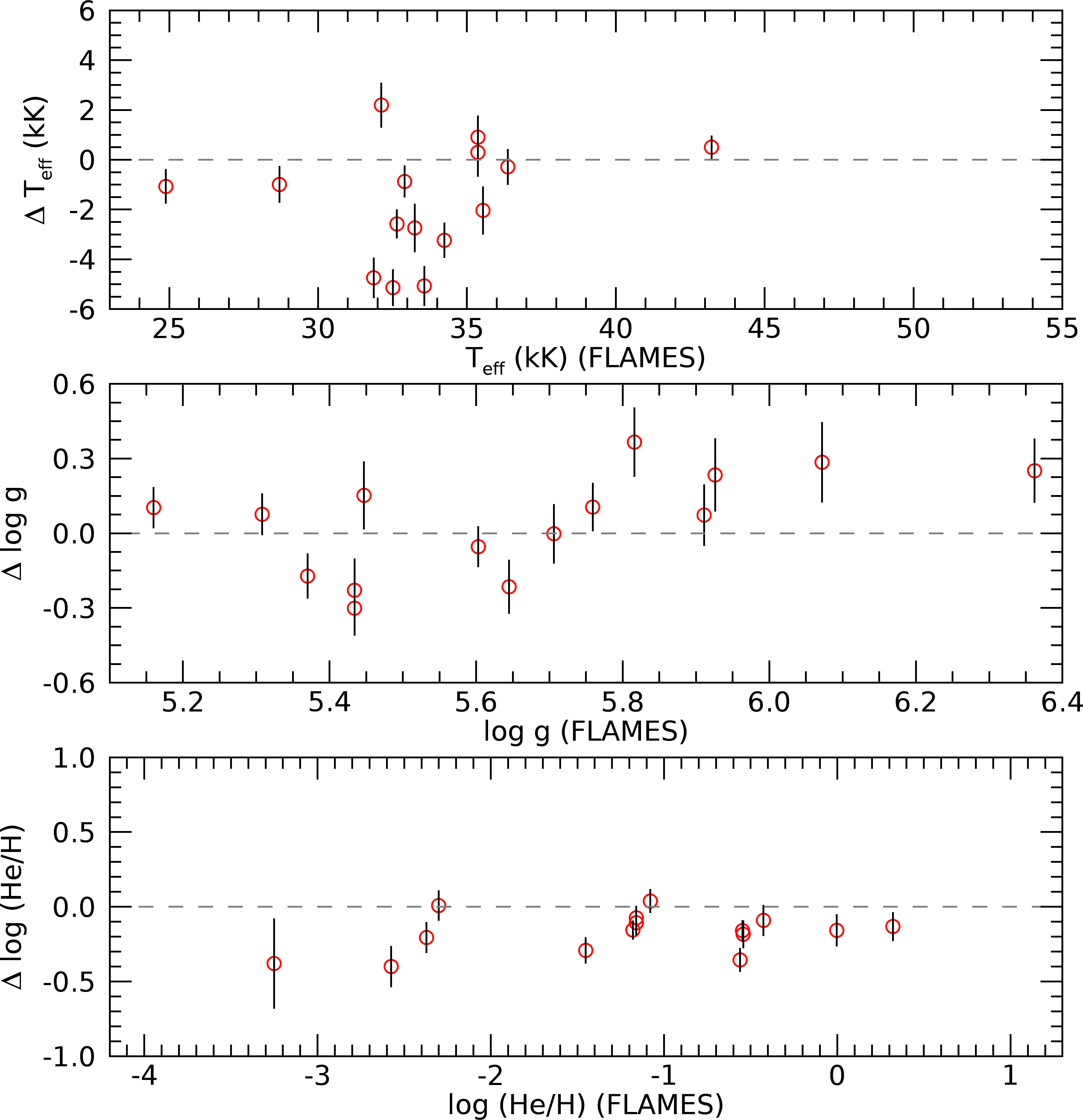}}
\caption{Same as Fig. \ref{compvimos} but for the stars in common between the FLAMES sample and the other samples. }
\label{compflames}
\end{center}
\end{figure}

For each observed sample we compared the resulting parameters in the following way: we selected all stars duplicated in any other sample and for each of these computed the difference between the \teff, log $g$, and log \nhe\ derived for the different samples. The results are illustrated for each of the samples in turn in Figs. \ref{compvimos} to \ref{compflames}. In addition, we computed the average differences (i.e., $\overline{\Delta T_{\rm eff}}$) 
for each parameter and we report the results in Table \ref{tabcomp}\footnote{The helium abundance determinations of the star 168035 (present in VIMOS and FORS2.6) were rejected in the computation of the average because of its very low helium abundance leading to an uncertain determination.}.  
The number of parameter pairs included in each sample is also indicated in the second column of Table \ref{tabcomp}\footnote{We note that the number of pairs can be higher than the number of common stars in a sample since a star found in three sample provides two pairs of parameters to compare.}. 

Our results indeed reveal a few systematics shifts. Concerning the temperature, the FLAMES and \forsmb\ samples (Fig. \ref{compflames} and \ref{compmb}) show the largest deviations, with the FLAMES spectra returning lower than average \teff\ values,  while the \forsmb\ spectra return higher than average \teff values. The fact that the FLAMES spectra return lower \teff values, especially for the stars in the range 30$-$35 kK, agrees with the finding of \citet{lat14} that the He-rich stars in the FLAMES sample of \citet{moe11} have lower temperatures than in their FORS sample. We think that this apparent shift toward lower \teff\ is induced by the absence of \ion{He}{ii} lines in the FLAMES spectral range, preventing the ionization equilibrium of helium to be considered (implicitly) in the fitting procedure. Regarding the \forsmb\ sample, more than one-third of the target pairings are with FLAMES spectra, which enhances the apparent \teff\ shifts\footnote{Indeed, $\overline{\Delta T_{\rm eff}}$ decreases to 700 K when removing these six pairs.}.
A look at Fig. \ref{compvimos} to \ref{compflames} reveals that for many stars, the parameters derived using the different spectra do not agree with each other within the statistical uncertainties. This suggests the presence of additional uncertainties that are likely related to the observational data, for example, resolution, wavelength coverage, data reduction. In case of correct uncertainties, the ratio of the temperature difference and its uncertainty ($\sigma T$) 
   \begin{equation} \label{eq:A1}
   \frac{T_2 - T_1} {\sqrt{\sigma T_1^2 + \sigma T_2^2 }}
   \end{equation}  
should be normally distributed with a standard deviation of one. However, the distribution of the Eq. \ref{eq:A1} values for all pair of stars has a standard deviation of 2.5. To account for the additional observational uncertainties, we can multiply, for every star, the statistical uncertainty returned by the fitting procedure by a factor of 2.5. Considering that the average statistical uncertainty on \teff\ is $\sim$600 K, the average total uncertainty on \teff\ would be $\sim$1500 K.


\begin{table*}[h]
\center
\caption{Parameter differences for the stars in common between different samples.}
\label{tabcomp}
\small
\begin{tabular}{lcrrrrrr}
\hline
\hline
Sample & \# of pairs &  $\overline{\Delta T_{\rm eff}}$ (K) &  $\overline{\Delta log g}$ &  $\overline{\Delta He/H}$   \\
\hline
VIMOS & 28  & 3.9& 0.035  & 0.080 \\
FORS2.6 & 21  & 78.7 & $-$0.057 & $-$0.011  \\
FORS1.6 & 18 & $-$96.7 & 0.022 & $-$0.061  \\
\forsmb\ & 15 & 1464.1 & $-$0.088 & 0.095 \\
FLAMES & 15 & $-$1657.7 & 0.045 & $-$0.180 \\
\hline
\end{tabular}
\end{table*}

Concerning the surface gravity, the two samples showing the largest shifts are the two lowest resolution FORS samples, which provide lower log $g$ values (by 0.057 and 0.088 dex) than the other samples. These shifts are however not significantly larger than the average of the statistical errors on the surface gravity (0.08 dex). Using the same method as for \teff, we estimated a correction factor of 1.6 to apply to the statistical uncertainties in order to account for the observational component. This results in an average total uncertainty of 0.13 dex for the log $g$.

For the helium abundance, there is an obvious shift in the parameters derived from the FLAMES sample whereby they are systematically lower (by an average of 0.18 dex) than those obtained from the other spectra. This was already noted by \citet{moni12}, who suggested that the systematic differences in helium abundance could be due to the different spectral resolution, measurements based on lower resolution spectra resulting in higher helium abundances (by 0.2 $-$ 0.25 dex in their comparison). Using the same method as for the other parameters, we obtain a correction factor of 1.8 for the uncertainties on the helium abundance, leading to an average total uncertainty of 0.2 dex for log \nhe.



\section{Additional material}\label{appB}

\begin{table}[h]
\centering
\caption{Log of the VIMOS observations} \label{obsvimos}
\small
\begin{tabular}{cclc}
\hline\hline
Exposure & Epoch & MJD & No. of spectra \\
\hline
rv\textunderscore1 & 1 & 56813.9970847 & 67 \\
rv\textunderscore2 & 1 & 56814.0046662 & 64 \\
rv\textunderscore3 & 1 & 56814.01224724 & 61 \\
rv\textunderscore4 & 2 & 56832.00142376 & 62 \\
rv\textunderscore5 & 2 & 56832.0090002 & 62 \\
rv\textunderscore6 & 2 & 56832.01658093 & 62 \\
rv\textunderscore7 & 3 & 56840.05717539 & 75 \\
rv\textunderscore8 & 3 & 56840.06475274 & 75 \\
rv\textunderscore9 & 3 & 56840.07233286 & 74 \\
rv\textunderscore10 & 4 & 56840.99593468 & 58 \\
rv\textunderscore11 & 4 & 56841.0035155 & 58 \\
rv\textunderscore12 & 4 & 56841.01109309 & 61 \\
rv\textunderscore13 & 5 & 57159.07066157 & 54 \\
rv\textunderscore14 & 5 & 57159.07824265 & 54 \\
rv\textunderscore15 & 5 & 57159.08582304 & 54 \\
rv\textunderscore16 & 6 & 57160.12323767 & 69 \\
rv\textunderscore17 & 6 & 57160.13082294 & 66 \\
rv\textunderscore18 & 6 & 57160.13841248 & 66 \\
rv\textunderscore19 & 7 & 57165.11673919 & 65 \\
rv\textunderscore20 & 7 & 57165.12432017 & 64 \\
rv\textunderscore21 & 7 & 57165.13190114 & 63 \\
rv\textunderscore22 & 8 & 57166.0314756 & 60 \\
rv\textunderscore23 & 8 & 57166.03905219 & 57 \\
rv\textunderscore24 & 8 & 57166.04663305 & 60 \\
rv\textunderscore25 & 9 & 57166.07363816 & 65 \\
rv\textunderscore26 & 9 & 57166.08121647 & 44 \\
rv\textunderscore27 & 9 & 57166.08879755 & 35 \\
rv\textunderscore28 & 10 & 57166.1137898 & 24 \\
rv\textunderscore29 & 10 & 57166.12136683 & 41 \\
rv\textunderscore30 & 10 & 57166.12894791 & 48 \\
rv\textunderscore31 & 11 & 57194.06389226 & 42 \\
rv\textunderscore32 & 11 & 57194.07146679 & 45 \\
rv\textunderscore33 & 11 & 57194.07904432 & 42 \\
rv\textunderscore34 & 12 & 57215.99973784 & 35 \\
rv\textunderscore35 & 12 & 57216.00731525 & 20 \\
rv\textunderscore36 & 12 & 57216.01489586 & 27 \\
rv\textunderscore37 & 13 & 57428.31673457 & 48 \\
rv\textunderscore38 & 13 & 57428.32431726 & 57 \\
rv\textunderscore39 & 13 & 57428.33189659 & 64 \\
rv\textunderscore40 & 14 & 57428.3548221 & 65 \\
rv\textunderscore41 & 14 & 57428.36239923 & 72 \\
rv\textunderscore42 & 14 & 57428.36998041 & 72 \\
\hline
\end{tabular} \\

\end{table}

\begin{table}[h]
\centering
\caption{Log of the FORS observations} \label{obsfors}
\small
\begin{tabular}{lccc}
\hline\hline
Exposure & MJD & Airmass & No. of spectra \\
\hline
1 & 54562.065786 & 1.4015 & 15 \\
2 & 54562.202272 & 1.0895 & 15 \\
3 & 54613.089094 & 1.1020 & 11 \\
\hline
\end{tabular} \\

\end{table}

\begin{figure*}[h]
\begin{center}
\includegraphics[width=0.48\linewidth]{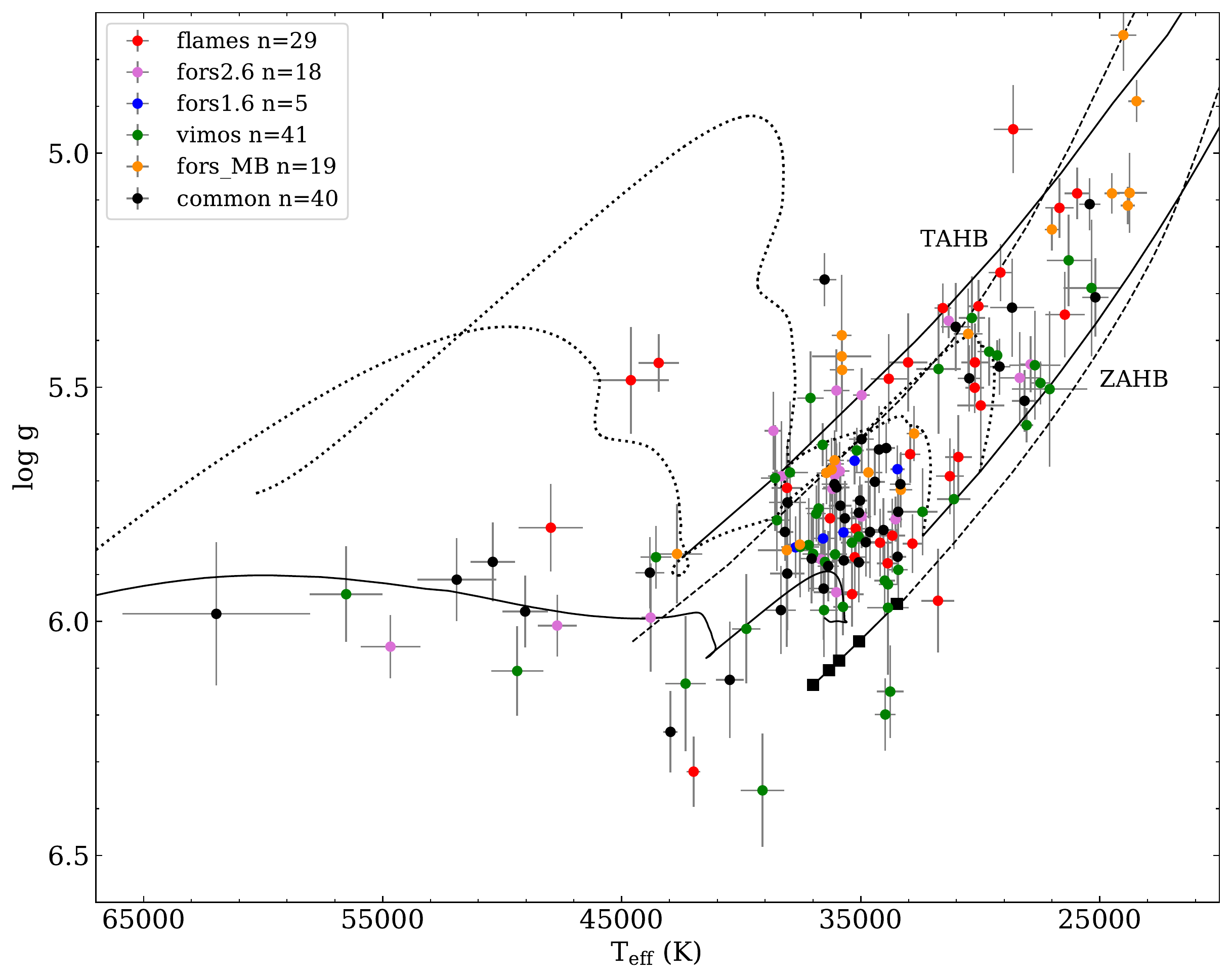}
\includegraphics[width=0.48\linewidth]{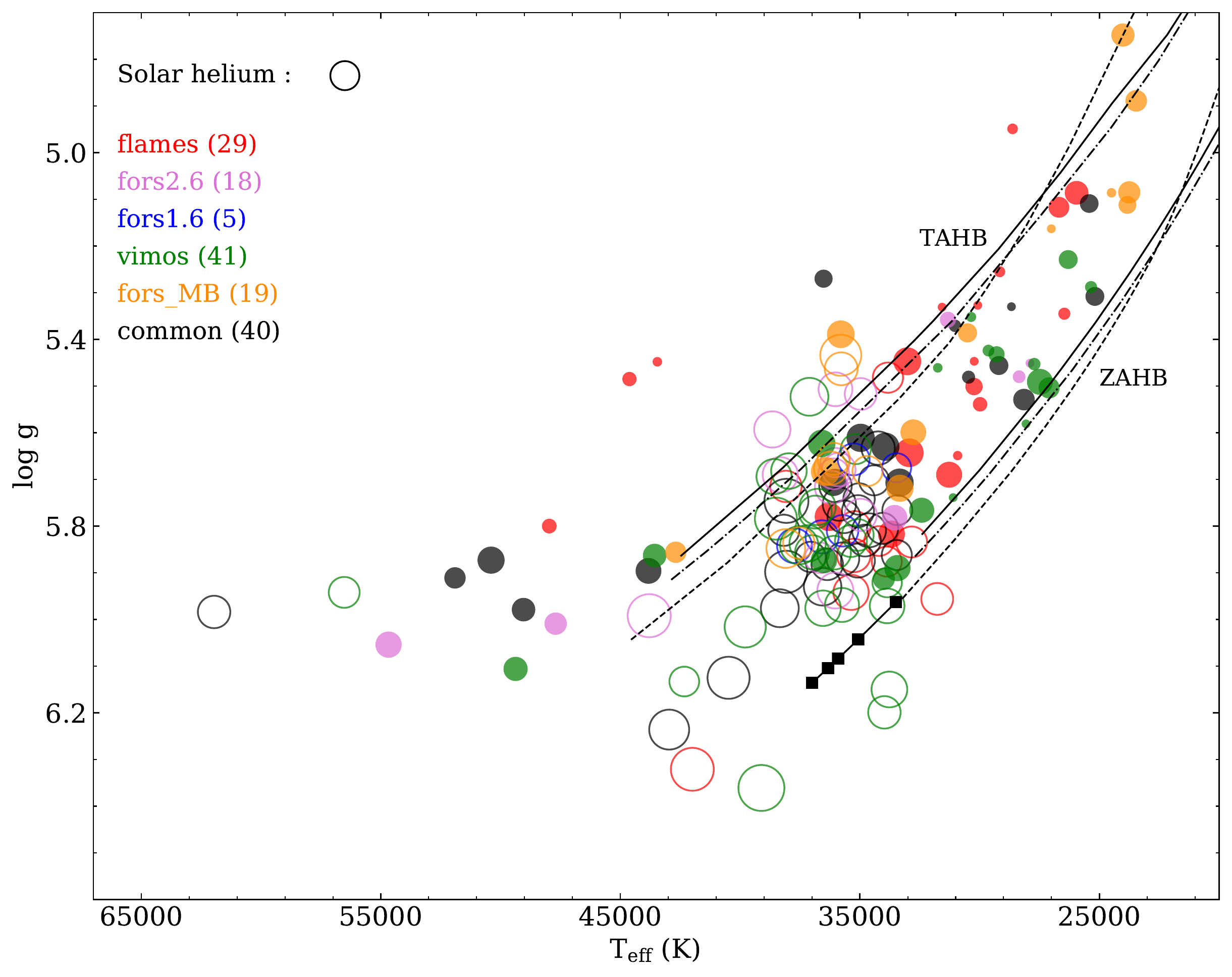}
\caption{Left$-$ Position of the 152 stars in the log $g$ $-$ \teff\ diagram.  The samples are indicated by different colors and the common group includes the stars present in two or three samples. The error bars used for individual stars are the statistical uncertainties returned by the fitting procedure. 
The ZAEHB, TAEHB and the evolutionary tracks are as in Fig.~\ref{teffgravhe}. Right$-$ Same as the left panel but with the  logarithmic helium abundance illustrated by the size of each circles, where super-solar and sub-solar abundances are represented by open and filled circles respectively. The circle size for a solar abundance is shown as an indication. }
\label{app_gteff}
\end{center}
\end{figure*}

\begin{figure*}[h]
\begin{center}
\includegraphics[width=0.48\linewidth]{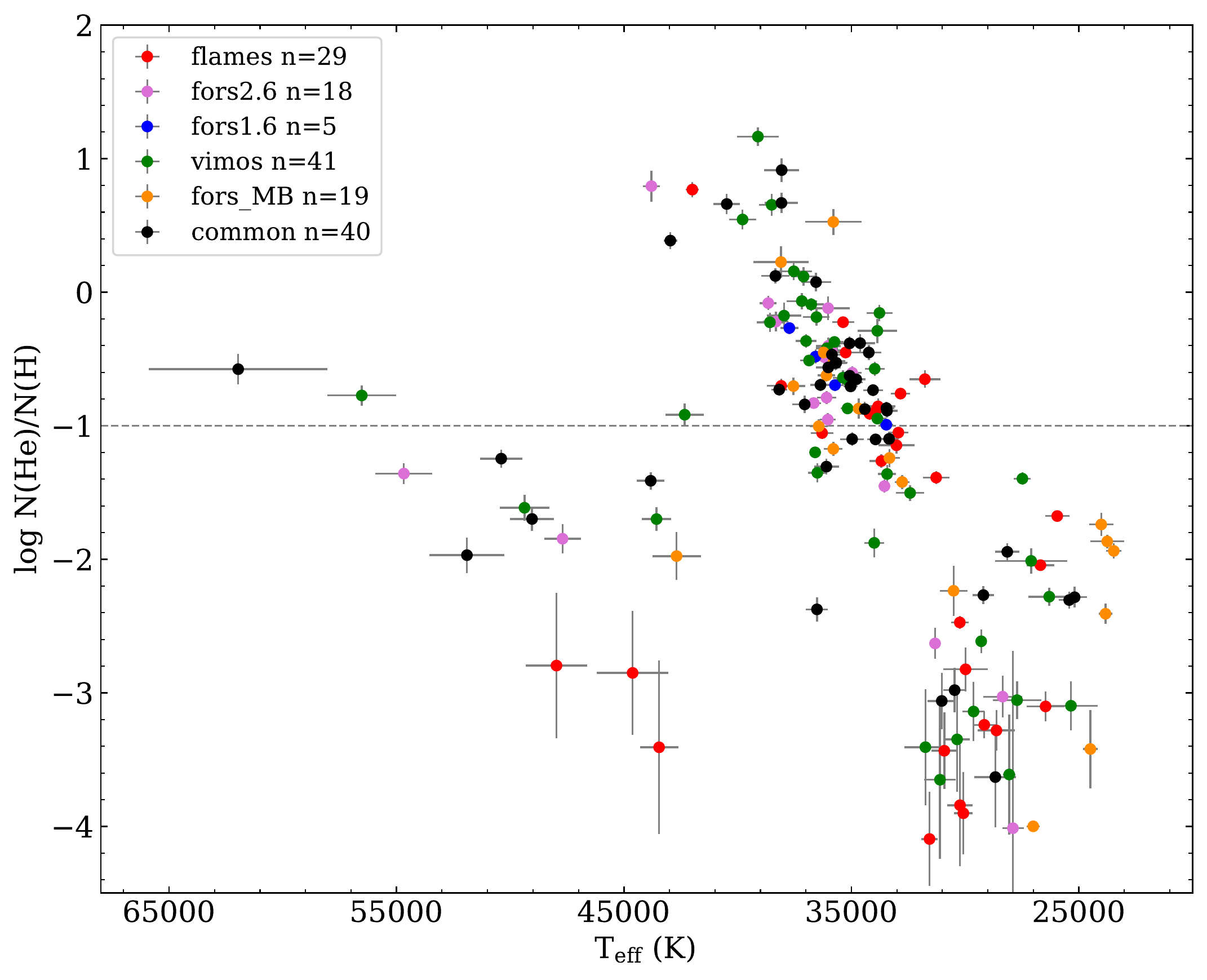}
\includegraphics[width=0.48\linewidth]{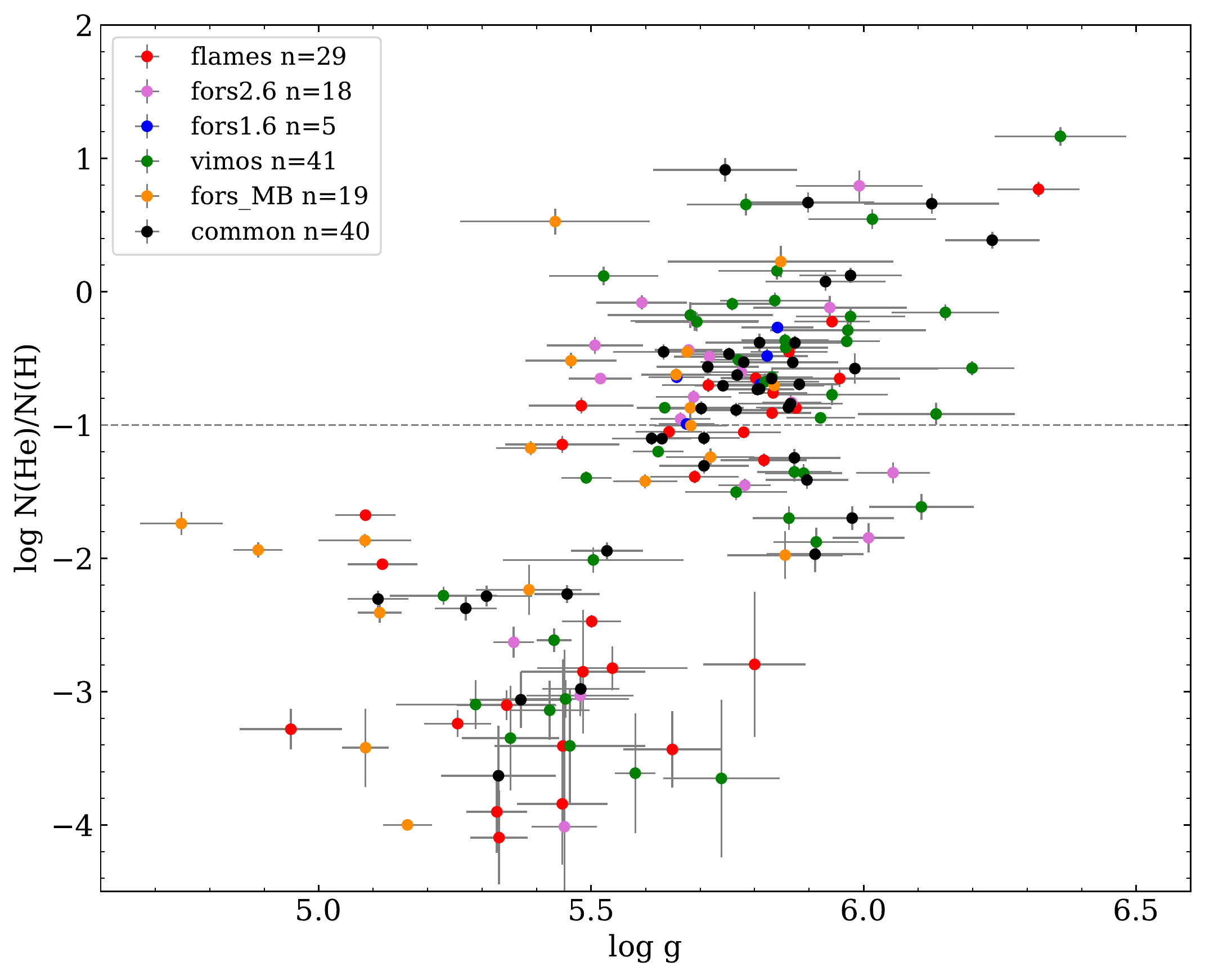}
\caption{Left$-$ Helium abundance as a function of effective temperature for the 152 stars of our sample. The samples and uncertainties are presented as in Fig.~\ref{app_gteff}. Right$-$ Helium abundance as a function of the surface gravity.  }
\label{app_he}
\end{center}
\end{figure*}

\begin{figure}[h]
\begin{center}
\resizebox{\hsize}{!}{\includegraphics{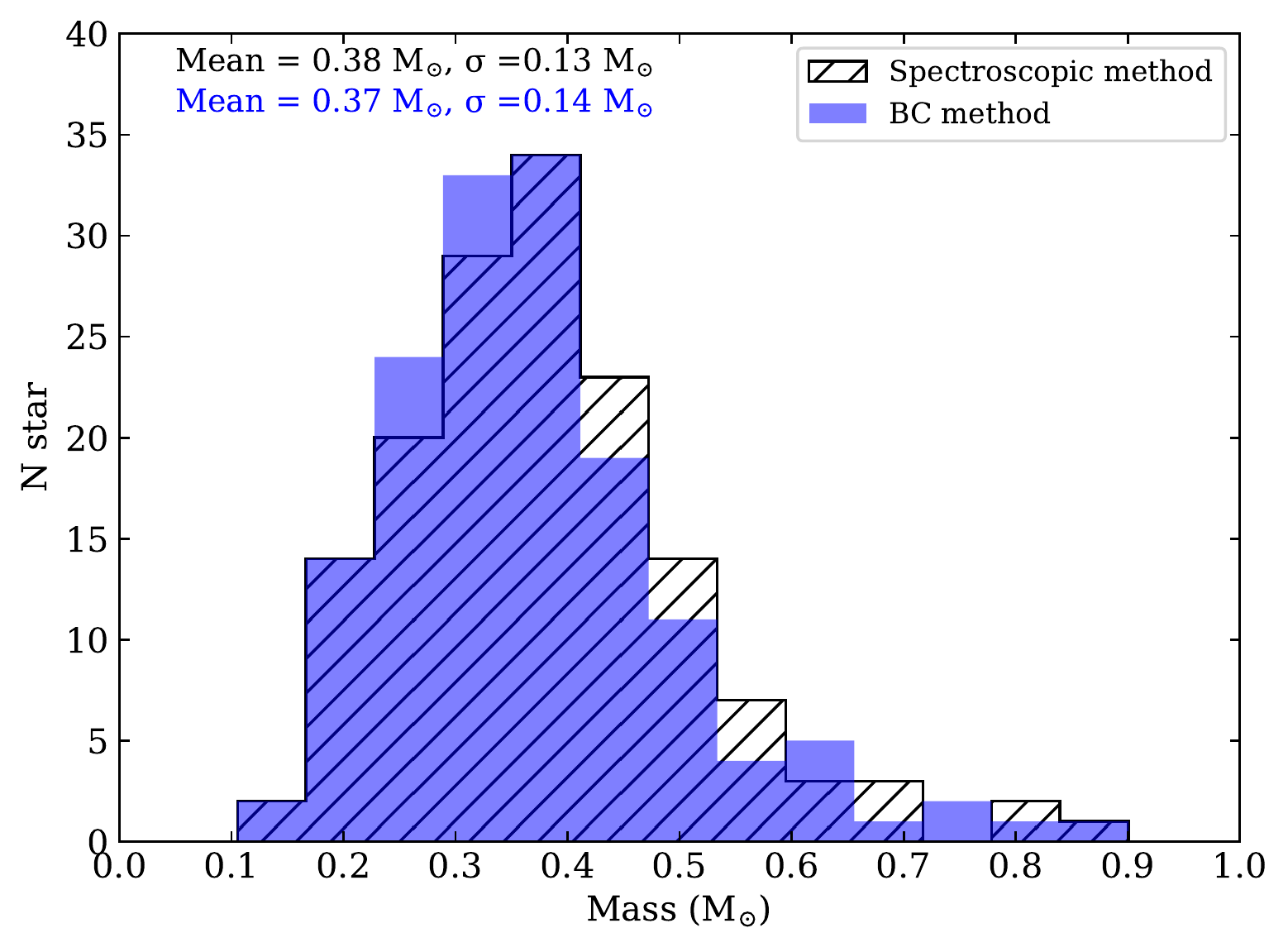}}
\caption{Distribution of the masses derived using the method described in this paper versus the method of \citet{moe17} using the bolometric correction. }
\label{dmasscomp}
\end{center}
\end{figure}

\end{appendix}

\end{document}